\documentclass{preprint}

\title{Generative Inversion of Spectroscopic Data for Amorphous Structure Elucidation}

\newcommand{\supptext}{\ref{sec:stext}}
\newcommand{\suppinfo}{Supplementary Information}

\newcommand{\methods}{\ref{sec:methods}}
\usepackage{tocloft}
\usepackage{etoc}

\author{
    Jiawei Guo,$^{1}$
    Daniel Schwalbe-Koda$^{1,*}$\\
    \vspace{1em}
    \normalfont{
        \small
        $^{1}$
        Department of Materials Science and Engineering, University of California, Los Angeles, CA, USA\\
        $^{*}$
        E-mail: dskoda@ucla.edu
    }
}

\newcommand{\gcm}{g/cm$^3$}

\usepackage[resetlabels]{multibib}
\newcites{Supp}{Supplementary References}
\newcites{Main}{References}

\begin{document}

\maketitle
\thispagestyle{firstpagestyle} 

\begin{abstract}
Determining atomistic structures from characterization data is one of the most common yet intricate problems in materials science.
Particularly in amorphous materials, proposing structures that balance realism and agreement with experiments requires expert guidance, good interatomic potentials, or both.
Here, we introduce GLASS, a generative framework that inverts multi-modal spectroscopic measurements into realistic atomistic structures without knowledge of the potential energy surface.
A score-based model learns a structural prior from low-fidelity data and samples out-of-distribution structures conditioned on differentiable spectral targets.
Reconstructions using pair distribution functions (PDFs), X-ray absorption spectroscopy, and diffraction measurements quantify the complementarity between spectral modalities and demonstrate that PDFs is the most informative probe for our framework.
We use GLASS to rationalize three contested experimental problems: paracrystallinity in amorphous silicon, a liquid-liquid phase transition in sulfur, and ball-milled amorphous ice.
In each case, generated structures reproduce experimental measurements and reveal mechanisms inaccessible to diffraction analysis alone.
\end{abstract}

\vspace{0.5cm}

Determining the three-dimensional atomic structure of materials from spectroscopic or diffraction measurements is among the oldest and most persistent inverse problems in materials science.\citeMain{Elliott1990Amorphous,Billinge2007Science}
For ordered materials, one century of progress in X-ray crystallography has enabled reliable and reasonably automated structure determination from diffraction data alone.\citeMain{kaduk2021Powder}
For amorphous materials, on the other hand, structure elucidation from experimental data remains challenging.
First, disorder smears spectroscopic and diffraction features, reducing the number of independent variables that can be inferred from the data.\citeMain{Billinge2007Science}
Consequently, even subtle spectral features at medium-range length scales become primary discriminators between competing structural models.\citeMain{Treacy1996ActaCryst,Voyles2001Nature,Sheng2006Nature}
Second, comparing simulated with experimental data is often difficult in amorphous materials.
For instance, molecular dynamics (MD) simulations, especially with machine learning interatomic potentials (MLIPs), sample physically realistic amorphous models at scale.\citeMain{deringer2021origins}
However, they do not guarantee reproduction of actual experimental spectra and rely on extensive development of potentials or simulation protocols.
On the other hand, reverse Monte Carlo (RMC)\citeMain{McGreevy1988RMC,McGreevy2001RMCReview} can generate structures consistent with experimental data, but requires extensive expert oversight, favors maximally disordered solutions, and often fails to capture local coordination or medium-range order.\citeMain{Cliffe2010RMC,Billinge2007Science}
Hybrid approaches such as empirical potential structure refinement\citeMain{Soper2001EPSR} or hybrid RMC\citeMain{opletal2008hrmc} partially address these limitations, but at high computational cost or relying on simplified potentials.
Third, meaningfully distinct structures can explain the same spectrum, making the inversion of spectral data an ill-posed problem.
Differentiable simulation and machine learning (ML)-based approaches have been applied to recover structures from diffraction data for nanocrystalline and disordered materials,\citeMain{Kwon2024SpectroscopyGuided,guo2025ab,anker2025autonomous} but struggle with solution uniqueness, particularly for mixed-phase systems or noisy data, and elevated computational cost.
Finally, combining spectroscopic modalities can resolve structural degeneracies (\supptext, Sec. \ref{sec:inversion-multi-modal}), but the resulting constraints on the solution space are often coupled and difficult to jointly invert.
Thus, no existing framework jointly inverts multi-modal spectroscopic data to recover amorphous or mixed-phase structures while ensuring physical plausibility without reliance on interatomic potentials or hand-crafted structural constraints.

Here, we introduce Generative Learning of Amorphous Structures from Spectra (GLASS), a framework that treats inversion of multi-modal spectroscopic data into three-dimensional atomistic structures as sampling from a physics-informed prior conditioned on differentiable simulations.
The framework learns a structural prior from low-fidelity data via a graph neural network (GNN)\citeMain{Kwon2024SpectroscopyGuided,yang2025generative} and generates structures based on six spectroscopic modalities: 2- and 3-body correlation functions, X-ray absorption near-edge spectroscopy (XANES), extended X-ray absorption fine structure (EXAFS), and X-ray or neutron diffraction.
Systematic tests across six modalities show that pair distribution functions (PDFs) are the most informative single probe for amorphous structure determination within GLASS, simultaneously constraining short- and medium-range order inaccessible to other individual modalities.
From these principles, we solve three challenging systems hindered by subtle interpretation of characterization data: paracrystallinity in amorphous silicon, a liquid–liquid phase transition in sulfur, and ball-milled amorphous ice.
In all cases, the generated structures rationalize the complex phenomena with statistical consistency, are validated against independent structural metrics, obey physical constraints, and reproduce the experimental spectra, all at minimal computational costs.
This demonstrates that generative modeling can automate the interpretation of spectroscopic data for amorphous materials, complementing simulations without compromising physical realism.

\section*{Generative elucidation of amorphous structures from differentiable spectroscopic modalities}

\begin{figure}[htb!]
   \centering
   \includegraphics[width=\textwidth]{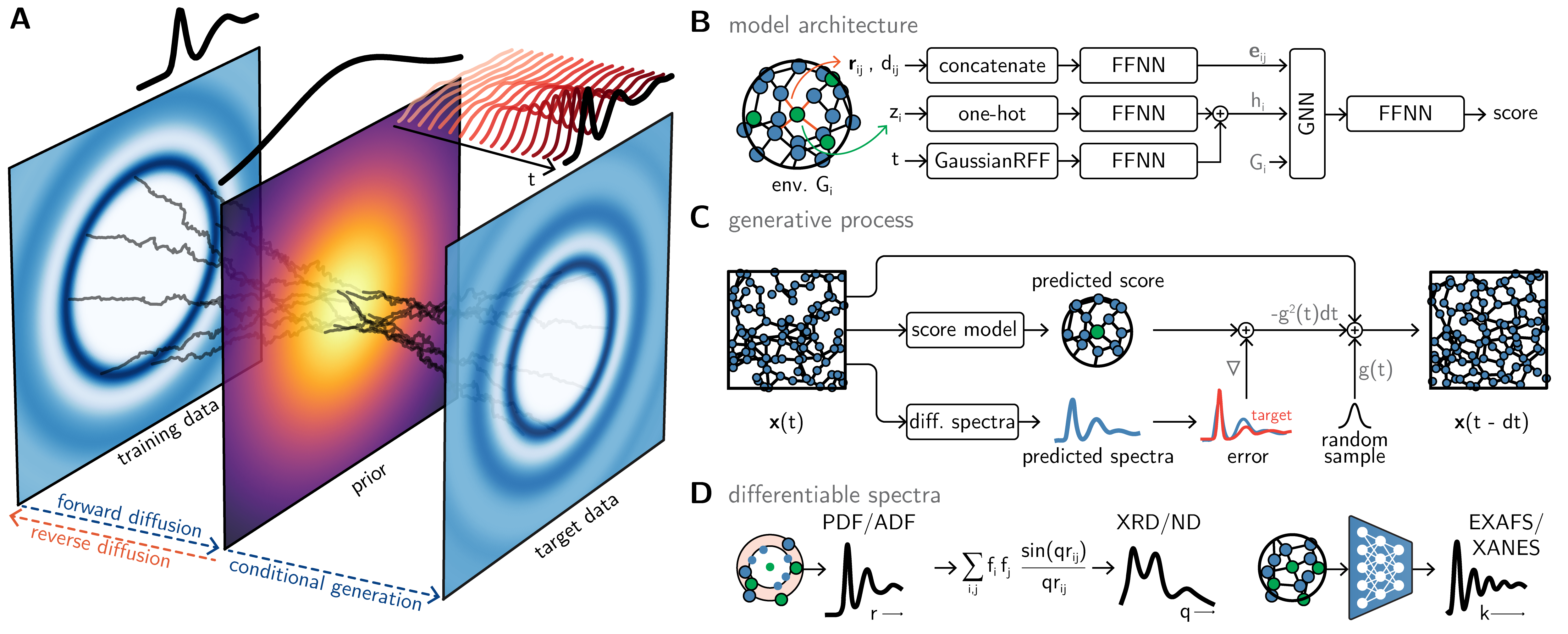}
   \caption{\textbf{Architecture of GLASS: Generative Learning of Amorphous Structures from Spectra.}
    \textbf{A.} Conceptual overview, where forward and backward diffusion map between training data and a prior. Conditional generation samples structures consistent with a given spectra by starting from the prior distribution.
    \textbf{B.} Score model used to learn the structural prior. Atomic environments $i$ are described by positions $\mathbf{r}_{ij}$ and distances $d_{ij}$ toward neighbors $j$ and atomic identity $z_i$.
    The time feature $t$ is encoded with Gaussian random Fourier features (RFF).
    Feed-forward neural networks (FFNN) and a graph neural network (GNN) are used to learn the final score.
    \textbf{C.} Conditional reverse-time denoising process. At each timestep, atomic coordinates are updated using a combination of the predicted score, gradients from differentiable spectroscopic observables, and stochastic noise, according to the score-based denoising approach.
    \textbf{D}, Differentiable spectroscopic modules enabling gradient-based conditional sampling.
    Real-space descriptors (PDF and ADF) are computed using smooth density estimators over two- and three-body terms.
    Diffraction observables (XRD and ND) are evaluated using differentiable Debye-style pair summations, and X-ray absorption spectra (XANES and EXAFS) are predicted using GNN-based surrogate models.
}
   \label{fig:overview}
\end{figure}

GLASS formulates amorphous structure determination as sampling from a learned structural prior conditioned on differentiable spectroscopic observables (Fig. \ref{fig:overview}A).
A generative model first learns the distribution of physically plausible atomic configurations with a stochastic differential equation implementing the score-matching denoising approach.\citeMain{song2021scorebased}
The structural prior is represented by a GNN that predicts a score for each atomic configuration (Fig. \ref{fig:overview}B).\citeMain{Kwon2024SpectroscopyGuided,yang2025generative}
This learned prior acts as a ``soft constraint'' for the data without hindering exploration of candidate structures, contrasting with hard constraints seen in RMC.\citeMain{yang2025generative}
Structure generation is implemented by reversing the diffusion process.
At each denoising step, the predicted score is combined with gradients from one or more differentiable spectroscopic modules and noise to update atomic coordinates (Fig. \ref{fig:overview}C).
This formulation separates the structural prior from the characterization data, with the score model regularizing sampling toward physically plausible configurations, while spectroscopic guidance drives the structure toward consistency with a specific measurement.
The relative weighting between prior and guidance controls this balance and demonstrates that neither score nor guidance can generate the targeted configurations without the other (Sec. \ref{sec:prior-guidance-init}).

\begin{figure}[htb!]
   \centering
   \includegraphics[width=0.5\textwidth]{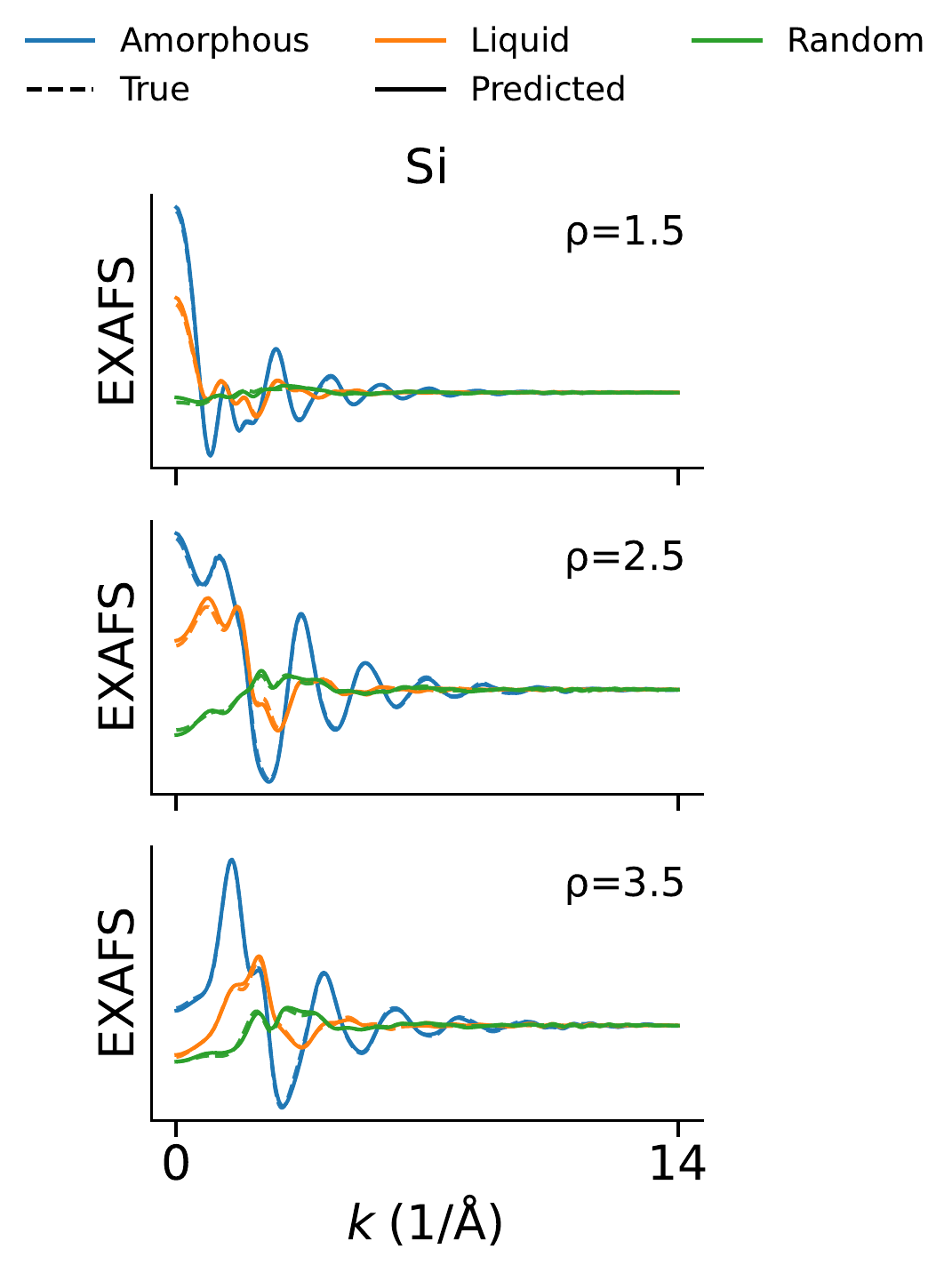}
   \caption{
    \textbf{Averaged EXAFS spectra for Si across densities and structural environments in the validation set.}
    Averaged EXAFS spectra $\chi(k)$ for Si at densities $\rho = 1.5$, $2.5$, and $3.5$ \gcm~ are shown, comparing neural network predictions against reference spectra. For each density, spectra are averaged over all atoms in the validation set belonging to a given structural class (amorphous, liquid, or randomly placed Si). Dashed lines denote the reference spectra; solid lines indicate the model predictions. The averages are computed over 582 validation atom-wise spectra at $1.5$ \gcm, 644 atom-wise spectra at $2.5$ \gcm, and 690 validation atom-wise-spectra at $3.5$ \gcm, with atoms randomly sampled from amorphous and liquid configurations generated by LAMMPS molecular dynamics simulations, as well as randomly placed Si structures. All predicted spectra are generated using a single model trained jointly across multiple densities and structural environments, including amorphous, liquid, and randomly initialized configurations.
    } \label{si:fig:01_Si_exafs_1.5_2.5_3.5}
\end{figure}

To enable gradient-based guidance, all spectroscopic observables are implemented as smooth, differentiable functions of atomic coordinates (Fig. \ref{fig:overview}D).
Real-space correlation functions, including PDFs and angular distribution functions (ADFs), constrain short-range order, while diffraction observables capture long-range correlations in reciprocal space.
Differentiable implementations of these observables reproduce the key spectral features of high-fidelity calculations while remaining computationally efficient, thus enabling evaluation of gradients during the denoising trajectory (Secs. \ref{sec:diff-spectral-scaling},\ref{sec:diff-spectra} Fig.~\ref{si:fig:profile_combined}).
X-ray absorption spectroscopy (XAS), on the other hand, depend on electronic structures and multiple-scattering pathways that cannot be captured by differentiable analytical equations.
We trained GNN surrogates to predict atom-resolved EXAFS and XANES spectra calculated with FEFF9 directly from local atomic environments (Secs. \ref{sec:xas-hifi},\ref{sec:diff-surrogate-models}, Figs. \ref{si:fig:01_Si_exafs_1.5_2.5_3.5},\ref{si:fig:01_Si_exafs_atomwise_best10_worst10_2.5}) using the same architecture shown in Fig. \ref{fig:overview}B.
Each surrogate is trained on datasets spanning amorphous, liquid, and random configurations to maximize transferability across the diverse environments sampled during denoising and replacing more expensive spectroscopic calculations.
Together, the GLASS framework and differentiable spectroscopic methods are agnostic to system, initialization, and provide multiple targets that can be connected to experimental outcomes.

\section*{Reliable generation of out-of-distribution amorphous structures}

\begin{figure}[htb!]
   \centering
   \includegraphics[width=0.7\textwidth]{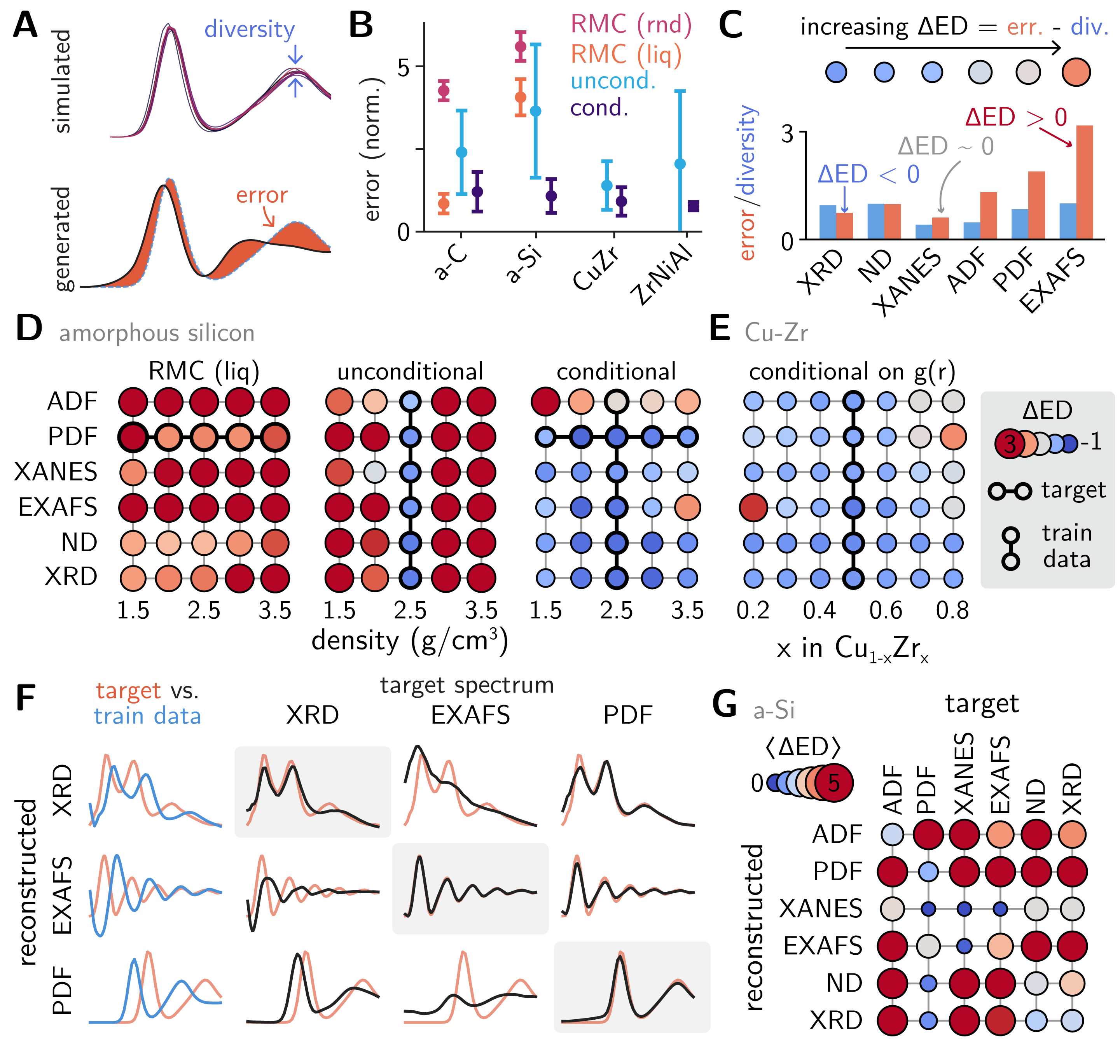}
   \caption{
   \textbf{Comparing reconstruction of amorphous structures with multi-modal spectroscopy.}
   \textbf{A.} Schematic illustration of the evaluation metrics. Spectral diversity quantifies the variability of spectra across independent replicas, while error measures the deviation between spectra of generated structures and the average reference spectrum.
   \textbf{B.} Average normalized spectral error for four amorphous systems and four methodologies: reverse Monte Carlo (RMC) initialized with random (rnd) or liquid (liq) structures, and GLASS (unconditional or conditional). Error bars indicate the standard deviation across spectral modalities and independent replicas.
   \textbf{C.} Schematic comparing the spectral error (orange) and reference diversity (blue) of a-Si across spectroscopic modalities. The difference $\Delta ED = \mathrm{error} - \mathrm{diversity}$ indicates whether generated spectra fall within the natural variability of the reference ensemble, and is represented with the color of the circle.
   \textbf{D.} Performance of RMC, unconditional, and conditional generation across densities and spectroscopic modalities for amorphous silicon (a-Si).
   Colors indicate $\Delta ED$ values, with red circles corresponding to worse reconstruction.
   \textbf{E.} Performance of conditional generation of Cu-Zr metallic glass across out-of-distribution compositions.
   \textbf{F.} Spectra of generated structures (black) obtained when different reference spectroscopic modalities (orange) are used as guidance (columns).
    Reference spectra of the training structures used for the score model are shown in blue.
   \textbf{G.} Average $\Delta ED$ of generated spectra for a-Si when conditioning on different spectroscopic observables.
   Benchmark datasets include amorphous carbon (a-C), amorphous silicon (a-Si), amorphous silica (SiO$_2$), Cu-Zr metallic glass, and Zr-Ni-Al alloys spanning variations in density, cooling rate, and composition (see Sec. \ref{sec:mol-dyn-sims} and  Figs.~\ref{si:fig:Si_C_all_features}--\ref{si:fig:comparison}).
   }
   \label{fig:benchmark}
\end{figure}

\begin{figure}[htb!]
   \centering
   \includegraphics[width=\columnwidth]{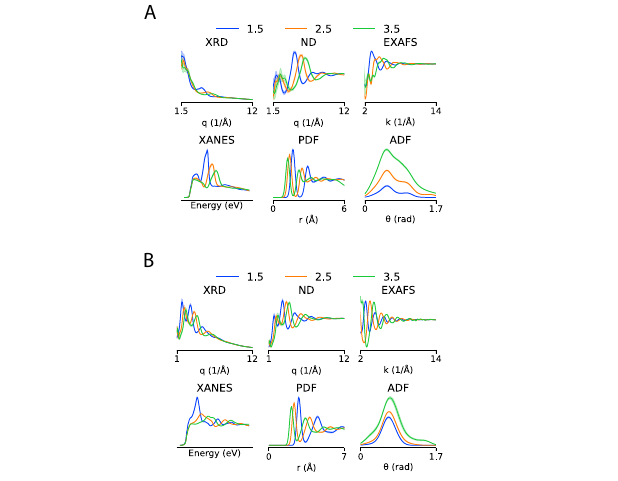}
   \caption{
    \textbf{Variability of reference spectroscopic signals with density in amorphous carbon and silicon.}
    \textbf{A.} Amorphous carbon (a-C) and \textbf{B.} amorphous silicon (a-Si).
    Representative spectra are shown for three densities (1.5, 2.5, and 3.5 \gcm), selected from the five densities used in the dataset, with 2.0 and 3.0 \gcm~ omitted for clarity. Solid lines denote the mean spectra averaged over 10 independent MD runs, while shaded regions indicate the corresponding standard deviations.}
   \label{si:fig:Si_C_all_features}
\end{figure}

\begin{figure}[htb!]
   \centering
   \includegraphics[width=\textwidth]{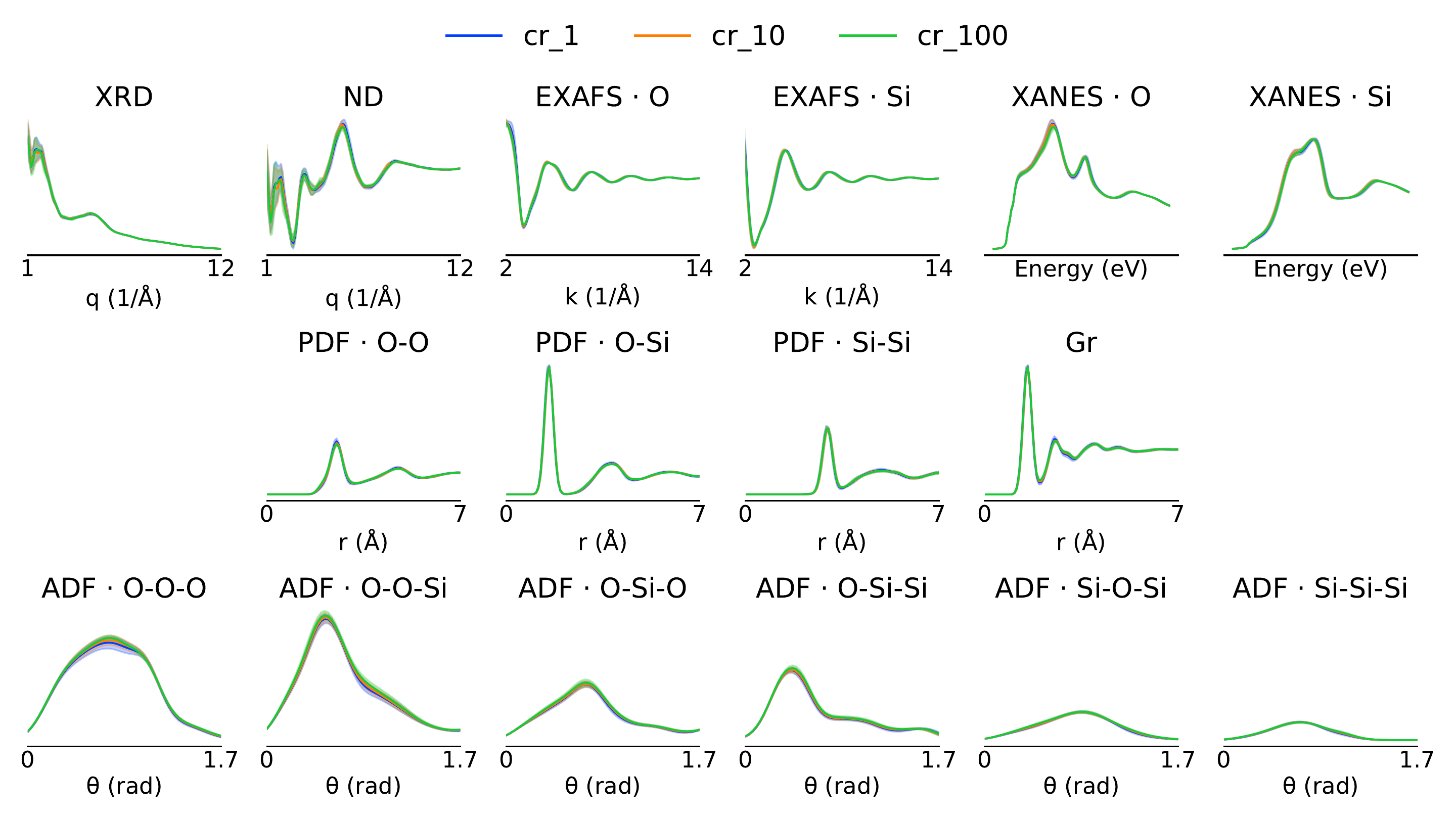}
   \caption{
    \textbf{Variability of reference spectroscopic signals with cooling rates in amorphous silica (a-SiO$_2$).}
    Representative spectra are shown for three cooling rates (1, 10, and 100 K\,ps$^{-1}$ shown in blue, orange, and green, respectively) from the reference dataset generated using LAMMPS molecular dynamics (MD) simulations. Solid lines denote the mean spectra averaged over 10 independent MD runs, while the shaded regions indicate the corresponding standard deviations. The spectra are highly similar across the different cooling rates, reflecting the relatively weak structural sensitivity to cooling rate in these simulations, with the small system size further contributing to the overlapping spectral signals.
   }
   \label{si:fig:SiO2_all_features}
\end{figure}

\begin{figure}[htb!]
   \centering
   \includegraphics[width=\textwidth]{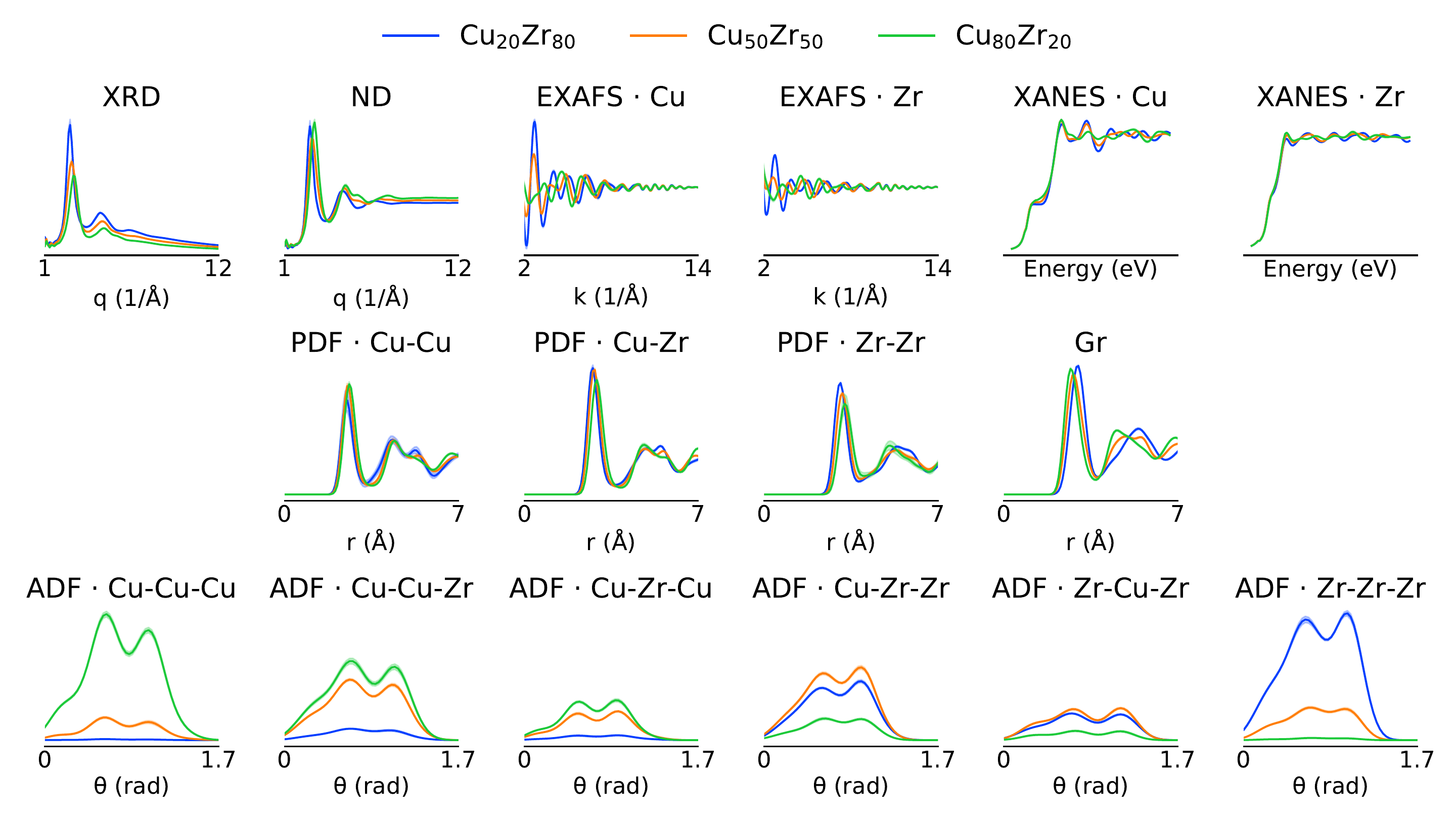}
   \caption{
   \textbf{Variability of reference spectroscopic signals with composition in amorphous Cu–Zr alloys.}
    Representative spectra are shown for three compositions (Cu$_{20}$Zr$_{80}$, Cu$_{50}$Zr$_{50}$, and Cu$_{80}$Zr$_{20}$), selected from the seven compositions included in the dataset. Solid lines denote the mean spectra averaged over 10 independent MD runs, while shaded regions indicate the corresponding standard deviations.
   } \label{si:fig:CuZr_all_features}
\end{figure}

\begin{figure}[htb!]
   \centering
   \includegraphics[width=\textwidth]{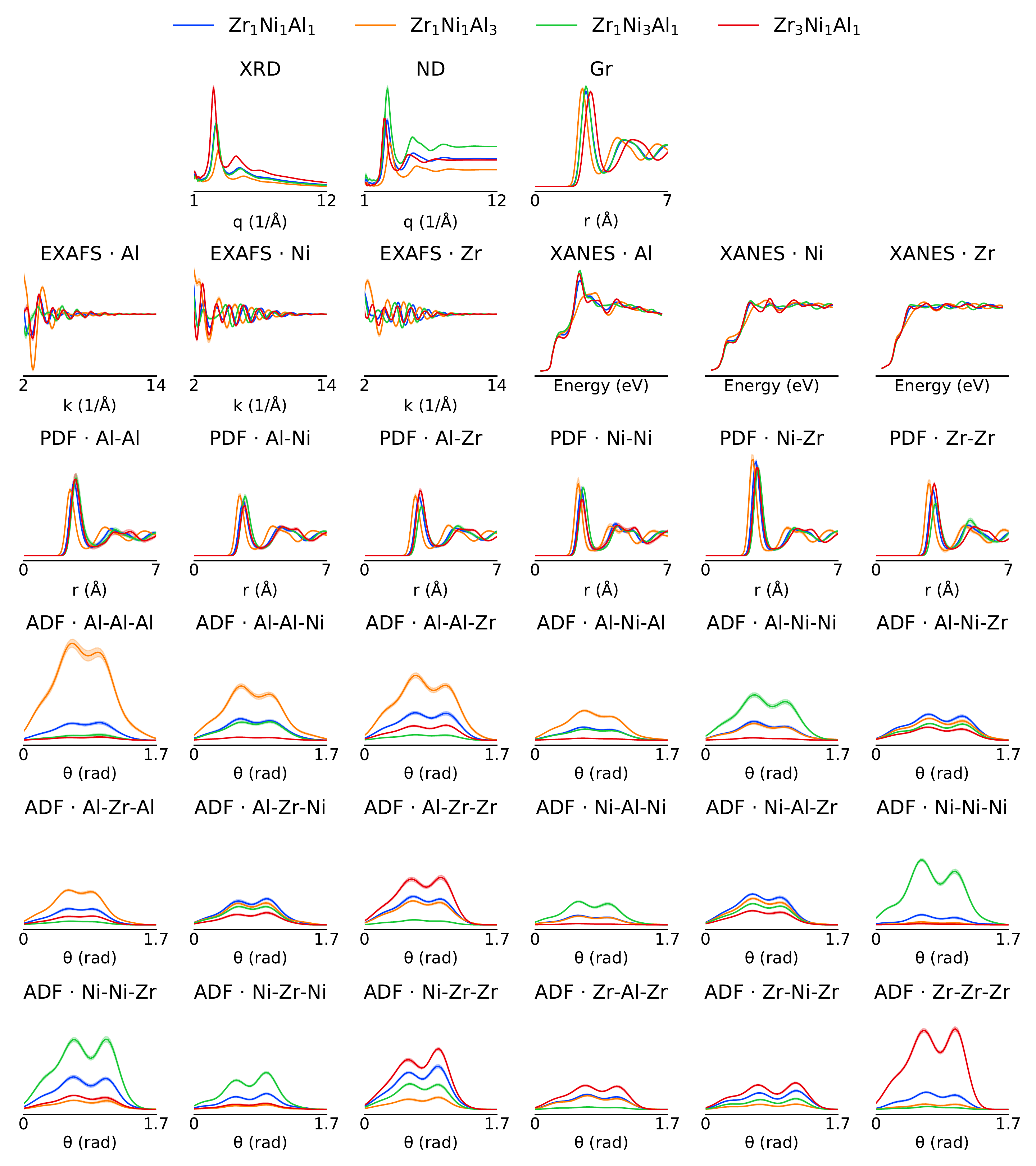}
   \caption{\textbf{Variability of reference spectroscopic signals with composition in amorphous Zr–Ni–Al alloys.}
    Representative spectra are shown for four compositions (Zr$_1$Ni$_1$Al$_1$, Zr$_1$Ni$_1$Al$_3$, Zr$_1$Ni$_3$Al$_1$, and Zr$_3$Ni$_1$Al$_1$), selected from 25 total compositions generated by varying the stoichiometric ratio of each element from 1 to 3. Compositions that are simple multiples of one another (e.g., 1:1:1, 2:2:2, and 3:3:3) are considered equivalent and therefore counted only once. Solid lines denote the mean spectra averaged over 10 independent MD runs, while shaded regions indicate the corresponding standard deviations.} \label{si:fig:ZrNiAl_all_features}
\end{figure}

\begin{figure}[htb!]
   \centering
    \includegraphics[width=0.8\textwidth]
    {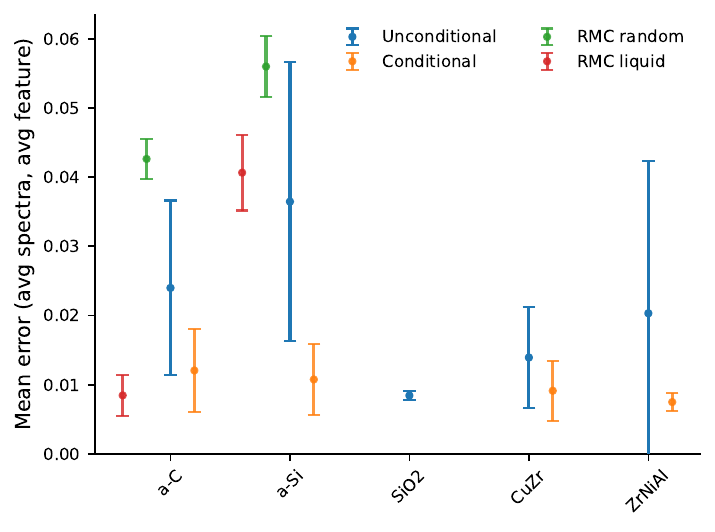}
    \caption{\textbf{Comparison of spectral reconstruction errors across amorphous materials, including SiO$_2$.}
    Average (normalized) spectral error of reverse Monte Carlo (RMC) and our generative model in reproducing the structure of five characteristic amorphous materials: a-C, a-Si, SiO$_2$, CuZr, and ZrNiAl. Error bars represent the standard deviation of errors aggregated across all spectral modalities and structural variations in the reference datasets, including 5 different  densities for a-C and a-Si, 3 different cooling rates for SiO$_2$, and different alloy compositions for CuZr (7 compositions) and ZrNiAl (25 compositions).
    For SiO$_2$, only unconditional denoising results are shown because the unconditional model already achieves consistently low errors with small variance across both spectral observables and cooling rates. This indicates that the spectroscopic signals themselves do not strongly distinguish between different cooling rates, and therefore additional conditional denoising provides little benefit in this case.
    }
\label{si:fig:comparison}
\end{figure}

A central requirement for any automatic structure-determination framework is the ability to generate accurate out-of-distribution samples, as experimental structures cannot be expected to match simulated ones.\citeMain{Elliott1991Amorphous,Zallen1983AmorphousStructure}
To evaluate this capability, we constructed a benchmark dataset using molecular dynamics (MD) simulations.
Ensembles of independently simulated amorphous structures were produced across five materials --- a-C, a-Si, a-SiO$_2$, Cu-Zr, and Zr-Ni-Al --- spanning systematic variations in density, cooling rate, and composition for archetypical amorphous systems (Sec. \ref{sec:mol-dyn-sims}, Figs. \ref{si:fig:Si_C_all_features}--\ref{si:fig:comparison}).
For each configuration, six spectroscopic modalities (PDF, ADF, XRD, ND, XANES, EXAFS) were computed using high-fidelity reference codes (Sec. \ref{sec:xas-hifi}).
From this dataset we define two complementary metrics (Fig. \ref{fig:benchmark}A):
reference diversity, the spectral variability across replicas sharing identical composition and processing parameters;
and generation error, quantified as the deviation between the spectrum of a generated structure and the average of the reference ensemble spectra.

\begin{figure}[htb!]
   \centering
    \includegraphics[width=0.7\textwidth]
    {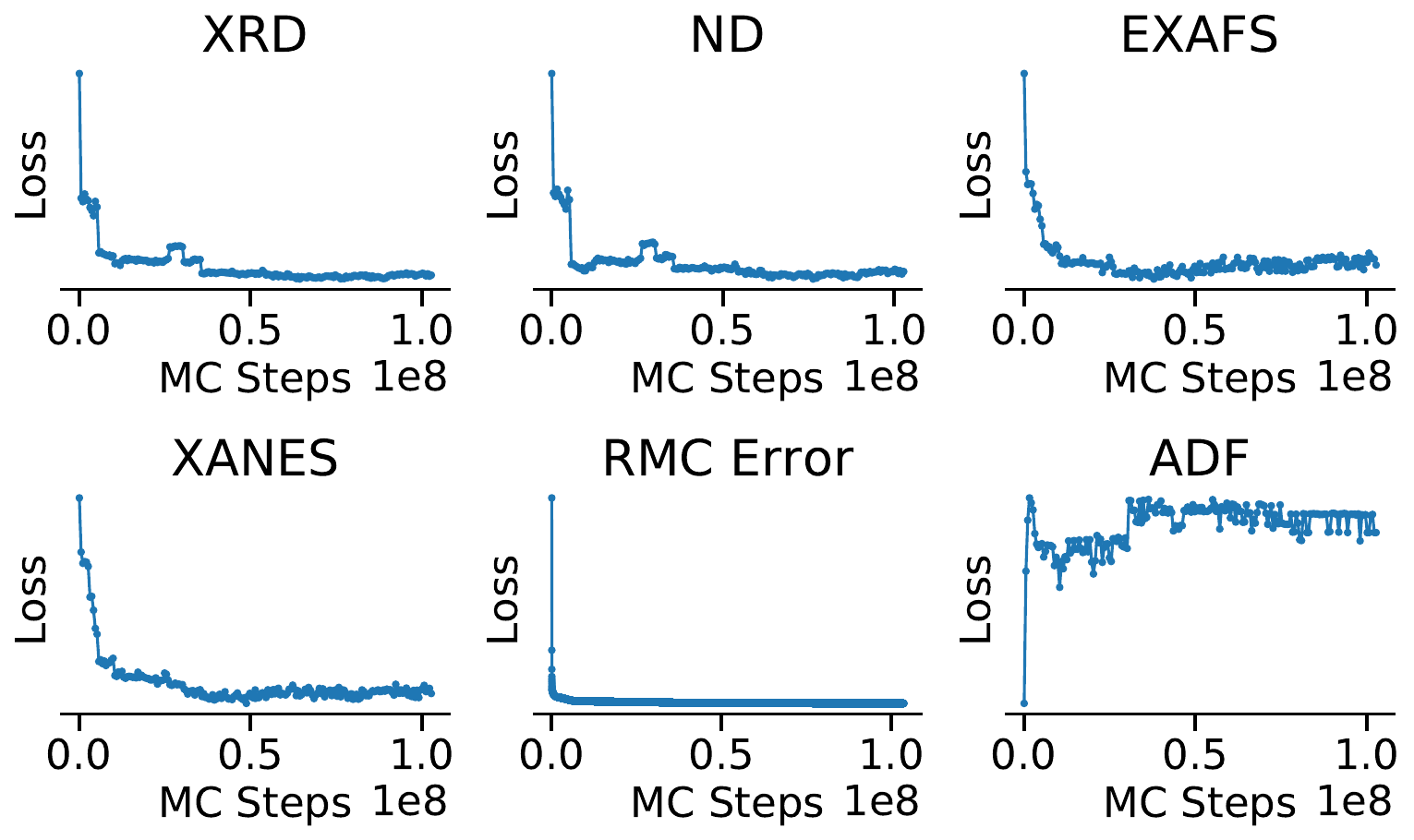}
    \caption{
    \textbf{PDF-guided RMC loss as a function of Monte Carlo atom steps.}
    A total of $10^8$ (104-million) atom-steps were performed. The figure shows the evolution of multiple spectroscopic losses for amorphous Si at 1.5 g/cm$^3$, one of the most challenging densities to recover.
    The \textbf{RMC error} corresponds to the PDF loss minimized during refinement using the \texttt{FULLRMC} package, and is distinct from the PDF curves reported elsewhere, which are computed using our differentiable framework.
    Within the $10^8$ steps, the first half corresponds to the exploration stage, followed by a refinement stage.
    The simulation is initialized from a liquid-derived structure, which is already closer to the target than random initialization.
    }
\label{si:fig:loss_vs_atomstep_with_rmc}
\end{figure}

\begin{figure}[htb!]
   \centering
   \includegraphics[width=0.8\textwidth]{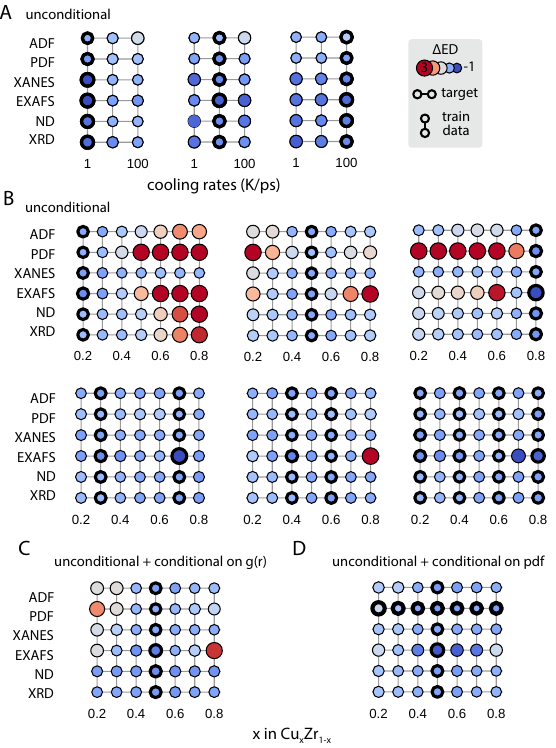}
   \caption{
    \textbf{Reconstruction performance across cooling rates and alloy compositions.}
    \textbf{A.} Performance of the unconditional generative model for amorphous silica (a-SiO$_2$) across different cooling rates.
    \textbf{B–D.} Reconstruction performance for Cu–Zr metallic glass across compositions $x$ in Cu$_x$Zr$_{1-x}$.
    \textbf{B.} Unconditional generative models with progressively larger model sizes.
    \textbf{C.} Two-step denoising strategy in which structures are first denoised using the unconditional model and then refined by conditional guidance using the total pair correlation function $g(r)$.
    \textbf{D.} Two-step denoising with guidance from partial pair distribution functions (PDF).
    Rows correspond to spectroscopic observables. Circle color encodes the reconstruction score $\Delta ED$, where progressively red circles indicate larger values and thus worse agreement with the reference spectra. Columns outlined in black denote conditions included in the training dataset.
    }
   \label{si:fig:SiO2_CuZr_heatmaps}
\end{figure}

\begin{figure}[htb!]
   \centering
    \includegraphics[width=0.8\textwidth]{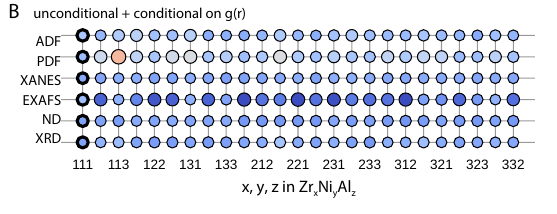}
   \caption{
    \textbf{Cross-composition reconstruction performance for amorphous Zr–Ni–Al alloys.}
    \textbf{A.} Performance of the unconditional generative model across compositions $x,y,z$ in Zr$_x$Ni$_y$Al$_z$.
    \textbf{B.} Performance of the two-step denoising strategy, where randomly initialized structures are first denoised using the unconditional model and then refined through conditional guidance using the total pair distribution function $g(r)$.
    Rows correspond to spectroscopic observables, while columns represent alloy compositions generated by varying $x,y,z \in \{1,2,3\}$ with equivalent compositions (e.g., 111, 222, 333) treated as identical and therefore omitted.
    Circle color encodes the reconstruction score $\Delta ED$, where progressively red circles indicate larger values and thus worse agreement with the reference spectra. Columns outlined in black denote compositions included in the training dataset.
    }
   \label{si:fig:ZrNiAl_heatmaps}
\end{figure}

As a baseline, we evaluated RMC reconstruction for a-C (Fig. \ref{si:fig:rmc_C}) and a-Si (Figs. \ref{fig:benchmark}B, \ref{si:fig:rmc_Si}), which are already non-trivial to reconstruct with this method.\citeMain{biswas2004reverse}
In both a-C and a-Si, RMC captures overall spectral trends, but exhibits systematically higher errors and strong sensitivity to initialization (Sec. \ref{sec:rmc-init-sensitivity}, Fig. \ref{si:fig:pdf_compare_Si_C}).
When initialized from random configurations, RMC-generated structures failed to converge towards low-error solutions even with a large number of steps (Fig. \ref{fig:benchmark}B, \ref{si:fig:loss_vs_atomstep_with_rmc}).
Even when initialized from liquid structures and selecting the best result across multiple runs, performance remained inconsistent, converging for a-C but not for a-Si.
Whereas methodological improvements over the baseline RMC performance can improve convergence and structural models,\citeMain{cliffe2017structural} they also rely on additional, expert-driven work that tends to be system-dependent, hindering automation.
Unconditional generation within GLASS is reasonable for Cu-Zr, SiO$_2$, and Zr-Ni-Al (Figs. \ref{fig:benchmark}B,\ref{si:fig:SiO2_CuZr_heatmaps},\ref{si:fig:ZrNiAl_heatmaps}), but lead to high reconstruction errors for a-Si and a-C.
In contrast, the conditional generation remains robust even when initialized from random structures (Fig. \ref{fig:benchmark}B), despite being trained on a small prior dataset consisting of a single ensemble of configurations, bypassing the need for expert-driven rules at structure generation.

\begin{figure}[htb!]
   \centering
   \includegraphics[width=0.757\textwidth]{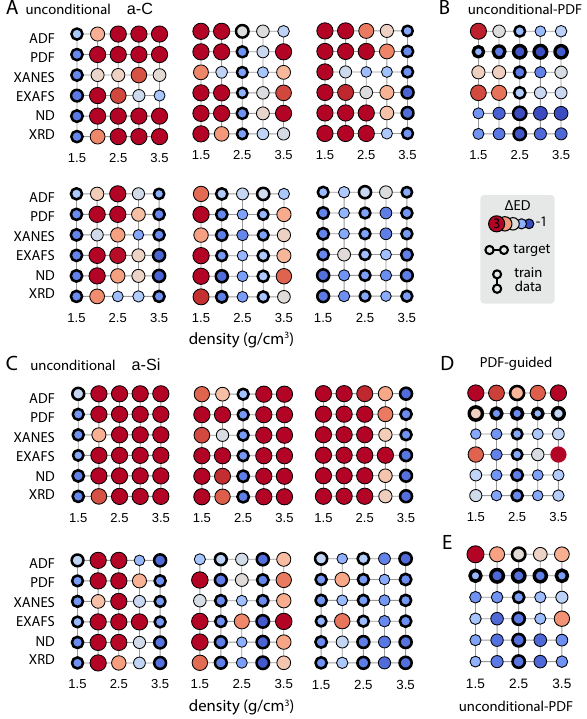}
   \caption{
    \textbf{Cross-density reconstruction performance for amorphous carbon and silicon.}
    \textbf{A.} Unconditional generative model performance for amorphous carbon (a-C).
    \textbf{B.} Best two-stage denoising strategy for a-C, where randomly placed atoms are first denoised with the unconditional model to recover training-like amorphous behavior, followed by PDF-guided conditional denoising at lower noise to reach the target density.
    \textbf{C.} Unconditional generative model performance for amorphous silicon (a-Si). \textbf{D.} Direct PDF-guided denoising starting from the initial random structure, without the unconditional pre-denoising step, resulting in consistently poorer reconstructions. \textbf{E.} Two-stage denoising strategy applied to a-Si, analogous to panel \textbf{B}.
    Rows correspond to spectroscopic observables, and columns indicate the target density used during denoising. Columns highlighted in black denote densities included in the training dataset. Circle color encodes the reconstruction score $\Delta ED$, where progressively red circles indicate larger values and thus worse agreement with the reference spectra.
    }
   \label{si:fig:Si_C_heatmaps}
\end{figure}

\begin{figure}[htb!]
   \centering
    \includegraphics[width=\textwidth] {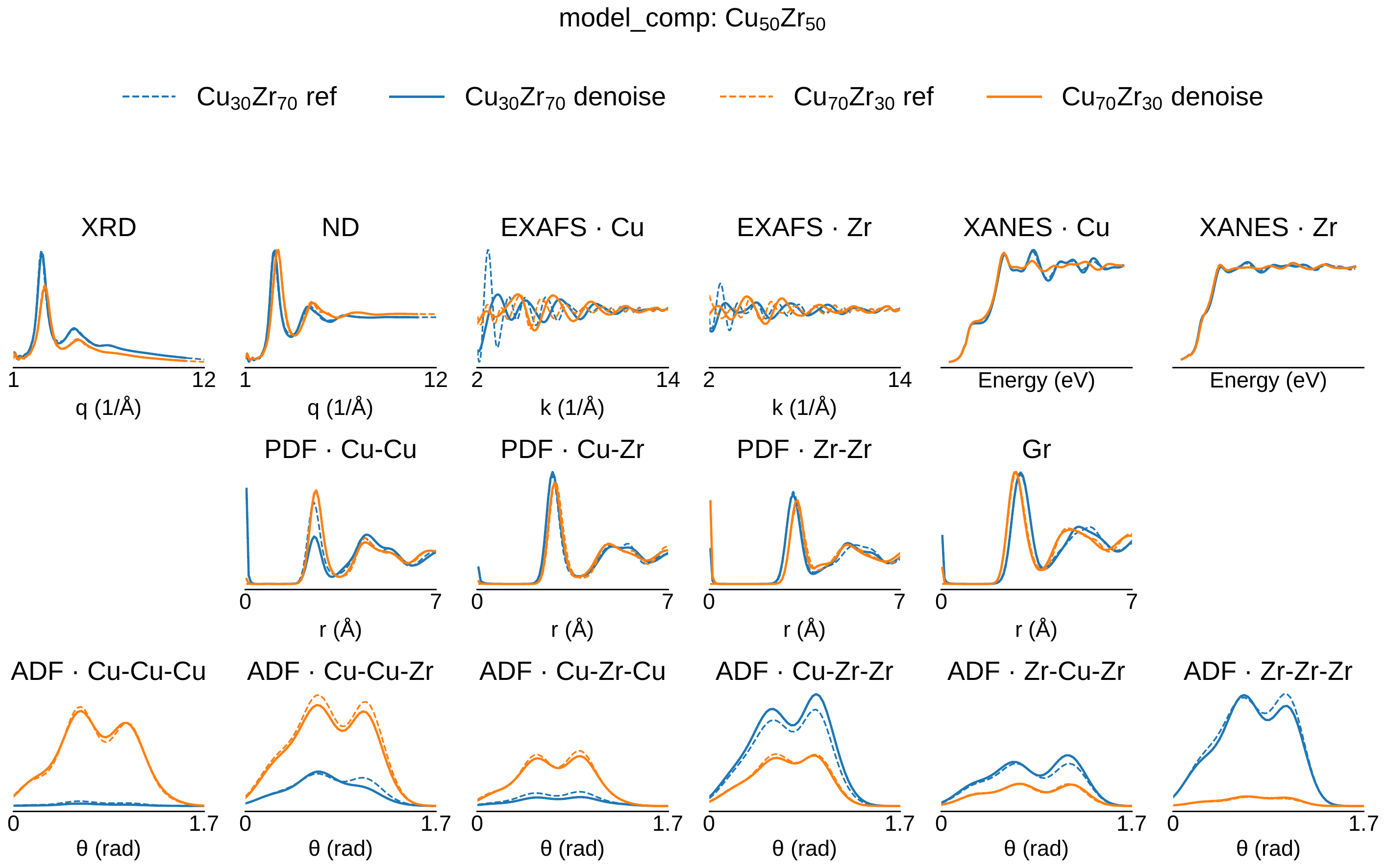}
    \caption{\textbf{Spectroscopic features of unconditional denoised Cu--Zr structures relative to reference signals.}
    The score model was trained only on the Cu$_{50}$Zr$_{50}$ composition and used without spectroscopic guidance to generate amorphous structures at Cu$_{30}$Zr$_{70}$ and Cu$_{70}$Zr$_{30}$. Despite the model being trained on a single composition, the generated structures reproduce several qualitative structural features of the target systems, with notable overlap observed in the partial pair distribution functions (PDFs).}

\label{si:fig:CuZr_unconditional_CuZr_7.4-50_specs}
\end{figure}

\begin{figure}[htb!]
   \centering
    \includegraphics[width=\textwidth]
    {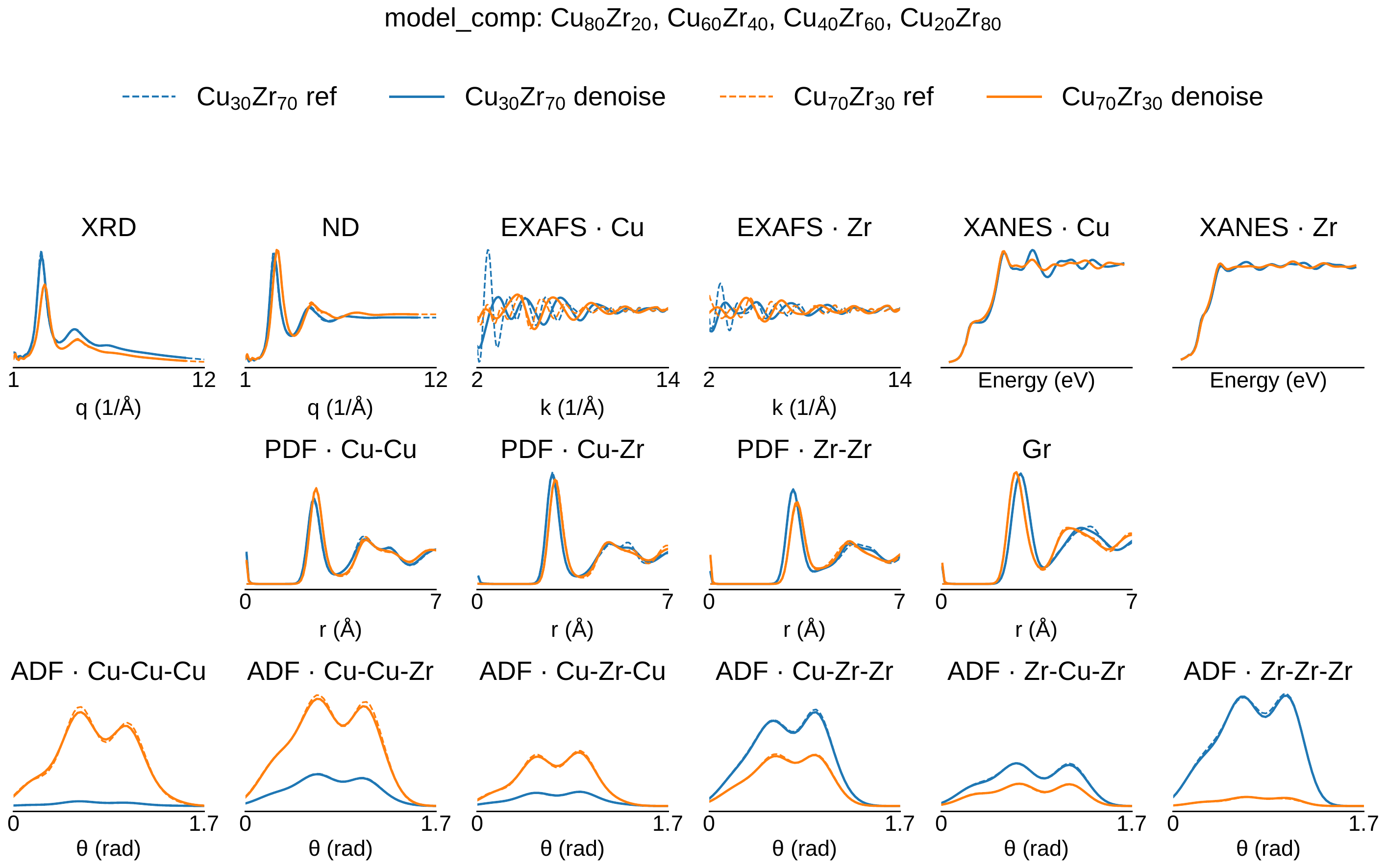}
    \caption{\textbf{Spectroscopic features of unconditional denoised Cu--Zr structures from a multi-composition prior.}
    The score model was trained on multiple Cu--Zr compositions (Cu$_{80}$Zr$_{20}$, Cu$_{60}$Zr$_{40}$, Cu$_{40}$Zr$_{60}$, and Cu$_{20}$Zr$_{80}$ and used without spectroscopic guidance to generate amorphous structures at Cu$_{30}$Zr$_{70}$ and Cu$_{70}$Zr$_{30}$. Despite the substantially larger training set, the generated spectra exhibit less overlap with the reference signals compared with the single-composition prior.}

\label{si:fig:CuZr_unconditional_CuZr_7.4-80_60_40_20_specs}
\end{figure}

\begin{figure}[htb!]
   \centering
    \includegraphics[width=\textwidth]
    {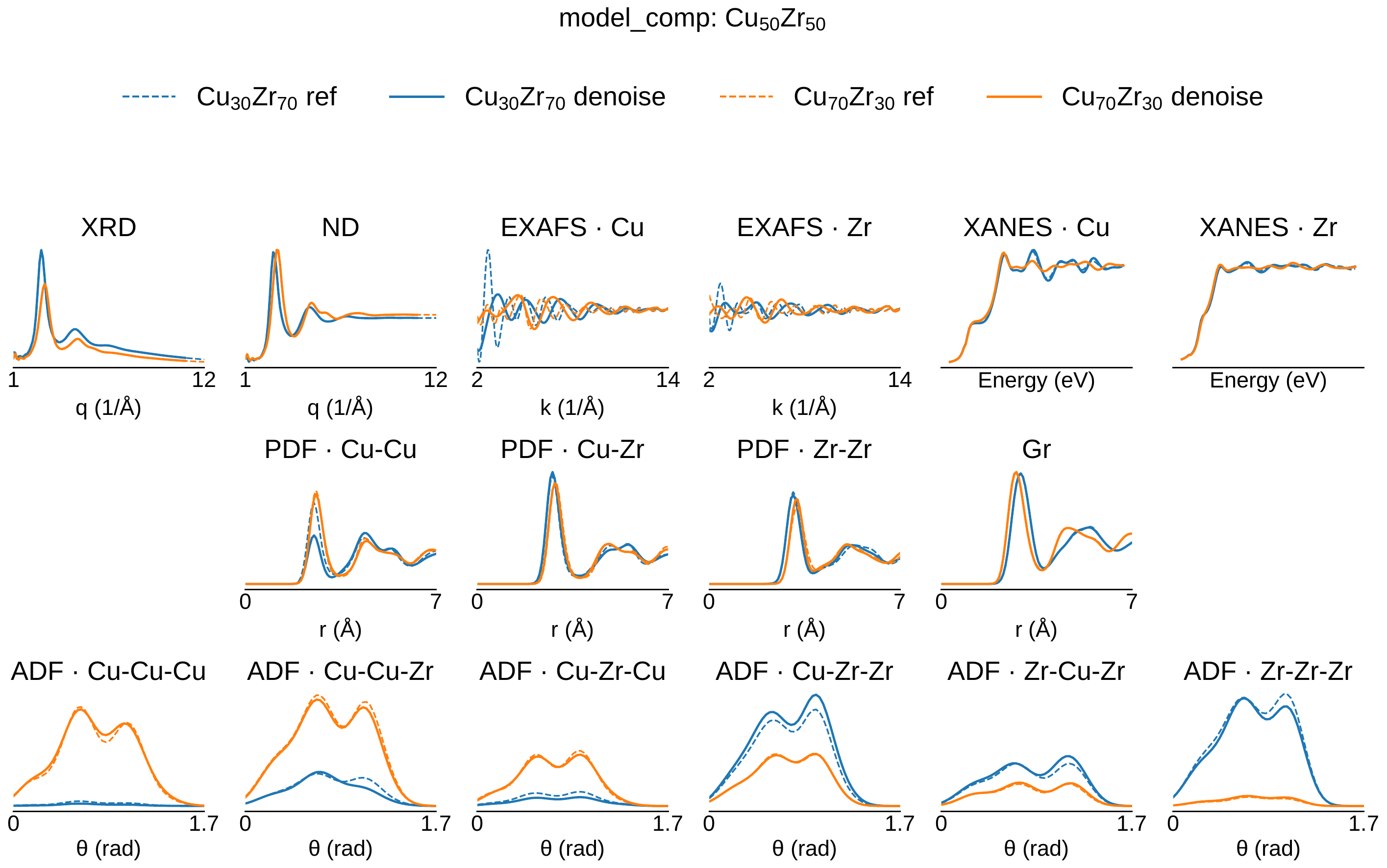}
    \caption{\textbf{Spectroscopic features of $g(r)$-guided Cu--Zr structure generation.}
    The score model was trained on Cu$_{50}$Zr$_{50}$ and used to generate structures at Cu$_{30}$Zr$_{70}$ and Cu$_{70}$Zr$_{30}$ with total PDF $g(r)$ guidance during denoising. In contrast to unconditional generation, the guided structures show stronger agreement with the reference signals, and the previously observed spectral overlaps disappear even when using the smallest prior model. Nevertheless, small variations in EXAFS remain noticeable even after the near-perfect fits.
    }

\label{si:fig:CuZr_unconditional_Gr_t-0.4_CuZr_7.4-50_specs}
\end{figure}

To evaluate reconstruction quality of GLASS against absolute metrics, we compared generated spectral errors against the intrinsic diversity of the reference ensemble within the same units (Fig.~\ref{fig:benchmark}C).
The difference $\Delta ED$ = error - diversity determines whether a generated structure falls within the natural variability of the reference ensemble ($\Delta ED \leq 0$) or deviates from the expected signals ($\Delta ED > 0$) (Sec. \ref{sec:error-diversity-metrics}).
Figure \ref{fig:benchmark}D summarizes the performance of each generation strategy across densities and spectral modalities for a-Si (see Figs. \ref{si:fig:ZrNiAl_heatmaps},\ref{si:fig:Si_C_heatmaps} for the complete benchmark), and extends the insights of Fig. \ref{fig:benchmark}B to an out-of-distribution analysis.
RMC consistently produces structures with large spectral errors across all modalities even when initialized from liquid configurations.
Similar results have also been reported for metallic glass reconstruction.\citeMain{liu2023reliability}
Unconditional sampling from a prior model trained at 2.5 \gcm~ generates reasonable structures only at the training density but fails to generalize across other densities.
In contrast, conditional denoising with PDF guidance within GLASS reproduces structures across all tested densities, including, to an extent, ADFs that capture 3-body correlation functions, even though the score model was trained at a single density.
These results also extend to multi-element metallic glasses, where unconditional generation is simpler, but unable to generate reasonable structures in out-of-distribution compositions (Sec. \ref{sec:phys-constraints-score}, Figs. \ref{si:fig:CuZr_unconditional_CuZr_7.4-50_specs}--\ref{si:fig:CuZr_unconditional_Gr_t-0.4_CuZr_7.4-50_specs}).
On the other hand, conditional generation exhibits excellent reconstruction performance even without interatomic potentials (Fig. \ref{fig:benchmark}E).
A complete discussion of these results is found in the \suppinfo.

\section*{Completeness of spectroscopic techniques for structural reconstruction}

\begin{figure}[htb!]
   \centering
    \includegraphics[width=0.6\textwidth]
    {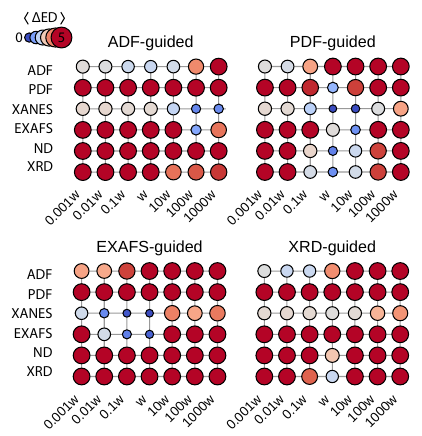}
    \caption{\textbf{Effect of prior–guidance weighting on conditional denoising.}
    Heatmap summary of reconstruction performance as a function of the weighting parameter $w$, which controls the balance between the structural prior (score model) and the spectroscopic guidance term during denoising (\methods). Smaller weights (moving left) correspond to stronger reliance on the score-model. Because the prior was trained on a limited dataset and only at 2.5 g/cm$^3$, excessive reliance on the prior leads to poor agreement with the target spectra when reconstructing structures at 1.5 g/cm$^3$. Conversely, larger weights (moving right) emphasize the guidance loss and can drive the optimization outside the physically reasonable configuration space learned by the prior, leading to unphysical structures and degraded agreement across other observables due to overfitting. Optimal performance occurs at intermediate weights, where the prior constrains the structural manifold while the guidance term steers the configuration toward agreement with the target spectrum. This figure presents the full weighting benchmark across multiple spectroscopic guidance signals (ADF, PDF, EXAFS, and XRD). The plots are identical to those shown in Fig. \ref{fig:benchmark}B of the main text, except that the main text displays only the optimal weighting for each guidance signal, whereas here the full sweep over $w$ is shown.
    All benchmarks were performed on amorphous Si, where the score model was trained exclusively on structures at 2.5 g/cm$^3$ and evaluated on reconstruction of amorphous structures at 1.5 g/cm$^3$.
    Selection of $w$ can be performed in a self-consistent way, where a good value of $w$ minimizes reconstruction error for in-distribution validation data.
    }
    \label{si:fig:rho_bench}
\end{figure}

Multi-modal spectroscopy is routinely employed to resolve structural ambiguities in materials characterization.
Though techniques vary broadly in experimental cost and accessibility, their relative information content for structure reconstruction has not yet been determined.
To address this gap, we generated structures conditioned on each of the six spectroscopic modalities considered here, and later evaluated cross-modal reconstruction quality.
Figure \ref{fig:benchmark}F illustrates spectra for GLASS-generated and reference a-Si structures when three different modalities are used as targets.
Following our out-of-distribution analysis (Fig. \ref{fig:benchmark}D), the score model was trained on amorphous structures at a density of 2.5 \gcm and evaluated at a density of 1.5 \gcm.
When XRD or EXAFS are used as guidance, the generated structures reproduce their target spectrum with near-perfect agreement but fail to match the remaining modalities.
In contrast, when PDF is used as guidance, both EXAFS and XRD are accurately recovered without explicitly being used in the generation stage.
Figure \ref{fig:benchmark}G further shows that, across systems, densities, and multiple runs, PDF-guided structure generation outperforms all other modalities in structural quality (full results in Fig. \ref{si:fig:rho_bench}).
On the other hand, EXAFS or XANES guidance\citeMain{carbone2020machine,Kwon2024SpectroscopyGuided} fails to produce structures that recover the other modalities, which is not an artifact of the NN nor $k$-space guidance (Sec. \ref{sec:spectral-comp}).
Moreover, the results recover expected redundancies within spectroscopic pairs: EXAFS and XANES share local structural and electronic sensitivity, such that reproducing one generally yields the other, and similarly for ND and XRD.

These results quantify two findings:
agreement in one spectral modality does not guarantee consistency with others; and multi-modality does not immediately imply orthogonality between characterization methods (Sec. \ref{sec:spectral-comp}).
Cross-modal transferability instead emerges from the interplay between the guidance signal and the learned structural prior that restricts sampling to physically plausible configurations.
Without such a prior, matching one observable does not ensure agreement with others, as seen in RMC reconstructions that reproduce PDFs while violating unconstrained observables (Sec. \ref{sec:rmc-init-sensitivity}, Figs. \ref{si:fig:rmc_C},\ref{si:fig:rmc_Si}).
Furthermore, when only one measurement is available, PDFs are the most informative for conditional generation within GLASS; when multiple measurements can be acquired, diffraction and absorption data provide complementary information, but are insufficient to automatically reconstruct structures without real-space pair correlation functions.

\section*{Paracrystallinity in amorphous silicon generated from experimental data}

\begin{figure}[htb!]
   \centering
   \includegraphics[width=0.33\textwidth]{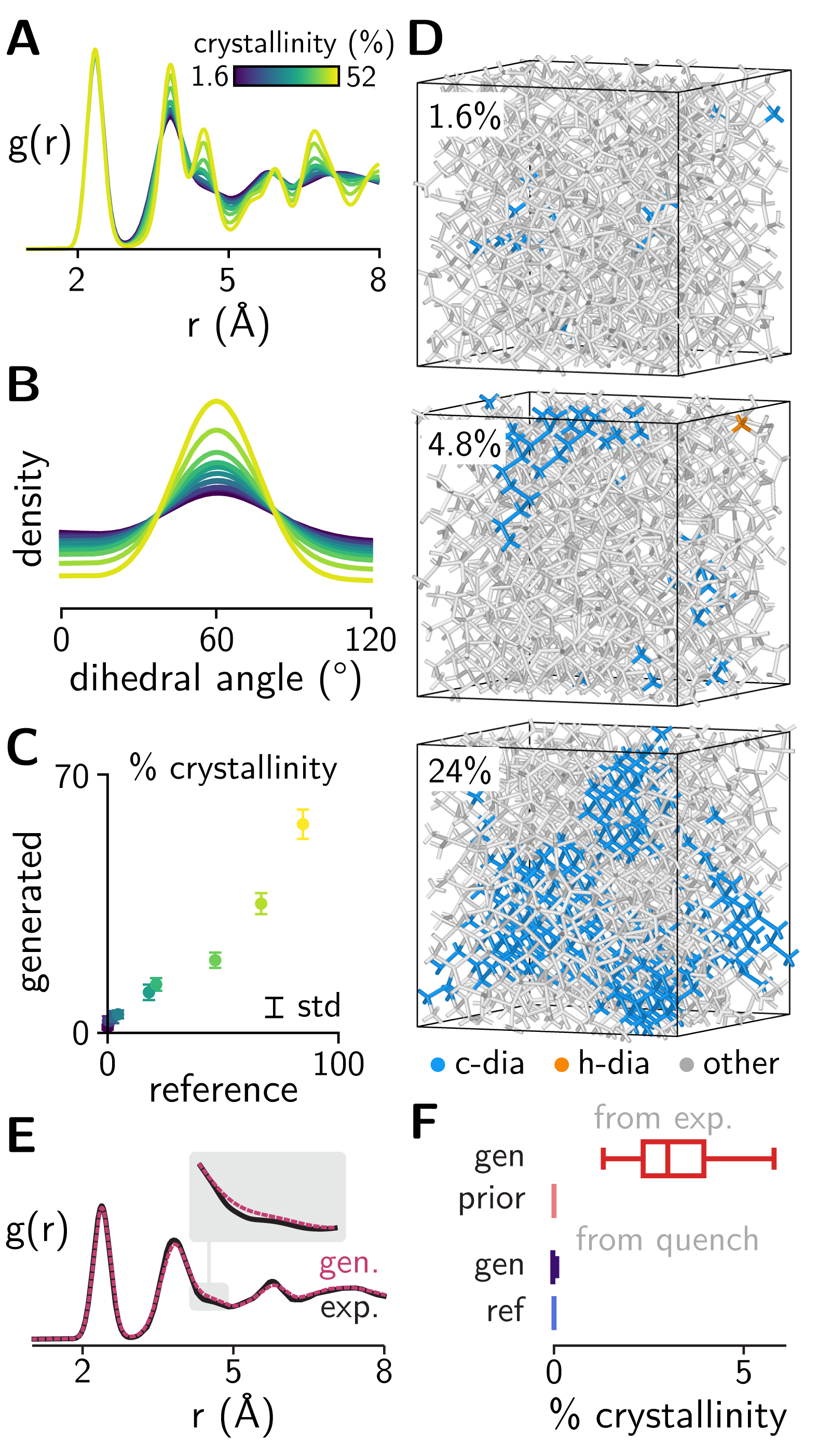}
   \caption{
    \textbf{Generating amorphous-crystalline mixtures in silicon via PDF-guided denoising.}
    \textbf{A.} Pair distribution functions $g(r)$ of mixed crystalline-amorphous configurations as a function of increasing crystallinity (colors).
    \textbf{B.} Dihedral-angle distributions of generated structures from different crystallinity (colors), which are not used in GLASS and provide partial validation on the structures.
    \textbf{C.} Crystallinity of generated structures correctly capture the trends of reference values despite underpredictions. Each point represents the mean over independent denoising runs and error bars indicate the standard deviation.
    \textbf{D.} Representative generated structures at different crystallinity levels.
    Atoms are colored according to local environment: cubic diamond (blue), hexagonal diamond (orange), and other environments (gray).
    \textbf{E.} GLASS-generated structures (red) reproduce the experimental $g(r)$ (black) used as target, including subtle, yet important variations at medium-range (inset).
    \textbf{F.} Crystallinity in samples generated (gen) conditioned on the experimental PDF are consistently above zero. In contrast, the crystallinity of samples from unconditional generation or conditioned on PDFs from melt-quench reference configurations is consistently zero. The box and whiskers depict the interquartile range and range of the distribution, respectively, and the central line depicts the median.
   }
    \label{fig:paracrystalline}
\end{figure}

The coexistence of crystalline and disordered domains at nanometer scales has long been debated as a model for amorphous silicon.\citeMain{treacy1998paracrystallites,treacy2012local,roorda2012comment,lewis2022fifty}
Recent MD studies have shown that melt-quench simulations at varying cooling rates produce configurations spanning a continuum from fully disordered to partially crystallized states.\citeMain{rosset2025signatures}
However, this simulation-based approach requires carefully training an interatomic potential and establishing a processing protocol, and the resulting structures depend sensitively on both.
The question of statistical consistency of paracrystalline models with experimental measurements therefore remains open.

We address this problem by using GLASS to invert experimental and simulated PDFs into atomistic configurations, then analyzing the existence of paracrystallinity in the data.
The score model was trained on a minimal dataset consisting of only fully amorphous or fully crystalline Si configurations (Sec. \ref{sec:si:paracrystallinity}), thus without knowledge of paracrystalline models.
Nevertheless, when conditioned on PDFs from Rosset \textit{et al.}\citeMain{rosset2025signatures} whose crystallinity is known, GLASS correctly generates structures exhibiting a mixture of crystalline and amorphous environments across a range of 1\% to 52\% crystallinity (Fig. \ref{fig:paracrystalline}A) while also accurately recovering the behavior of dihedral-angle distributions (Fig. \ref{fig:paracrystalline}B).
The framework also exhibits a monotonic trend in crystallinity compared to the reference structures (Fig. \ref{fig:paracrystalline}C), though the number of crystalline domains is systematically underpredicted (Sec. \ref{sec:paracrystallinity-modeling-a-si}).
Thus, generated structures are likely a lower bound in crystallinity and do not invalidate the model's ability to generate structures with a mixture of ordered and disordered phases (see examples at Fig. \ref{fig:paracrystalline}D).

\begin{figure}[htb!]
   \centering
    \includegraphics[width=0.6\textwidth]
    {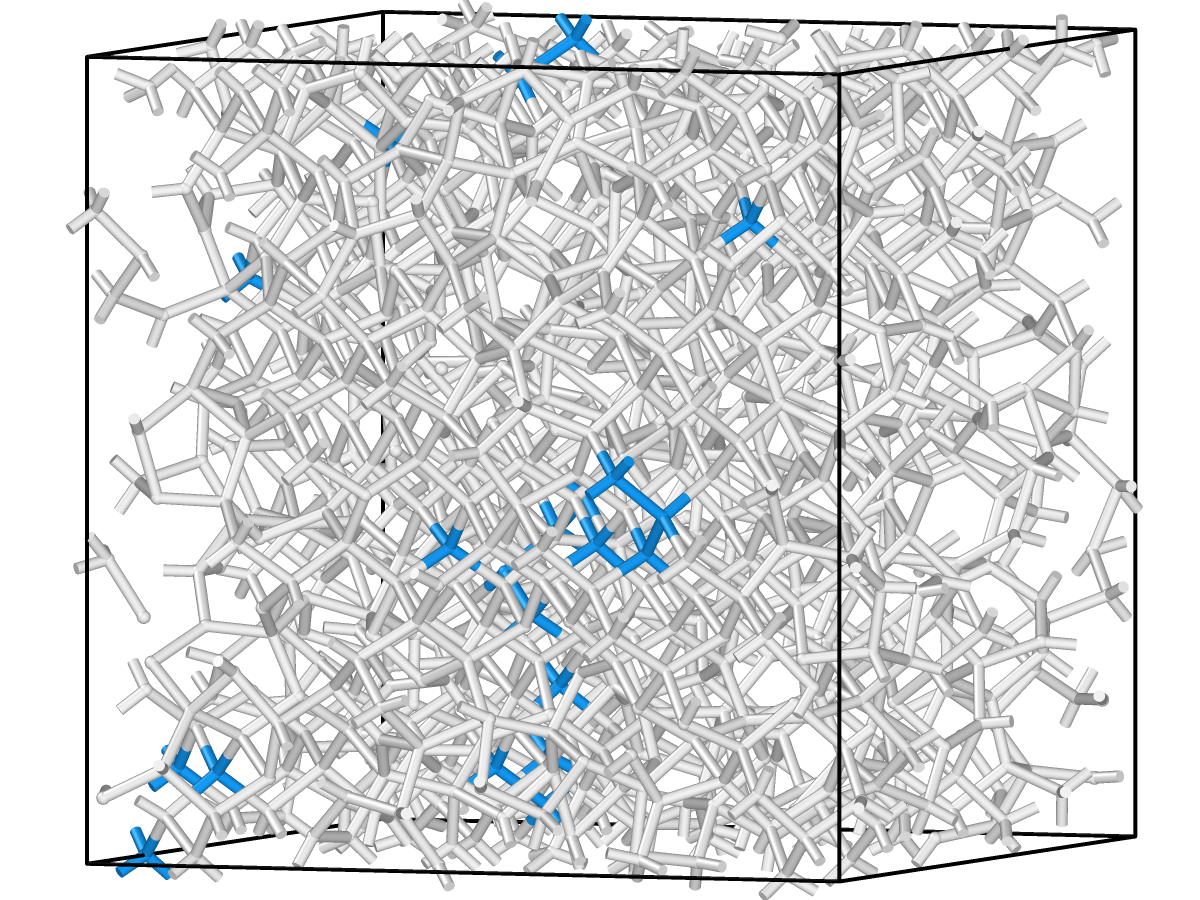}
    \caption{\textbf{Structure of a-Si generated from experimental pair distribution function (PDF).}
    Si-Si bonds are determined using a cutoff of 2.8 \AA. Blue, orange, and white colors represent environments similar to cubic diamond, hexagonal diamond, and none, respectively, as determined by the polyhedral template matching method (Sec. \ref{sec:si:paracrystallinity}).
    }
\label{si:fig:si-exp}
\end{figure}

\begin{figure}[htb!]
   \centering
    \includegraphics[width=0.6\textwidth]
    {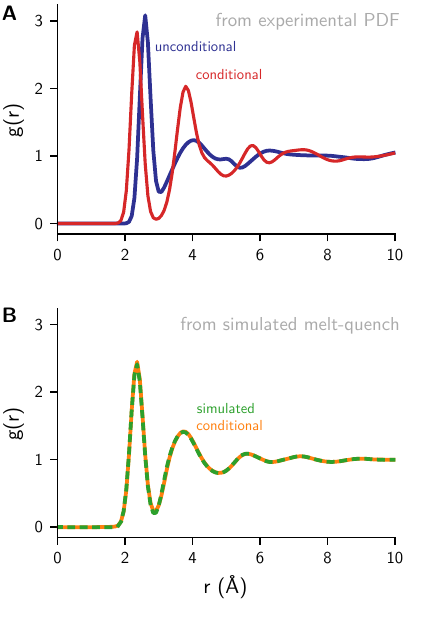}
    \caption{\textbf{Pair distribution functions (PDFs) for amorphous silicon.}
    \textbf{A.} PDFs for generated a-Si directly from the prior model (unconditional) and conditioned to the experimental PDF (conditional).
    \textbf{B.} PDFs for generated a-Si conditioned on the reference PDF from melt-quench simulations.
    }
\label{si:fig:si-pdfs}
\end{figure}

\begin{figure}[htb!]
   \centering
    \includegraphics[width=0.6\textwidth]
    {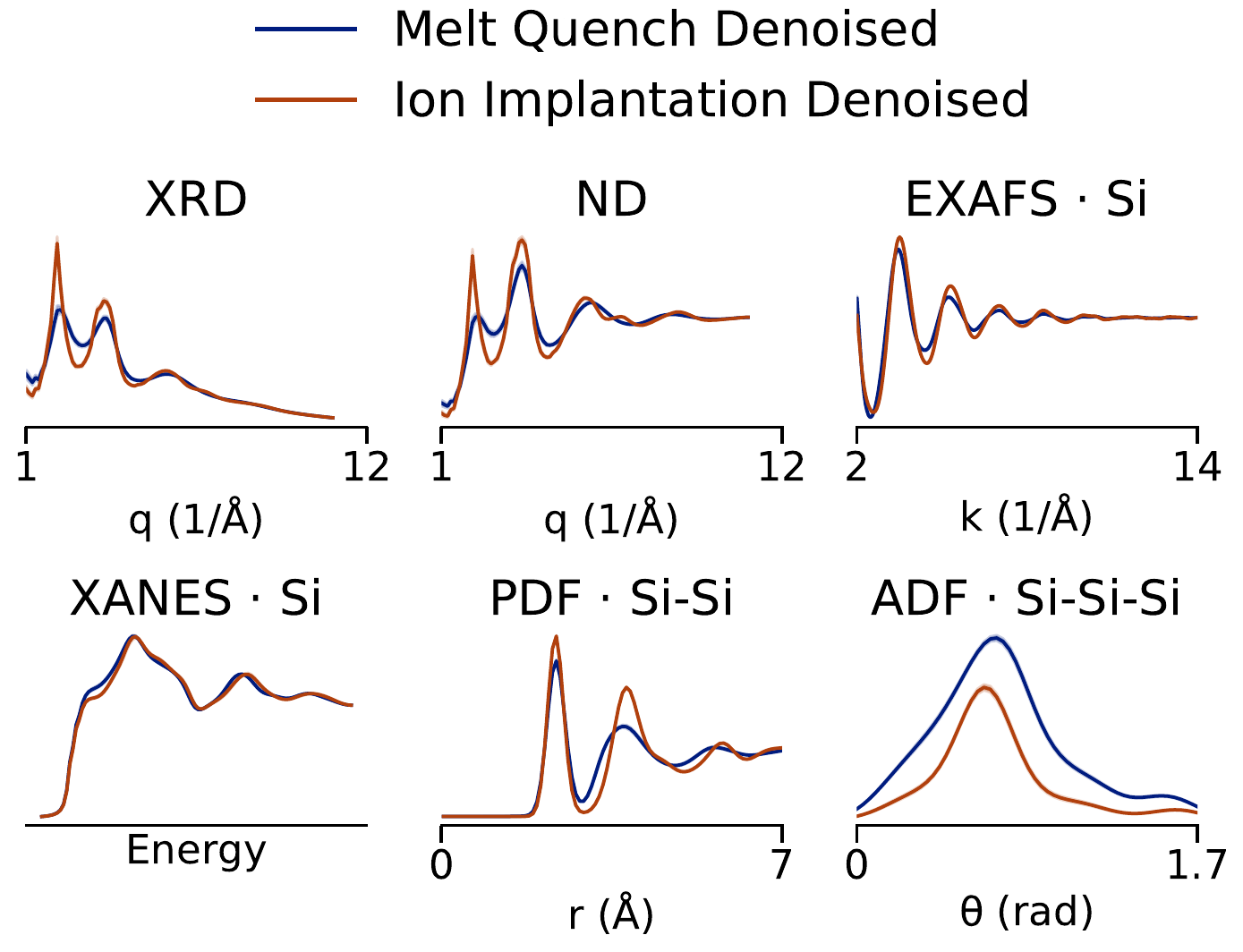}
    \caption{\textbf{Multi-modality validation comparing denoised structures guided by simulated melt-quench versus experimental ion-implantation PDFs.} Spectra computed from the two denoised structures are compared across XRD, ND, EXAFS, XANES, PDF (Si–Si), and ADF (Si–Si–Si). The melt–quench–guided structure exhibits fully amorphous character with no detectable crystallinity, whereas the ion-implantation–guided structure shows signatures of partial crystallinity (see main text). This distinction is evident in the sharper and more pronounced peaks in XRD and ND for the ion-implantation case. In contrast, EXAFS and XANES remain largely similar between the two structures, indicating that local coordination environments and short-range physics are preserved despite differences in medium- and long-range order.
    }
\label{si:fig:compare_two_denoisers}
\end{figure}

Given the debate on whether paracrystalline structures are compatible with experimental X-ray diffraction measurements,\citeMain{treacy2012local,roorda2012comment} we used the experimental PDF of high-quality a-Si prepared from self-ion implantation from Laaziri \textit{et al.}\citeMain{laaziri1999high} as guidance for our generative model.
While generating replicas with RMC can be computationally costly, here we used GLASS to produce 270 structures of 1000 atoms each to obtain robust statistics.
Across different initializations and random seeds, all generated structures reproduced the experimental $g(r)$ almost exactly at both short- and medium-ranges (Fig. \ref{fig:paracrystalline}E).
Analysis of generated structures revealed a crystallinity of $ 3.2 \pm 1.4$ \% (Fig. \ref{fig:paracrystalline}F), consistent with previous findings using fluctuation electron microscopy and MD simulations.\citeMain{treacy2012local,rosset2025signatures}
The generated crystalline domains are often a few atoms large and are dispersed into an amorphous matrix (Fig. \ref{si:fig:si-exp}).\citeMain{borisenko2012medium}
In contrast, when the same prior model and initializations are used, but PDFs from fully amorphous structures are used as targets (Figs. \ref{si:fig:si-pdfs},\ref{si:fig:compare_two_denoisers}), the denoised trajectory systematically converges to zero crystallinity (Fig. \ref{fig:paracrystalline}F), with only 4 out of 100 independent runs exhibiting $<$0.1\% diamond-like environments each.
This demonstrates that, within large ensembles of generated structures, a-Si models exhibiting paracrystallinity are statistically the most consistent with the experimental X-ray diffraction measurements according to GLASS, and are not artifacts of the reconstruction methods.

\section*{Mechanism for the liquid-liquid phase transformation in sulfur}

\begin{figure}[htb!]
   \centering
   \includegraphics[width=0.7\textwidth]{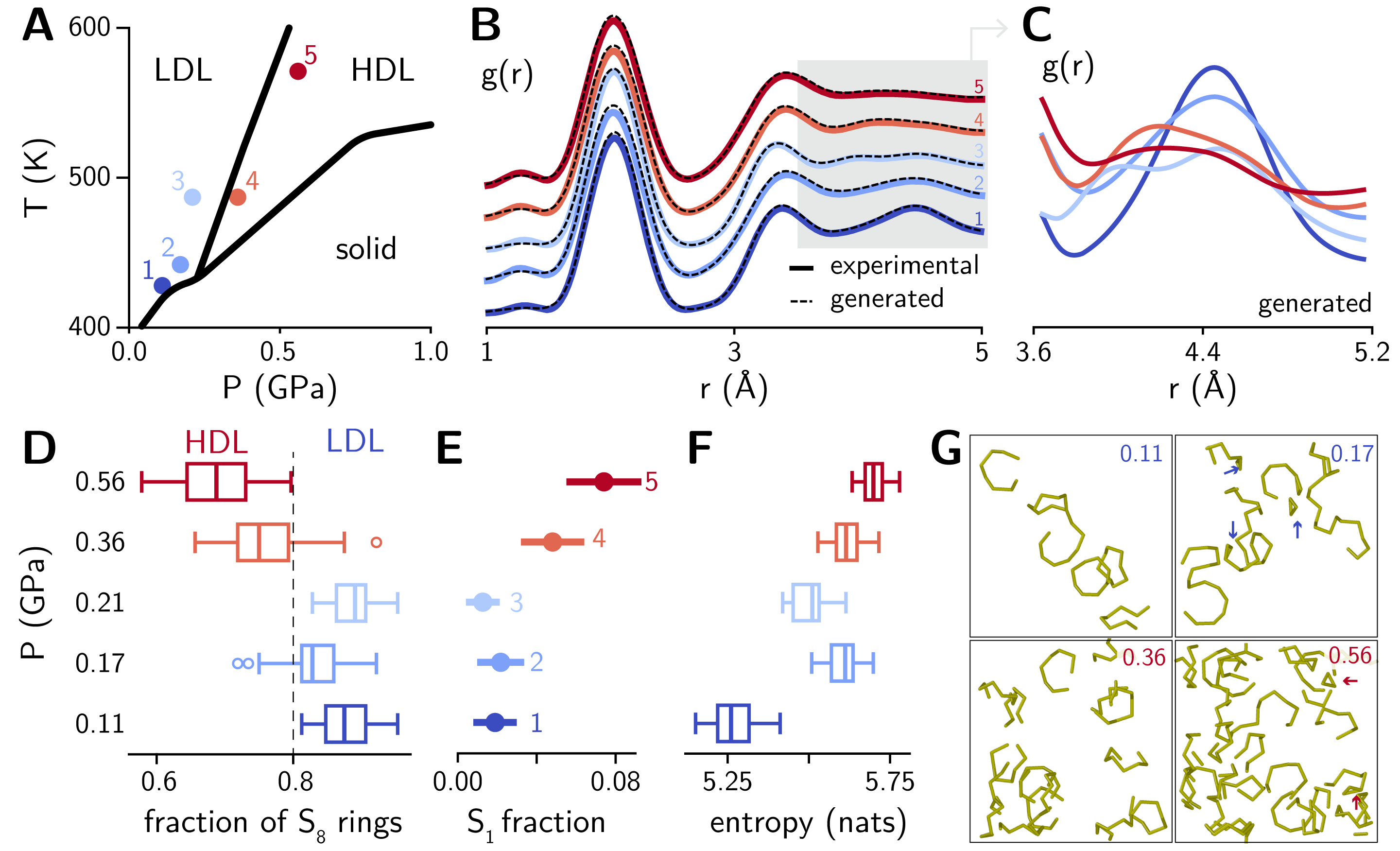}
    \caption{
    \textbf{Generating structures for liquid–liquid phase transition (LLPT) in sulfur from experimental data.}
    \textbf{A.} Pressure-temperature phase diagram of sulfur illustrating the LLPT between low-density liquid (LDL) and high-density liquid (HDL). Colored points correspond to the experimental results used as guidance, with colors matching panels \textbf{B}-\textbf{G}. Black curves indicate the experimentally reported phase boundaries.
    \textbf{B.} Experimental (solid) and generated (dashed) $g(r)$ across the LLPT.
    \textbf{C.} Signatures of the LLPT at the medium-range of generated $g(r)$. The peak near 4.45 \AA\ associated with S$_8$ rings is suppressed, while a feature near 4.0 \AA\ shifts toward $\sim4.15$~\AA\ upon entering the HDL phase.
    Structural descriptors extracted from reconstructed configurations: \textbf{D}, fraction of intact S$_8$ rings, \textbf{E}, fraction of chain terminations (S$_1$), and \textbf{F}, atomistic information entropy. Box-and-whisker points represent the interquartile range, range, and median of the statistics of independently generated structures.
    \textbf{G.} Example of generated configurations at pressures of 0.11, 0.17, 0.36, and 0.56~GPa. S$_8$ rings are removed for clarity. Small triangular motifs (arrows) arise from artifacts in the experimental $g(r)$ at very short distances ($\sim1.2$~\AA).}
   \label{fig:sulfur}
\end{figure}

Recent \textit{in situ} XRD and Raman measurements demonstrated that liquid sulfur undergoes a first-order phase transition between a low-density liquid (LDL) composed of S$_8$ molecular rings and a high-density liquid (HDL) with increased polymeric chain content.\citeMain{henry2020Liquid}
Aside from density discontinuities, the structural signatures of such transitions are subtle.
Short-range order is essentially invariant across the transition, with identical S-S bond lengths and S-S-S angles in both phases.
Only medium-range order changes measurably, making it challenging to interpret the experimental spectra and topological reorganization near the liquid-liquid phase transition (LLPT).
Thus, the mechanism of the LDL-HDL transformation in sulfur is difficult to rationalize without structural models.\citeMain{tanaka2020Liquid}

We used conditional denoising with GLASS to propose atomistic structures matching the experimental $g(r)$ measured across the LDL–HDL transition\citeMain{henry2020Liquid} (Fig. \ref{fig:sulfur}A,B, Sec. \ref{sec:si:sulfur}).
The score model was trained on a minimal dataset containing only two limiting structural states: a pure S$_8$-ring liquid and a fully polymeric network, each sampled at three temperatures (Sec. \ref{sec:training-dataset-sulfur}, Fig. \ref{si:fig:S_train_eval}).
Initial configurations were constructed as randomly distributed S$_8$ rings in a box with periodic boundary conditions, accelerating convergence during early denoising (Sec. \ref{sec:spectral-opt-limitations}, Fig. \ref{si:fig:S_vary_model_init}).
Because the experimental density was not reported for these specific points, it was treated as a hyperparameter and selected as the density that minimizes the $g(r)$ reconstruction error around the third- and fourth-shells (Sec. \ref{sec:grid-search-shell-weighted-pdf}).
Despite the absence of energy calculations, the denoised structures reproduce the subtle spectral changes observed experimentally (Fig. \ref{fig:sulfur}B,C).
The reconstructed structures reveal a structural mechanism for the transition: a gradual increase in ring opening with increased pressure on each side of the phase boundary, but a discontinuity in the ring concentration at the first-order phase transition.
As S$_8$ rings break into polymeric segments, medium-range correlations are changed while local bonding geometry remains intact, consistent with the invariance of the first two PDF peaks.
Topological analysis of generated structures shows a statistically significant decrease in intact S$_8$ rings across the LDL-HDL transition (Fig. \ref{fig:sulfur}D) accompanied by an increase in polymeric chains (S$_1$ fraction in Fig. \ref{fig:sulfur}E), consistent with trends inferred from Raman spectra.\citeMain{henry2020Liquid}
To demonstrate that the atomic reorganization is exclusive to the neighborhood of the phase boundary, we computed the atomistic information entropy\citeMain{schwalbekoda2025information} of generated structures for points 1--5, which quantifies the diversity of environments in each structure.
Figure \ref{fig:sulfur}F shows that point 1 in Fig. \ref{fig:sulfur}A, far from the LDL-HDL transition, exhibits more uniform environments (lower information entropy) compared to structures near the phase boundary.
Representative reconstructed configurations illustrate the progressive growth of polymeric sulfur chains with increasing pressure (Fig.~\ref{fig:sulfur}G).
Inspection shows that small artifacts are seen in the structures (e.g., a few S$_3$ triangles marked with arrows in Fig.~\ref{fig:sulfur}G), as the generative model proposed a match to the small peak near 1.2 \AA~ in Fig. \ref{fig:sulfur}B, which is also a known artifact of the Fourier transform from $S(Q)$ to $g(r)$.
Nevertheless, the method explains collective topological reorganization in a LLPT, demonstrating its use for fast interpretation of complex experimental data without the need for realistic simulations or large datasets.

\section*{Proposing an amorphous structure for ball-milled ice}

\begin{figure}[htb!]
   \centering
   \includegraphics[width=\textwidth]{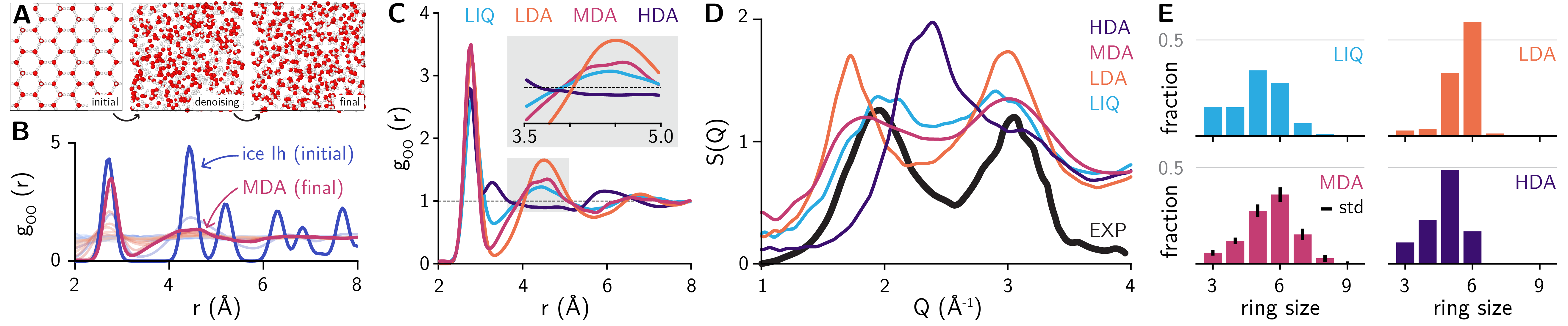}
    \caption{
    \textbf{Generative reconstruction of medium-density amorphous (MDA) ice.}
    \textbf{A.} The generation trajectory starts from crystalline hexagonal ice (I$h$) and converges to an amorphous MDA structure while passing through random intermediate states, as also shown in \textbf{B}, $g_{\mathrm{OO}}(r)$. Oxygen atoms are shown in red and hydrogen atoms in white.
    \textbf{C.} Comparison of $g_\mathrm{OO}(r)$ across amorphous ice, with reference low-density amorphous ice (LDA, blue), liquid water (LIQ, cyan), and high-density amorphous ice (HDA, purple), and generated MDA (pink). The inset highlights the trends of the second coordination shell near $\sim 4.5$~\AA.
    \textbf{D.} Comparison of calculated structure factors $S(Q)$ of reference and generated structures and experimental measurements (black).
    \textbf{E.} Ring-size distributions of the hydrogen-bond network for LIQ, LDA, MDA, and HDA. The reconstructed MDA shows fewer six-membered rings relative to LDA and broader ring distribution resembling the liquid state.
    }
   \label{fig:water}
\end{figure}

Medium-density amorphous (MDA) ice occupies a structural gap between the established low-density amorphous ice (LDA) and high-density amorphous ice (HDA) phases.
MDA was recently discovered by ball-milling crystalline ice I$h$ at 77 K and resembled a glassy state of ambient liquid water.\citeMain{rosu2023medium}
Molecular simulations suggested MDA is a nonequilibrium shear-driven amorphous state whose density spans the full LDA-to-HDA range depending on shear rate and pressure, with structures proposed through shearing.\citeMain{rosu2023medium,almeidaribeiro2024medium}
However, resolving the structure of MDA directly from experiment is difficult due to the limited range of the experimental structure factor $S(Q)$ and difficulties deconvoluting intramolecular O-H and H-H contributions from them.
To reconstruct MDA from scattering data, we adapted our differentiable PDF to include scattering factors approximating neutron diffraction measurements (Secs. \ref{sec:si:water},\ref{sec:scattering-weights-correlations}, Fig. \ref{si:fig:mda_different_weighting}).
Then, using I$h$ initializations and a guess $g(r)$ as the target, GLASS eliminates long-range order while reconstructing hydrogen-bond networks without predefined constraints on individual O-H bonds  (Fig.~\ref{fig:water}A,B).
Comparison of O-O partial PDFs $g_{OO}(r)$ shows the evolution of the medium-range order along LDA $\rightarrow$ MDA $\rightarrow$ liquid $\rightarrow$ HDA sequence (Fig.~\ref{fig:water}C).
Independent validation from reciprocal-space measurements also confirm that generated structures are appropriate (Fig.~\ref{fig:water}D).
Ring statistics of the hydrogen-bond network further show a reduction of six-membered rings and an increase in five-membered rings relative to LDA (Fig.~\ref{fig:water}E), but also an expansion of the network towards 7 and 8-membered rings, approximating an intermediate between liquid and LDA/HDA structures.
Thus, even though the score model was trained only on a single LDA configuration, generated structures show excellent agreement with experimental and independent computational results, further demonstrating the utility of GLASS in resolving elusive amorphous configurations directly from measurements.

\section*{Discussion}

\begin{figure}[htb!]
   \centering
    \includegraphics[width=0.5\textwidth]
    {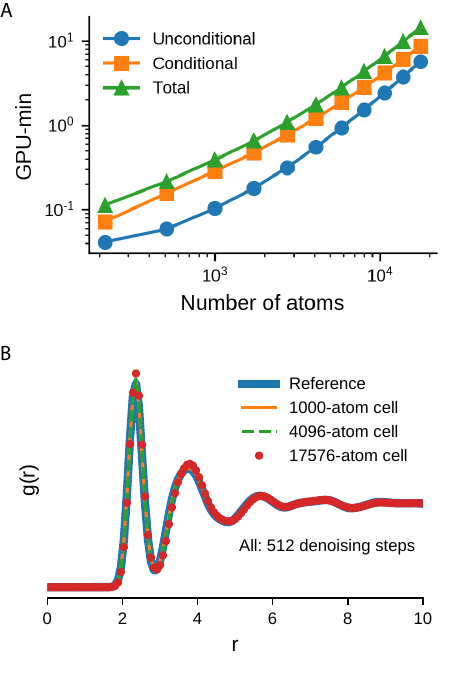}
    \caption{
    \textbf{Scaling behavior of unconditional and PDF-guided conditional denoising.}
    \textbf{A.} GPU runtime as a function of system size for unconditional (prior-only), conditional (PDF-guided), and total denoising. All simulations were performed on a single NVIDIA RTX A6000 GPU using identical denoising hyperparameters. The computational cost increases systematically with the number of atoms, reflecting both the growing cost of graph construction and score evaluation at each denoising step.
    \textbf{B.} Comparison of the resulting pair distribution functions $g(r)$ for systems ranging from 1,000 to 17,576 atoms against the reference structure. All runs used an identical denoising schedule of 512 steps for both part \textbf{A} and \textbf{B}. Despite the increase in system size, the agreement with the reference is consistently maintained without requiring additional denoising steps, indicating that a properly benchmarked and optimized set of hyperparameters enables size-transferable convergence.
    }
    \label{si:fig:scaliing}
\end{figure}

Reliably inverting spectroscopic data into amorphous structures can solve longstanding challenges in condensed matter, from phase transition mechanisms to the rational engineering of disorder.
Our GLASS framework addresses convergence, cost, and degeneracy issues, which have historically limited refinement-based methods, while maintaining physically plausible configurations.
As opposed to Monte Carlo-type approaches or even recent differentiable methods,\citeMain{anker2025autonomous} GLASS solves the inversion problem in up to 1 GPU-minute for 5000 atom-systems (Fig. \ref{si:fig:scaliing}) in an NVIDIA RTX A6000 GPU across all spectroscopic modalities.
Application to three experimentally motivated systems confirms that the framework generalizes beyond its low-fidelity training data and remains robust to noise in experimental spectra or mixtures of phases.
In all cases, multi-modality in spectroscopic data can be used for reconstruction, but is mostly used here for validation of GLASS-proposed structural models.
Decoupling generation from energy calculations enables further independent validation of the results through simulations while also producing training data for future MLIP development.

Limitations of GLASS such as occasional artifacts and underprediction of crystallinity are less severe than those of existing refinement methods, and can likely be addressed by active learning of prior models and hybrid simulation-generative approaches.
Scaling to arbitrary amorphous systems in a foundation-model paradigm, however, may require datasets that do not yet exist, but may be straightforward to build considering the data efficiency of GLASS (Sec. \ref{sec:training-dataset-sulfur}).
Our benchmarks provide a starting point for evaluating generative reconstruction of amorphous materials, particularly under multi-modal spectroscopic data, informing further development of newer generative models for amorphous materials.
Natural extensions of the work include integration of additional characterization modalities, improved generative architectures for atomistic data, and deployment within autonomous experimental design.

\section*{Methods}
\customlabel{sec:methods}{Methods}

\subsection*{Generative Learning of Amorphous Structures from Spectra (GLASS) architecture}

\textbf{Score model training.}
Score-based generative models\citeMain{song2021scorebased} were trained to learn a structural prior over atomic configurations using denoising score matching in periodic atomistic systems.
Atomic structures were represented as periodic graphs $\mathcal{G}=(\mathcal{V},\mathcal{E})$ constructed with a fixed neighbor cutoff of 5~\AA.
Nodes correspond to atoms and edges connect neighboring atom pairs within a cutoff.
Node features consisted of species encodings, while edge attributes encoded local geometry using the minimum-image displacement vector and interatomic distance $(\Delta x,\Delta y,\Delta z, r)$.

The score network $\hat{\mathbf{s}}_\theta(\tilde{\mathbf{x}},t)$ was parameterized as a message-passing graph neural network following a MeshGraphNets-style encoder--processor--decoder architecture.\citeMain{pfaff2020learning,Kwon2024SpectroscopyGuided}
Node and edge features were first embedded using multilayer perceptrons with SiLU activations and per-layer normalization.
The diffusion time step $t$, which determines the noise level, was incorporated through a learned time embedding $\phi_t(t)$ based on Gaussian random Fourier features,\citeMain{Tancik2020FourierFeatures} which was added to the node embeddings (Fig. \ref{fig:overview}B).
Latent features were propagated through five message-passing layers with hidden dimension 200, and a decoder predicted a three-dimensional, per-atom score vector scaled by the inverse noise level as $\hat{\mathbf{s}}_\theta(\tilde{\mathbf{x}},t)/\sigma$.

The training process followed the denoising score matching framework.\citeMain{Kwon2024SpectroscopyGuided,yang2025generative}
Given atomic coordinates $\mathbf{x}$, noisy configurations were generated as
\[
\tilde{\mathbf{x}} = \mathbf{x} + \sigma \boldsymbol{\epsilon}, \quad
\boldsymbol{\epsilon} \sim \mathcal{N}(0,\mathbf{I}),
\]
and the model was trained by minimizing the loss function $\mathcal{L}$,
\[
\mathcal{L} =
\mathbb{E}\left[
\left\|
\sigma \hat{\mathbf{s}}_\theta(\tilde{\mathbf{x}},t) + \boldsymbol{\epsilon}
\right\|_2^2
\right],
\]
which allows the network $\hat{\mathbf{s}}_\theta$ to approximate the score function $\nabla_{\tilde{\mathbf{x}}}\log p_\sigma(\tilde{\mathbf{x}})$ across noise levels.
Models were trained using the Adam optimizer with learning rate $10^{-3}$.\citeMain{Kingma2015Adam}
An exponential moving average of model parameters (decay 0.9999) was maintained during training.

\medskip
\noindent\textbf{Unconditional denoising (score dynamics).}
After training, the learned score model is used within a stochastic dynamic strategy to denoise atomic coordinates.
Starting from an initial configuration $\mathbf{x}_T$ at the highest noise level $T$, structures were generated or refined by integrating the reverse-time stochastic differential equation over a sequence of decreasing noise levels ($t: T \rightarrow 0$).
At each step, coordinates were updated according to
\[
\mathbf{x}_{t-\Delta t}
=
\mathbf{x}_t
+
\left[
f(t)\mathbf{x}_t - g(t)^2 \hat{\mathbf{s}}_\theta(\mathbf{x}_t,t)
\right]\Delta t
+
g(t)\sqrt{|\Delta t|}\,\mathbf{z}_t,
\qquad
\mathbf{z}_t \sim \mathcal{N}(0,\mathbf{I}),
\]
where $f(t)$ and $g(t)$ define the noise schedule.
The score $\hat{\mathbf{s}}_\theta(\mathbf{x}_t,t)$ was evaluated on the periodic graph constructed from the current atomic coordinates at each step, $\mathbf{x}_t$.
This score-driven dynamics defines a structural prior that denoises configurations toward physically plausible regions of configuration space.

\medskip
\noindent\textbf{Conditional denoising with spectroscopic guidance.}
To incorporate target structural information, conditional generation balances the prior score shown above with a guidance term derived from a differentiable observable.
Given the current noisy coordinates $\mathbf{x}_t$ at noise level $t$, a clean-structure estimate was first formed as
\[
\hat{\mathbf{x}}_0 = \mathbf{x}_t + \sigma(t)^2 \hat{\mathbf{s}}_\theta(\mathbf{x}_t,t).
\]
A differentiable forward model $F(\hat{\mathbf{x}}_0)$, such as the pair distribution function (PDF) or any of the other spectroscopic modalities, was then evaluated and compared to a target signal $\mathbf{y}$.
The guidance term was defined from the gradient of the spectral mismatch,
\[
\hat{\mathbf{s}}_{\mathrm{guide}}
\propto
-\nabla_{\hat{\mathbf{x}}_0}
\,\mathcal{L}_{\mathrm{spec}}\!\left(F(\hat{\mathbf{x}}_0),\mathbf{y}\right),
\]
where $\mathcal{L}$ is the squared $L_2$ loss between the predicted and target signals, and combined with the learned prior as
\[
\hat{\mathbf{s}}_{\mathrm{tot}}(\mathbf{x}_t,t)
=
\hat{\mathbf{s}}_\theta(\mathbf{x}_t,t)
+
w\,\hat{\mathbf{s}}_{\mathrm{guide}}(\mathbf{x}_t,t),
\]
where $w$ controls the trade-off between spectroscopic guidance and prior score.
The combined score was used in the same stochastic denoising dynamics described before, enabling structure generation that simultaneously respects the learned prior and matches the target observable.

\subsection*{Differentiable spectroscopic modalities}

To enable spectroscopic guidance within GLASS, we implemented differentiable surrogate models for multiple experimental observables.
All observables were formulated as smooth functions of atomic coordinates in \texttt{PyTorch}, enabling gradients with respect to atomic positions to be computed via automatic differentiation.
In addition, differentiable surrogate models were trained to predict X-ray absorption spectra from atomic structures using graph neural networks.

\medskip
\noindent\textbf{Real-space observables.}
Element-resolved pair distribution functions (PDF) were computed under periodic boundary conditions using the minimum-image convention.
Pairwise distances were accumulated on a fixed radial grid using Gaussian kernel smoothing rather than discrete histogram binning, ensuring smooth dependence on atomic coordinates and enabling differentiation.
Angular distribution functions (ADF) were computed from three-body correlations by evaluating bond angles between neighboring atoms within a cutoff radius (typically $5$~\AA).
Angular histograms were similarly constructed using Gaussian smoothing on a fixed angular grid.

\medskip
\noindent\textbf{Reciprocal-space observables.}
Diffraction intensities were computed using a differentiable Debye scattering formulation based on pairwise atomic distances.
X-ray diffraction (XRD) intensities were evaluated on a uniform reciprocal-space grid using element-dependent atomic form factors\citeMain{Waasmaier1995} and included an isotropic Debye-Waller damping factor\citeMain{Egami2003Underneath} to account for thermal and static disorder.
Neutron diffraction (ND) intensities were computed within the same Debye framework but with constant element-specific coherent neutron scattering lengths obtained from standard tabulations,\citeMain{Sears1992} replacing the X-ray form factors. X-ray atomic form factors were evaluated using the analytic parameterization of Waasmaier and Kirfel,\citeMain{Waasmaier1995} in which the non-dispersive scattering factor is expressed as a sum of Gaussian functions of $\sin(\theta)/\lambda$, consistent with DebyeCalculator.

\medskip
\noindent\textbf{X-ray absorption spectroscopy.}
Differentiable surrogate models were trained to predict X-ray absorption spectra (XAS) from atomic structures.
For each material system, a single model was trained on pooled datasets containing amorphous structures, liquid configurations sampled from melt simulations, and randomly initialized structures.
Reference spectra were generated using site-resolved \texttt{FEFF9} (version 9.6.4)\citeMain{Rehr2010FEFF9} calculations, producing atom-resolved spectra used as supervised learning targets.
Atomic configurations were represented as periodic graphs with a $5$~\AA\ neighbor cutoff, with species encodings as node features and $(\Delta x,\Delta y,\Delta z,r)$ as edge attributes.

EXAFS models were trained directly on $k$-space spectra rather than their Fourier-transformed $R$-space representations. This choice avoids information loss and artifacts introduced by finite-window Fourier transforms, preserves the native physics of the \texttt{FEFF} formalism, and provides smoother, more localized gradients with respect to structural parameters. In contrast, $R$-space representations entangle phase and amplitude information and depend sensitively on preprocessing choices, making them less suitable for learning and differentiable optimization.

Both XANES and EXAFS predictors used a \textit{SpecNet} architecture\citeMain{song2021scorebased, Kwon2024SpectroscopyGuided} implemented in the graphite package,\citeMain{hsu2024score} based on a MeshGraphNets-style message-passing neural network with an encoder-processor-decoder structure.
Node and edge embeddings were computed using multilayer perceptrons with SiLU activations and LayerNorm, followed by five message-passing layers with hidden dimension $200$.
A final decoder mapped node embeddings to predicted spectra, enabling atom-wise spectral predictions.
The XANES model predicted spectra on a grid of $100$ points, while the EXAFS model predicted $400$ points.
Models were trained using a mean-squared error loss and optimized using Adam with learning rate $10^{-3}$.
An exponential moving average of model parameters (decay $0.9999$) was maintained during training to improve stability during downstream inference and spectroscopic guidance.

\noindent Together, these differentiable observables provide fast, differentiable approximations of experimental measurements, enabling spectroscopic guidance during denoising while remaining consistent with high-fidelity reference calculations.

\subsection*{Molecular dynamics simulations}

Molecular dynamics (MD) simulations were performed using \texttt{LAMMPS} (2 Aug 2023 -- Update 3).\citeMain{thompson2022lammps}
Interatomic interactions were described using the Tersoff potential for amorphous carbon (a-C),\citeMain{Tersoff1988_1, Tersoff1988_2} amorphous silicon (a-Si),\citeMain{Tersoff1988_1, Tersoff1988_2} and amorphous silica (a-SiO$_2$),\citeMain{Munetoh2007SiO} and the embedded-atom method (EAM) potential for metallic alloys including Cu-Zr\citeMain{Mendelev2009CuZr} and Zr-Ni-Al.\citeMain{Cheng2009BMG}
For each structural variation (density, cooling rate, or composition), ten independent replicas were generated to ensure statistical sampling.
System sizes were 216 atoms for a-Si and a-C, 300 atoms for a-SiO$_2$, and 500 atoms for the alloy systems.
Density variations for elemental systems were introduced by uniformly scaling the cubic simulation cell prior and after equilibration.
All initial configurations were first relaxed using conjugate-gradient energy minimization.
Systems were then equilibrated at 300~K in the canonical (NVT) ensemble using a Nos\'e-Hoover thermostat,\citeMain{Nose1984, Hoover1985} followed by equilibration in the isothermal-isobaric (NPT) ensemble at 300~K and zero external pressure to allow relaxation of the simulation cell.
Amorphous structures were generated using a melt-quench protocol.
Equilibrated systems were heated to 5000~K under NPT conditions at zero pressure and maintained for 50~ps to ensure complete melting.
The resulting liquid was subsequently cooled to 300~K under NPT conditions.
For all systems except a-SiO$_2$, cooling was performed over 47~ps, corresponding to an effective cooling rate of approximately 100~K/ps.
For a-SiO$_2$, additional slower cooling rates of 1~K/ps and 10~K/ps were explored to assess quench-rate dependence.
Following the quench, systems were rescaled to the initial density and further equilibrated at 300~K and zero pressure to remove residual stresses.

\subsection*{Reverse Monte Carlo simulations}

Reverse Monte Carlo (RMC)\citeMain{McGreevy1988RMC,McGreevy2001RMCReview} simulations were performed using the \texttt{FULLRMC} package (version 4.1.0),\citeMain{aoun2016fullrmc} which refines atomistic configurations by iteratively proposing stochastic coordinate perturbations and accepting or rejecting trial moves to minimize the discrepancy between simulated and target structural observables.
For each structural variation (e.g., different densities for a-Si and a-C), two initialization strategies were considered: (1) random initial configurations, and (2) liquid-derived configurations extracted from melt snapshots generated by molecular dynamics.
For each initialization type, ten independent starting structures were generated.
Each starting configuration was then refined through ten independent RMC runs with distinct random seeds, producing an ensemble of 100 statistically independent refinement trajectories.

All refinements were performed with periodic boundary conditions using the same simulation cell for the input structure at the targeted density.
The refinement objective was defined using a pair-correlation constraint constructed from the target pair distribution function provided in an experimental-style \texttt{.exp} file.
Pair-correlation contributions were weighted by atomic number to remain consistent with the weighting scheme used in the PDF guidance benchmarks.
No additional geometric constraints were applied for the amorphous Si and C systems considered here, allowing the behavior of PDF-only refinement to be evaluated.

Refinement proceeded in two sequential stages consisting of an exploration phase followed by a refinement phase.
In the exploration phase, trial moves emphasized broader configurational sampling to rapidly reduce large discrepancies with the target pair correlation function, whereas the refinement phase employed more conservative moves to improve local agreement with the constraint.
For all simulations reported in the main text, each stage was executed for 60,000 outer iterations with four recursive selections per iteration.
Because atoms were grouped as individual move units, each outer iteration attempted $4 \times N_{\mathrm{atoms}}$ Monte Carlo trial moves.
Unlike fully random atom selection, the recursive selection scheme in \texttt{FULLRMC} ensures that nearly all atoms are visited within each iteration cycle, resulting in more uniform coverage of configurational degrees of freedom.
This improves sampling efficiency and reduces the number of Monte Carlo steps required to reach convergence compared to purely random selection strategies.

For the 216-atom systems studied here, this corresponds to a total of $60{,}000 \times 2 \times 4 \times 216 = 103{,}680{,}000$ (104 million) Monte Carlo trial moves per trajectory.
The resulting ensemble of refined structures was analyzed to evaluate the dependence of RMC outcomes on initialization and to identify the best-fitting configurations across independent runs.

While \texttt{FULLRMC} provides efficient sampling through its recursive selection strategy, it has practical limitations, including reduced scalability to larger system sizes and no use of gradients from available guidance types.
These limitations were not restrictive for the systems and observables considered in this study.

\section*{Data Availability}

The datasets generated in this work will be released upon publication of the manuscript.

\section*{Code Availability}

The code to reproduce the results/plots from this work will be released at publication time.

\section*{Acknowledgements}

This work was supported by Toyota Research Institute under the Synthesis Advanced Research Challenge.
This work used computational and storage services associated with the Hoffman2 Shared Cluster provided by UCLA Office of Advanced Research Computing’s Research Technology Group.
Additional computational resources from Delta GPU at NCSA were used through allocation MAT240040 from the Advanced Cyberinfrastructure Coordination Ecosystem: Services \& Support (ACCESS) program, which is supported by National Science Foundation grants \#2138259, \#2138286, \#2138307, \#2137603, and \#2138296.
We thank Bruce Dunn, Kai Yang, Amalie Trewartha, Linda Hung, Steven Torrisi, Amanda Volk, Koki Nakano, and Ryo Asakura for discussions on this work.
We also thank Marcos Calegari Andrade and Hsin-Yu Ko for the support with $S(Q)$ calculations for water.

\section*{Conflicts of Interest}

The authors have no conflicts to disclose.

\section*{Author Contributions}

\noindent\textbf{Jiawei Guo:} Formal Analysis; Data Curation; Investigation; Software; Validation; Visualization; Writing - Original Draft; Writing - Review \& Editing.
\noindent\textbf{Daniel Schwalbe-Koda:} Conceptualization; Formal Analysis; Investigation; Methodology; Project Administration;  Visualization; Writing - Original Draft; Writing - Review \& Editing; Funding Acquisition; Supervision.

\clearpage
\beginsupplement

\beginsuppinfo
\customlabel{sec:sinfo}{Supplementary Information}

\etocdepthtag.toc{supplementary}
\etocsettagdepth{default}{none}
\etocsettagdepth{supplementary}{subsection}

{\cftsetindents{section}{0em}{3.0em}
 \cftsetindents{subsection}{0em}{3.0em}
 \tableofcontents}

\clearpage
\section{Supplementary Methods}
\customlabel{sec:smeth}{Supplementary Methods}

The Supplementary Methods elaborate on the results from the main paper, as well as additional results shown in the \suppinfo.

\subsection{High-fidelity X-ray absorption spectroscopy calculations}
\label{sec:xas-hifi}
X-ray absorption near-edge structure (XANES) and extended X-ray absorption fine structure (EXAFS) spectra were computed using the real-space multiple-scattering code \textsc{FEFF9}\citeSupp{Rehr2010FEFF9-SI}.
For each atomistic configuration, absorber-centered atomic clusters were constructed from the full periodic structures using periodic boundary conditions and the minimum-image convention.
For a given absorbing atom, all neighboring atoms within a predefined cutoff radius were selected and expressed in a local coordinate frame relative to the absorber.
The index of the absorbing atom within each cluster was stored to enable automated generation of site-resolved \texttt{feff.inp} files.

FEFF input files were generated automatically for each absorber-centered cluster.
Chemical species present in the cluster were mapped to FEFF potential indices (\texttt{ipot}) via the \texttt{POTENTIALS} card, with \texttt{ipot} $= 0$ reserved for the absorber and incrementing indices assigned to each distinct neighboring element.
The \texttt{ATOMS} list contained Cartesian coordinates (in \AA) together with their corresponding \texttt{ipot} labels.
Standard FEFF control and output settings were employed (\texttt{CONTROL 1 1 1 1 1 1}; \texttt{PRINT 1 0 0 0}).

For EXAFS calculations, scattering paths were restricted using a maximum half-path length cutoff of
\[
r_{\mathrm{max}} = 5.0~\text{\AA},
\]
ensuring inclusion of physically relevant single- and multiple-scattering contributions.
For XANES calculations, full multiple scattering (FMS) and self-consistent field (SCF) calculations were performed within a common radius of $5.0~\text{\AA}$ (\texttt{FMS 5.0 1}; \texttt{SCF 5.0 1}), and near-edge spectra were explicitly requested via the \texttt{XANES} card.
Unless otherwise specified, all calculations were carried out at the $k$-edge of the absorbing element.

To improve the fidelity of near-edge features in disordered systems, the core-hole effect was neglected (\texttt{COREHOLE none}), and the imaginary component of the self-energy was disabled using the \texttt{RSIGMA} option.
These settings reduce artificial broadening and enhance the transferability of computed spectra across structurally heterogeneous environments.

FEFF calculations were executed in batch mode following a standardized module sequence.
After input parsing (\texttt{rdinp}), electronic structure quantities and scattering phase shifts were generated using \texttt{atomic}, \texttt{pot}, \texttt{screen}, and \texttt{xsph}.
Multiple-scattering contributions and path generation were performed via \texttt{fms}, \texttt{mkgtr}, \texttt{path}, and \texttt{genfmt}.
The absorption spectra were then evaluated using \texttt{ff2x}, followed by spectral convolution and inelastic/background corrections using \texttt{sfconv}, \texttt{compton}, and \texttt{eels}.
All EXAFS and XANES calculations were organized in site-resolved directories to facilitate large-scale automated spectral generation.

The resulting site-resolved EXAFS and XANES spectra were used as reference data for downstream structural analysis, benchmarking, and training of spectroscopic surrogate models.

\subsection{Training of differentiable XANES and EXAFS surrogate models}
\label{sec:diff-surrogate-models}
Differentiable surrogate models for XANES and EXAFS were trained to enable spectroscopic guidance during score-based structure generation.
For each material system, a single surrogate was trained using a pooled dataset containing all structural variations relevant to that system (e.g., multiple densities for a-Si and a-C, multiple cooling rates for a-SiO$_2$, and multiple compositions for Cu--Zr and Zr--Ni--Al alloys).
To improve robustness to the broad range of atomic environments encountered during denoising, the training sets included not only fully converged amorphous structures, but also liquid-phase configurations sampled from the MD melt stage and random initial structures that are known to exhibit high and complementary structural diversity.\citeSupp{schwalbekoda2025information-SI}
Reference XANES/EXAFS spectra for each atomic environment were generated using site-resolved \texttt{FEFF9}\citeSupp{Rehr2010FEFF9-SI} calculations (Section \ref{sec:xas-hifi}), producing atom-wise spectra that serve as supervised learning targets.

Atomic configurations were represented as periodic graphs constructed with a fixed neighbor cutoff of 5~\AA.
Each frame was converted into a graph where nodes represent atoms and edges connect all neighbor pairs within the cutoff under periodic boundary conditions.
Node attributes consisted of one-hot encoded species with $N_{\mathrm{species}}$ channels, while edge attributes encoded local geometry using a 4D descriptor containing the Cartesian components of the minimum-image displacement vector together with the interatomic distance, i.e., $(\Delta x,\Delta y,\Delta z, r)$.
The dataset was randomly split into 90\% training and 10\% validation subsets.

To increase stability with respect to noisy intermediate structures, Gaussian perturbations were applied to atomic coordinates during training with a maximum noise level of $0.8$ \AA, consistent with the perturbation that was successful used in our previous work.\citeSupp{yang2025generative-SI}
Spectra were used without additional rescaling of intensities (\texttt{scale\_y}=1.0).
To increase the effective number of gradient steps per epoch and reduce variance in the training dynamics, independent, decorrelated noise was applied to identical 128 replicas of dataset.
Models were trained with batch size 32 using 8 data-loading workers on GPUs with distributed data parallelism (DDP), and training progress was monitored using TensorBoard logging.
Instead of relying on a fixed number of training epochs, training was carried out for sufficiently long schedules to ensure full convergence, as verified by agreement between predicted and reference spectra and the absence of further improvement in validation error.
For XANES and EXAFS model, this was usually around 3,500 epochs, and for score models, we trained the models for up to 12,000 epochs.

Both XANES and EXAFS surrogate models employed the same SpecNet architecture\citeSupp{song2021scorebased-SI, Kwon2024SpectroscopyGuided-SI, graphite_llnl-SI} based on a MeshGraphNets-style message-passing neural network.\citeSupp{pfaff2020learning-SI}
The model followed an encoder-processor-decoder structure:
(1) node and edge embeddings were produced by two-layer MLPs with SiLU activations followed by LayerNorm;
(2) latent features were propagated through a Processor module comprising five message-passing blocks with hidden dimension 200;
and (3) a final MLP decoder mapped per-node latent features to predicted spectra.
In this formulation, each node outputs a spectrum, enabling atom-wise supervision and allowing spectra to be aggregated or averaged over absorbing sites as needed.
The XANES model predicted spectra on a grid of 100 points, while the EXAFS model predicted 400 points.
Models were trained using a mean-squared error (MSE) loss between predicted and reference spectra, evaluated separately on training and validation subsets via a per-graph mask.
Optimization used Adam\citeSupp{Kingma2015Adam-SI} with an initial learning rate of $10^{-3}$.
An exponential moving average (EMA) of model parameters was maintained throughout training (decay 0.9999).

\subsection{Molecular dynamics simulations}
\label{sec:mol-dyn-sims}

Classical molecular dynamics (MD) simulations were performed using \texttt{LAMMPS}\citeSupp{Plimpton1995LAMMPS-SI,thompson2022lammps-SI} (2 Aug 2023 -- Update 3) under periodic boundary conditions in all three dimensions.
Simulations were conducted using \texttt{units metal}, with temperature in Kelvin, time in picoseconds, pressure in bars, and distances in \AA.
Atomic interactions were described using the Tersoff potential\citeSupp{Tersoff1988_1-SI, Tersoff1988_2-SI} for a-C, a-Si, and a-SiO$_2$ systems,\citeSupp{Munetoh2007SiO-SI} and the embedded-atom method (EAM/fs) potential for metallic alloys, including Cu--Zr\citeSupp{Mendelev2009CuZr-SI} and Zr--Ni--Al.\citeSupp{Cheng2009BMG-SI}

For each structural variation (density, cooling rate, or composition), 10 independent replicas were generated to ensure statistical sampling.
System sizes were chosen as follows: 216 atoms for a-Si and a-C, 300 atoms for a-SiO$_2$, and 500 atoms for alloy systems.
Density variations for elemental systems were introduced by uniformly scaling the cubic simulation cell prior to NPT run, and after the final NPT relaxation.

All initial configurations were first relaxed using a conjugate-gradient energy minimization with force and energy tolerances of $10^{-10}$ eV/\AA\ and $10^{-10}$ eV, respectively, and maximum iteration limits of $10^5$ steps.
Atomic velocities were sampled from a Maxwell-Boltzmann distribution at 300~K.

Equilibration proceeded in two stages.
First, systems were equilibrated in the canonical (NVT) ensemble at 300~K for 50~ps using a Nosé-Hoover thermostat\citeSupp{Nose1984-SI, Hoover1985-SI} with a temperature damping parameter of 100~fs.
This stage allowed initial equilibration without volume fluctuations.
Subsequently, equilibration continued in the isothermal-isobaric (NPT) ensemble at 300~K and zero external pressure for an additional 50~ps using Nosé-Hoover temperature and pressure control, with thermostat and barostat relaxation times of 100~fs and 1000~fs, respectively.
Isotropic pressure coupling was applied to allow uniform relaxation of the simulation cell.

To generate amorphous structures, equilibrated configurations were melted by heating to 5000~K under NPT conditions at zero pressure and held for 50~ps to ensure complete loss of structural memory.
During this stage, atomic configurations were periodically written to trajectory files, and liquid-phase configurations were stored for subsequent analysis and initialization comparisons.

Amorphization was achieved by linearly cooling the liquid from 5000~K to 300~K under NPT conditions at zero pressure.
For all systems except a-SiO$_2$, cooling was performed over 47~ps, corresponding to an effective cooling rate of approximately 100~K/ps.
For a-SiO$_2$, additional cooling rates of 1~K/ps and 10~K/ps were explored to assess quench-rate dependence.
Isotropic pressure control was maintained throughout cooling to allow density relaxation and eliminate residual volumetric stresses.
While these cooling rates are high compared to experimental conditions, they are sufficient to consistently generate datasets of amorphous configurations that are used in our benchmark study.

Following the quench, systems were further equilibrated at 300~K and zero pressure in the NPT ensemble for an additional 10~ps to remove residual stresses and ensure mechanical equilibration.
Final atomic configurations were written in both data and restart formats for subsequent structural and spectroscopic analysis.
Thermodynamic quantities including temperature, potential energy, pressure, volume, box dimensions, and density were monitored throughout using custom thermo output.
All melt-quench simulations were performed using GPU-accelerated builds of \texttt{LAMMPS} using Kokkos.\citeSupp{Trott2022Kokkos}

\subsection{Score model training}
\label{sec:score-model-training}

We trained score-based generative models to learn a prior over atomic configurations via denoising score matching in periodic systems.
The score network was implemented as a message-passing graph neural network (MeshGraphNets-style)\citeSupp{Kwon2024SpectroscopyGuided-SI, yang2025generative-SI} in \texttt{PyTorch Lightning} and trained using stochastic perturbations of atomic coordinates.
Each atomic configuration was represented as a periodic graph constructed with a fixed neighbor cutoff (default $r_{\mathrm{cut}}=5$~\AA).
Nodes correspond to atoms and edges connect all neighbor pairs within the cutoff; node features consisted of species encodings with $N_{\mathrm{species}}$ channels, and edge attributes contained the Cartesian displacement vector components and distance, i.e., $(\Delta x,\Delta y,\Delta z, r)$.

The score model used an encoder-processor-decoder architecture.
Node and edge features were first embedded using two-layer MLPs with SiLU activations followed by LayerNorm.
Conditioning on the diffusion time/noise level was incorporated through a learnable time embedding applied additively to the node embeddings.
Specifically, scalar time inputs $t$ were mapped to a $d$-dimensional embedding using Gaussian random Fourier features\citeSupp{Tancik2020FourierFeatures-SI} followed by an MLP and LayerNorm, and the resulting embedding was added to each node representation.
The processor consisted of $N_{\mathrm{conv}}=5$ message-passing blocks with hidden dimension $d=200$, and a decoder mapped the final node latents to a 3D per-atom score prediction.
Following common diffusion-model parameterizations, the network output was scaled by the inverse noise level, returning $\hat{\mathbf{s}}_\theta(\mathbf{x},t)/\sigma$.

Training data were loaded using a custom \texttt{StructureSpecDataModule}, configured in ``prior-training'' mode (\texttt{train\_prior=True}).
For each training example, atomic coordinates were perturbed with additive Gaussian noise up to a maximum noise level $k$ (default $k=0.8$), producing noisy coordinates $\tilde{\mathbf{x}} = \mathbf{x} + \sigma \boldsymbol{\epsilon}$ where $\boldsymbol{\epsilon}\sim\mathcal{N}(0,\mathbf{I})$.
The model was trained to predict the negative noise consistent with denoising score matching.\citeSupp{song2021scorebased-SI, yang2025generative-SI}
Specifically, given the model output $\hat{\mathbf{s}}_\theta(\tilde{\mathbf{x}},t)$ and the injected noise $\boldsymbol{\epsilon}$, the loss $\mathcal{L}$ that was minimized was the $L_2$ norm
\[
\mathcal{L}
=
\mathbb{E}\left[
\left\|
\hat{\mathbf{s}}_\theta(\tilde{\mathbf{x}},t)\,\sigma + \boldsymbol{\epsilon}
\right\|_2^2
\right],
\]
implemented as a per-atom squared error averaged over the batch.
This objective allows $\hat{\mathbf{s}}_\theta(\tilde{\mathbf{x}},t)$ to approximate the score of the perturbed data distribution at each noise level.

Models were trained using the Adam optimizer\citeSupp{Kingma2015Adam-SI} with learning rate $10^{-3}$ and distributed data parallelism (DDP).
We used a batch size of 32 and 8 dataloader workers.
An exponential moving average (EMA) of model parameters was maintained during training with decay 0.9999 and used as a more stable estimator of the learned score function for downstream sampling and guidance experiments.
Training was continued until full convergence, as assessed by stabilization of the training loss across epochs.

\subsection{High-Fidelity diffraction calculations}
\label{sec:high-fidelity-diffraction}

X-ray diffraction (XRD) and neutron diffraction (ND) patterns were computed using the Debye scattering formalism as implemented in the \texttt{DebyeCalculator} package.\citeSupp{Trizio2023DebyeCalculator}
Diffraction intensities were evaluated directly from atomic coordinates without assuming periodic lattice symmetry, which is appropriate for amorphous and disordered systems.

For each configuration, atomic species were mapped to their corresponding atomic numbers (for XRD) and neutron scattering lengths (for ND). Diffraction intensities were evaluated on a uniform $q$-grid defined by $(q_{\min}, q_{\max}, \Delta q)$ over the experimentally relevant scattering range.
Pairwise scattering contributions were included up to a real-space cutoff of $20~\text{\AA}$ with a radial resolution of $0.01~\text{\AA}$.
To suppress finite-size artifacts and termination ripples arising from the finite cutoff, an exponential damping factor ($\texttt{qdamp} = 0.04$) was applied.

Thermal and static disorder were incorporated through an isotropic Debye-Waller factor using a constant displacement parameter $B_{\mathrm{iso}} = 1.5~\text{\AA}^2$, representative of amorphous structures under ambient conditions.
No additional structural averaging or symmetry constraints were imposed.

XRD and ND intensities were computed independently by switching the radiation type within the \texttt{DebyeCalculator} (i.e., \texttt{rad\_type = `xray'} or \texttt{`neutron'}) while keeping all other numerical parameters fixed.
This ensures that differences between X-ray and neutron diffraction patterns arise solely from the underlying scattering factors rather than changes in computational settings.

The resulting Debye-calculated diffraction patterns were treated as high-fidelity reference spectra for analysis and validation.
Owing to the $\mathcal{O}(N^2)$ scaling of pairwise Debye summation and the absence of analytic gradients, these calculations are computationally expensive and not suitable for per-step evaluation during denoising.
Instead, they were used to benchmark a set of low-cost, differentiable surrogate diffraction models designed to reproduce the key features of the Debye results (peak positions, relative intensities, and overall envelope) while enabling efficient gradient-based optimization during structure generation.

\subsection{Reverse Monte Carlo simulations}
\label{sec:reverse-mc}

Reverse Monte Carlo (RMC) refinements were performed using the \texttt{FULLRMC} package\citeSupp{aoun2016fullrmc-SI}, which provides a modular constraint-based Monte Carlo engine for refining atomistic configurations against experimental-style structural observables.
In \texttt{FULLRMC}, refinement proceeds by proposing stochastic coordinate perturbations (Monte Carlo trial moves) and accepting or rejecting them to reduce a user-defined objective composed of one or more constraints.
This enables direct refinement against target pair-correlation functions while maintaining periodic boundary conditions and configurable move-generation strategies.

For each structural variation (e.g., different densities for a-Si/a-C), two initialization types were considered: (1) random initial structures and (2) liquid-derived structures extracted from MD melt snapshots.
For each initialization type, 10 independent starting configurations were generated per condition.
Each starting configuration was then refined through 10 independent RMC runs using distinct random seeds, yielding an ensemble of statistically independent refinement trajectories.

Each \texttt{FULLRMC} run was initialized from a periodic structure provided in PDB format and refined against a target pair correlation function stored in an experimental-style \texttt{.exp} file.
Periodic boundary conditions were enforced by extracting the simulation cell vectors from the input structure using ASE and passing the corresponding box parameters to the \texttt{FULLRMC} engine (via \texttt{ENGINE.set\_boundary\_conditions}).
This ensured that all distance evaluations within constraints and all proposed trial displacements were interpreted under PBC consistent with the original simulation cell.

The refinement objective was defined using a \texttt{PairCorrelationConstraint} constructed from the target \texttt{.exp} file.
Unless otherwise specified, pair-correlation contributions were weighted by atomic number (\texttt{weighting="atomicNumber"}), matching the weighting used in our PDF guidance benchmarks.
No additional geometric constraints were applied for the amorphous Si and C systems considered here, in order to isolate the behavior of PDF-only refinement.
Although \texttt{FULLRMC} supports additional constraints such as minimum-distance or molecular integrity constraints, these were not enabled for a-Si/a-C in this work, allowing us to compare the performance of RMC for any arbitrary system without constraints that assume a particular behavior for the structure, which is a more realistic objective in practice.

Trial moves were generated using atom-wise grouping and recursive random selection.
Specifically, atoms were grouped as individual move units using \texttt{ENGINE.set\_groups\_as\_atoms}, and trial moves were drawn using a \texttt{RecursiveGroupSelector} with a \texttt{RandomSelector}.
The recursion parameter \texttt{RECUR} controls how many recursive selections are performed per Monte Carlo cycle and therefore sets the number of attempted moves per atom per outer iteration.
In our implementation, each outer iteration attempted
\[
N_{\mathrm{steps}} = \texttt{RECUR} \times N_{\mathrm{atoms}}
\]
Monte Carlo trial moves, where $N_{\mathrm{atoms}}$ is the number of atom-groups (one group per atom).
To ensure statistical independence across runs, each refinement was assigned a distinct random seed that was applied consistently to both NumPy and Python's built-in random number generator prior to sampling trial moves.

Refinement was carried out in two sequential phases following a common exploration-refinement strategy.
In the exploration phase, the selector emphasizes broader configurational sampling to escape poor local basins and rapidly reduce large mismatches to the target PDF.
In the subsequent refinement phase, trial moves become more conservative and focus on improving local agreement with the constraint.
Each phase was executed for \texttt{RANGE} outer iterations, and within each iteration the engine was advanced for $N_{\mathrm{steps}}$ trial moves, with state saved at the same frequency (\texttt{saveFrequency} = $N_{\mathrm{steps}}$).
Intermediate structures were periodically exported as PDB files (e.g., \texttt{explore\_step\_*.pdb} and \texttt{refine\_step\_*.pdb}) to enable monitoring and post-analysis of refinement trajectories.

\subsection{Paracrystallinity in amorphous silicon}
\label{sec:si:paracrystallinity}

\noindent\textbf{Training data for the score model:}
The structural prior (score model) was trained on a deliberately constrained dataset consisting of 20 LAMMPS melt–quenched amorphous silicon configurations at densities of 2.0 and 2.5 g/cm$^3$ (216-atom periodic cells), which are the same amorphous structures used in the previous benchmark.
In addition, a single 64-atom cubic diamond structure was included to expose the prior model to a crystalline local motif.

\noindent\textbf{Initial configurations:}
All paracrystalline reconstructions were initialized from periodic cubic diamond seeds, which were uniformly rescaled to match the target reference densities prior to denoising.
Periodic boundary conditions were enforced in all three directions, and the simulation cell was held fixed throughout sampling;
only atomic positions were updated during the denoising trajectory.

\noindent\textbf{Reference targets and guidance signal:}
Guidance targets were taken from the reference work from Rosset \textit{et al.}\citeSupp{rosset2025signatures-SI} and consist of 11 independent 1000-atom periodic cells spanning increasing degrees of crystallinity.
These structures were originally generated and sampled using the reference ML-driven MD workflow from that work and were selected here as representative states across the mixed amorphous-crystalline spectrum.
The conditioning observable is a differentiable real-space PDF, $g(r)$, with cutoff $r_\mathrm{max}=12$~\AA{}, 150 histogram bins, and Gaussian smearing $\sigma=0.15$~\AA{}.

\noindent\textbf{Denoising Procedure and guidance weighting:}
Denoising was performed in two stages. First, an \emph{unconditional} denoising phase was carried out using a schedule of $t_{\max}=1.0$, $t_{\min}=0.001$, and $N=512$ discretization steps to generate physically plausible structures under the learned prior. This was followed by a \emph{conditional} denoising phase, in which $g(r)$ guidance was applied with a schedule of $t_{\max}=0.6$, $t_{\min}=0.001$, and $N=512$ discretization steps.
The balance between the learned prior and the $g(r)$ guidance is controlled by a scalar weight $w=3000$.
This choice provided stable convergence toward the target correlation while maintaining physically plausible local environments under the prior model (see Section \ref{sec:prior-guidance-init} for a discussion on this topic).

\noindent\textbf{Sampling statistics and outputs:}
For each reference structure, 10 independent denoising trajectories were performed using 1000-atom periodic simulation cells.
For the experimental spectrum, where the density is not explicitly known, ten candidate densities were considered by uniformly rescaling the initial cubic diamond cells prior to denoising.
For each density, 27 independent trajectories were performed, yielding a total of 270 independent runs for the experimental guidance curve.

\noindent\textbf{Post-denoising analysis:}
No post-denoising structural relaxation was performed.
Even minor relaxation steps were found to induce measurable changes in $g(r)$, which can correspond to substantial shifts in crystallinity fraction and medium-range order.
To preserve strict consistency with the guided spectroscopic signal, all reported analyses are based directly on the final denoised configurations.

\noindent\textbf{Melt-quench comparison protocol:}
To compare against structures guided by experimental ion-implantation data, additional denoising simulations were performed using reference PDFs obtained from melt-quenched configurations.
These simulations used the same score model, identical 1000-atom simulation cells and density (2.29 \gcm), and the same denoising setup, including noise schedule, guidance weight, number of discretization steps, and initialization from cubic diamond seeds.
Unlike earlier benchmarks that used random initial configurations, all runs here were initialized identically to ensure consistency across comparisons.
Under these controlled conditions, the only difference between simulation (melt-quench) guided and experimentally (ion-implantation) guided denoising trajectories is the target PDF used for guidance, enabling a direct assessment of how differences in the conditioning signal influence the resulting structures.

\noindent\textbf{Structural classification via polyhedral template matching:}
Local atomic environments were characterized using polyhedral template matching (PTM)\citeSupp{Larsen2016PTM} as implemented in \texttt{OVITO},\citeSupp{Stukowski2010OVITO} which identifies crystal-like coordination motifs by comparing atomic neighborhoods to reference templates via a root-mean-square deviation (RMSD) metric. An RMSD cutoff of 0.1 was used to classify atoms as locally ``crystal-like,'' following the reference methodology.\citeSupp{rosset2025signatures-SI}

\subsection{Liquid-liquid phase transformation in sulfur}
\label{sec:si:sulfur}

\noindent\textbf{Initial configurations:}
Initial configurations were constructed as randomly distributed S$_8$ molecular rings under periodic boundary conditions, with random positions and orientations subject to non-overlap constraints.
Starting from intact S$_8$ rings preserves realistic intramolecular bonding topology and prevents unphysical bond formation during early denoising (see Section \ref{sec:spectral-opt-limitations} for a full discussion on this).
Because the experimental density was not explicitly reported for the samples in Fig. \ref{fig:sulfur}A of the main paper, density was treated as a hyperparameter and selected via targeted grid search prior to final denoising (Section \ref{sec:grid-search-shell-weighted-pdf}).

\noindent\textbf{Reference targets and guidance signal:}
Experimental pair correlation functions $g(r)$ were extracted from the work from Henry \textit{et al.}\citeSupp{SI-henry2020Liquid}
Since the data extend only to $\sim 7$~\AA, the high-$r$ tail was extrapolated assuming asymptotic decay to a constant value.
Reference curves were interpolated onto a uniform radial grid consistent with the differentiable PDF model.
To stabilize optimization, target curves were anchor-normalized by matching $g(r)$ near $r = 7.0$~\AA\ (window width $0.2$~\AA) to the value computed from the initial configuration.
This avoids trivial amplitude mismatches and stabilizes early-stage gradients.
The differentiable PDF module employs a Gaussian kernel with width $\sigma = 0.15$~\AA, cutoff $8.0$~\AA, and 100 radial bins.
Gradients are propagated to atomic coordinates via automatic differentiation.

\noindent\textbf{Sampling statistics and outputs:}
For each experimental sample, 10 independent random S$_8$ initializations were generated, and 5 independent denoising trajectories were run per initialization, resulting in 50 independent conditional denoising runs per sample.

\noindent\textbf{Guidance weighting:}
The relative weighting between the generative prior and PDF guidance was controlled by a scalar parameter $w$.
A value of $w = 5000$ was selected to balance structural validity against agreement with experimental $g(r)$, ensuring stable convergence without overwhelming the learned prior (Section \ref{sec:prior-guidance-init}).

\subsection{Structure of ball-milled medium-density amorphous ice}
\label{sec:si:water}

\noindent\textbf{Training Data for the Score Model:}
The generative prior used for MDA reconstruction via PDF-guided denoising was intentionally trained on a minimal dataset to isolate the role of spectroscopic guidance.
The training set consisted of a single low-density amorphous (LDA) ice configuration containing 2880 H$_2$O molecules at a density of $0.920~\mathrm{g\,cm^{-3}}$ from Rosu-Finsen \textit{et al.} \citeSupp{rosu2023medium-SI}
No additional amorphous or crystalline ice structures were included in training, ensuring that neither crystalline I$h$ nor MDA configurations were explicitly encoded in the model prior.

\noindent\textbf{Initial Configurations:}
All denoising runs were initialized from hexagonal ice (I$h$).
The structure was uniformly rescaled from its original density of $0.923$ \gcm~ to the target MDA density of $0.970$ \gcm.
The simulation cell dimensions were held fixed throughout the denoising process.

\noindent\textbf{Obtaining a reference target:}
PDFs from MDA configurations from the reference work (8640-atom supercells generated via repeated shear deformation and geometry optimization) were used as guidance targets.
This approach assumes that the initial $g(r)$ proposed in the original work are a good starting point, given that the limited range of the experimental $S(Q)$ does not allow an accurate reconstruction of $g(r)$ as a starting point.
In most cases, knowledge of $S(Q)$ is often sufficient to propose a trial $g(r)$ from the experimental data alone.
If the structural models were not present, a potential solution would be to use methods that perform regression over the space of $S(Q)$ to create a family of trial $g(r)$ curves,\citeSupp{Shanks2024Bayesian} which can be used to generate structures.
In this case, generated structures could propose new $S(Q)$ values, potentially leading to a self-consistent solution.

\noindent\textbf{Reference targets and guidance signal:}
PDFs were computed over a fitting range of $10~\mathrm{\AA}$ and smoothed using a Gaussian kernel with width $\sigma = 0.12~\mathrm{\AA}$ to ensure differentiability.
A total of 200 radial bins were used during guidance.
For visualization in the main text, partial PDFs were computed using our differentiable implementation with a reduced cutoff of $7~\mathrm{\AA}$ and 150 radial bins.
Structure factor curves were calculated using the DebyeCalculator,\citeSupp{} which differs from the implementation used in the reference work; this methodological difference accounts for minor discrepancies in reported $S(Q)$ curves.

\noindent\textbf{Sampling statistics and outputs:}
For each of the five reference configurations, eight independent denoising trajectories were performed using 1080-atom simulation cells to reduce computational cost.
In total, 40 denoising trajectories were generated and used for structural analysis.

\noindent\textbf{Denoising Procedure:}
Structure generation was performed using conditional denoising implemented in GLASS, guided by neutron-weighted PDFs.
An overlap constraint was enforced throughout the trajectory: any proposed atomic displacement resulting in an interatomic distance smaller than $0.95~\mathrm{\AA}$ was rejected on a per-atom basis while preserving atomic displacements that do not lead to overlap.
The denoising schedule consisted of 256 steps with linearly decreasing step sizes from 1.0 to 0.001, enabling large-scale rearrangements in early stages followed by fine-grained relaxation.

\noindent\textbf{Guidance Weighting:}
The balance between the generative prior and PDF guidance was controlled by a scalar parameter $w$.
A value of $w = 50$ was selected to maintain structural plausibility while achieving spectroscopic fidelity and stable convergence.
This value of $w$ differs from the previous sections due to the neutron-weighting rescaling the value of the PDF loss.

\noindent\textbf{Post-Denoising Relaxation:}
To strenghten the analysis, we sure that the final configurations and hydrogen-bond networks  represent an energy minimum in the potential energy surface of MDA.
As an approximation for this, we used the SevenNet foundational machine-learned interatomic potential\citeSupp{kim_sevennet_mf_2024} (version 7net-mf-ompa-omat24).
Energy minimization was performed for 200 steps using the FIRE optimizer with loose convergence thresholds: energy tolerance $E_{\mathrm{tol}} = 1\times10^{-5}$ kcal/mol and force tolerance $F_{\mathrm{tol}} = 1\times10^{-2}$ kcal/mol\,\AA$^{-1}$.

\clearpage

\section{Supplementary Text}
\customlabel{sec:stext}{Supplementary Text}

\subsection{Inversion of multi-modal spectroscopic data}
\label{sec:inversion-multi-modal}

Combining multiple spectroscopic modalities is a natural route to resolving structural degeneracies, as each modality operates on different physical observables, obey different selection rules, and is sensitive to different length scales and structural motifs.\citeSupp{Egami2003Underneath-SI, Rehr2000RMP, SI_Billinge2007Science}
For instance, pair distribution functions (PDF) describe pairwise distance correlations \citeSupp{Egami2003Underneath-SI, Terban2021ChemRev}, while X-ray absorption spectroscopy (XAS) focus on local coordination geometry and electronic environment. \citeSupp{Rehr2000RMP, SI_Koningsberger1988XAFS}
However, the dimensionality of the joint inverse problem grows with each added modality, making it challenging to refine structures given heterogeneous data streams. \citeSupp{SI_McGreevy2001RMCReview, SI_Soper2001EPSR}
Furthermore, there is little quantitative evidence of how different modalities lift degeneracies in the amorphous structure space.
As a result, multi-modal structure determination has largely relied on experts who manually select, weight, and reconcile inputs from different experiments on a case-by-case basis.\citeSupp{SI_McGreevy2001RMCReview, SI_Soper2001EPSR}

\subsection{Interplay between prior, guidance, and initialization in denoising}
\label{sec:prior-guidance-init}

The performance of score-based denoising is governed by a three-way interplay between the learned prior score, the guidance signal, and the choice of initialization.
Successful denoising requires a simultaneous balance of all three: a transferable prior, informative and self-consistent guidance, and an initialization that lies within the overlapping knowledge space spanned by both.

A transferable prior defines the physically plausible region of configuration space and prevents the sampler from exploring unphysical amorphous structures.
In practice, this includes suppressing unrealistic coordination environments, unphysical bond lengths, atomic overlaps, and implausible local motifs.
Priors tend to transfer better for short-range-dominated systems, such as a-Si, a-C, a-SiO$_2$, as well as dense metallic glasses (Cu-Zr, Zr-Ni-Al), where local coordination and packing constraints dominate the structural landscape.
In these cases, minimal prior knowledge is often sufficient to keep sampling within a reasonable amorphous manifold, explaining the successes in our previous work.\citeSupp{yang2025generative-SI}
In contrast, medium-range-dominated systems (e.g., liquid sulfur, MDA) exhibit much higher structural variability;
here, the prior must encode richer structural information or the initialization must be placed closer to the target for denoising to succeed.

Guidance, derived from structural or spectroscopic observables, actively drives the system toward configurations that reproduce a specific target signal, which often means pushing the structure away from the knowledge space learned by the prior model.
A good guidance signal must contain sufficient information content and, critically, be internally self-consistent.
Here, self-consistency means that the computational spectrum used for guidance must first be able to accurately reproduce the corresponding reference or experimental spectrum; otherwise, guidance forces may reflect model deficiencies rather than true structural discrepancies.
What constitutes sufficient information is strongly system-dependent.
For example, PDF guidance provides strong constraints on short-range structure and can effectively suppress unphysical short bonds, particularly when additional weight is applied at small $r$.
This makes PDF guidance especially robust for dense systems such as Cu-Zr alloys, even without explicitly enforcing physical constraints in the score model.

\begin{figure}[htb!]
   \centering
    \includegraphics[width=0.9\textwidth]
    {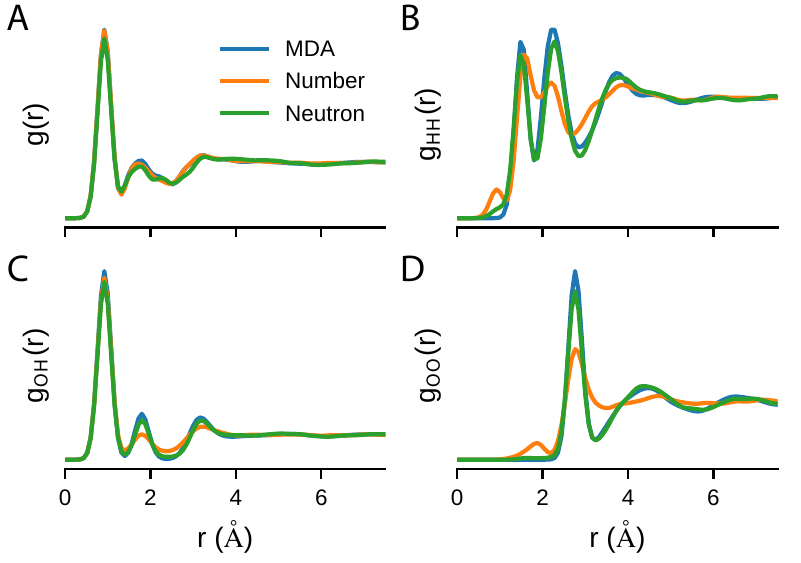}
    \caption{
    \textbf{Effect of guidance weighting on recovered structural correlations in MDA.}
    \textbf{A.} Number-weighted total pair distribution functions $g(r)$ obtained from denoising guided by either number-weighted or neutron-weighted $g(r)$. The resulting total $g(r)$ profiles are nearly indistinguishable relative to the reference, indicating that both guidance schemes can reproduce the aggregate structural signal when evaluated in a number-weighted form.
    Corresponding number-weighted partial pair distribution functions ($g_{\mathrm{HH}}(r)$ \textbf{B.}, $g_{\mathrm{OH}}(r)$ \textbf{C.}, and $g_{\mathrm{OO}}(r)$) \textbf{D.}  reveal clear discrepancies between the two guidance schemes. Despite similar agreement in the total $g(r)$, guidance based on number-weighted $g(r)$ fails to accurately recover the underlying partial correlations, particularly in the medium-range O--O network, due to the dominance of intramolecular O--H and H--H contributions. In contrast, neutron-weighted guidance better preserves physically relevant structural information.}
    \label{si:fig:mda_different_weighting}
\end{figure}

However, apparent agreement with a single guidance signal does not necessarily imply structural correctness.
In liquid water, a number-weighted PDF can reproduce the total PDF extremely well while yielding severely incorrect partial PDFs;
only neutron-weighted PDFs provide self-consistent constraints on O-O, O-H, and H-H correlations (see Fig. \ref{si:fig:mda_different_weighting}.
In contrast, guidance from observables such as ADF, XRD, ND, or EXAFS lacks direct sensitivity to short-range repulsion.
When used alone, these signals can drive the structure toward matching the target spectrum while allowing severe atomic overlaps or unphysical local environments.
In such cases, guidance can push the system outside the prior's learned knowledge space, destabilizing the denoising trajectory despite apparent spectral agreement.

A careful balance between prior and guidance is therefore required.
The prior must be strong enough to regularize sampling and suppress unphysical structures, while the guidance must be informative and self-consistent enough to steer the system toward the target observation without overwhelming the prior.
This balance is inherently system dependent: short-range-dominated systems generally tolerate broader initialization choices and weaker priors, whereas medium-range-dominated systems may require either richer priors, stronger guidance, or initializations placed closer to the target configuration.
In this work, we showed both cases.

Initialization mediates this balance by determining whether the starting configuration lies within the overlapping information space spanned by both the prior and the guidance.
When this overlap is large, denoising is robust and many starting points converge successfully.
When it is small, valid initializations become limited and outcomes become highly sensitive, emphasizing that denoising is not purely a function of the target observable but of the interaction between prior knowledge, guidance strength, and starting structure.

\subsection{Differential spectral profiling and computational scaling}
\label{sec:diff-spectral-scaling}

To enable spectroscopic guidance during structure denoising, we implemented a set of differentiable surrogate models for PDF, ADF, XRD, and ND that balance physical fidelity with computational efficiency.
Pair distribution functions (PDFs) and angular distribution functions (ADFs) were computed using fully differentiable internal \texttt{PyTorch} implementations, allowing gradients with respect to atomic positions to be obtained directly via automatic differentiation.
Neutron and X-ray diffraction intensities were initially benchmarked against \texttt{DebyeCalculator}\citeSupp{Trizio2023DebyeCalculator} to establish a high-fidelity reference, with gradients first computed using finite-difference schemes implemented through a custom automatic differentiation interface.

\begin{figure}[htb!]
   \centering
    \includegraphics[width=0.8\textwidth]
    {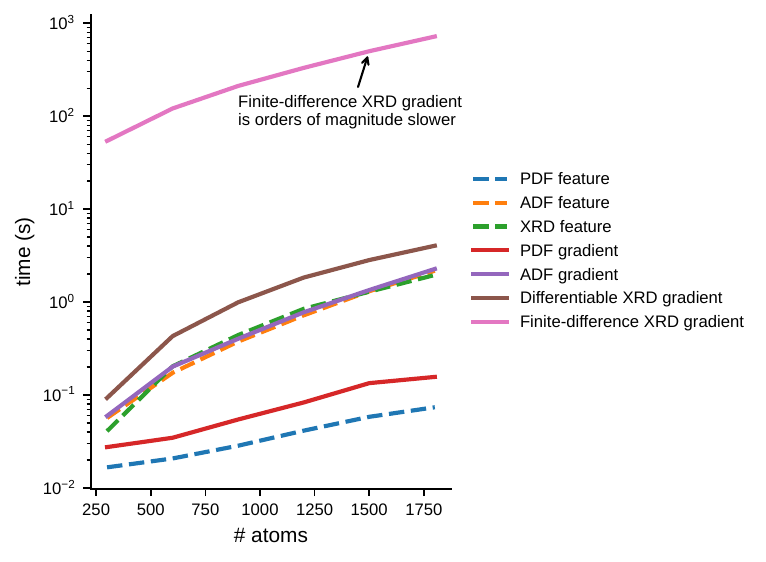}
    \caption{
    \textbf{Computational cost of spectroscopic feature extraction and gradient evaluation.}
    Wall-time for computing spectroscopic observables and their gradients as a function of system size (number of atoms). All timings correspond to wall-clock time measured on CPU hardware under default multi-threaded execution. Dashed lines denote the cost of feature extraction for the PDF, ADF, and XRD observables, while solid lines denote the corresponding gradient calculations used during conditional denoising. Gradients for PDF and ADF are computed using differentiable formulations with costs comparable to their forward evaluations, whereas differentiable XRD gradients are moderately more expensive. For comparison, the cost of computing XRD gradients using finite-difference approximations is also shown, which is orders of magnitude more expensive than the differentiable implementation. All benchmarks were performed on the same underlying atomic configurations while varying the number of atoms (300–1800) to evaluate scaling with system size.}
    \label{si:fig:profile_combined}
\end{figure}

We performed systematic scaling tests to quantify the computational cost of (1) feature extraction alone and (2) feature extraction combined with gradient evaluation as a function of system size (approximately $300$--$1800$ atoms), shown in Fig. \ref{si:fig:profile_combined}.
For PDF and ADF, both forward evaluation and gradient computation scale efficiently, with total runtimes remaining on the order of $10^{-2}$--$10^{-1}$ seconds even for the largest systems tested.
These costs are dominated by local neighbor list construction and smooth kernel evaluations, making PDF and ADF well suited for use at every denoising step.

In contrast, XRD gradient evaluation using finite-difference differentiation is substantially more expensive (Fig. \ref{si:fig:profile_combined}).
While forward XRD feature extraction remains comparatively inexpensive, the cost of gradient computation exceeds $50$ seconds for systems of $\sim1500$ atoms and increases steeply with system size.
This dramatic slowdown arises from the combination of quadratic pairwise distance scaling inherent to Debye summations and the repeated spectrum evaluations required by finite-difference gradients.
Direct profiling shows that finite-difference XRD gradients are more than an order of magnitude more expensive than their fully differentiable PDF and ADF counterparts.

These scaling results motivated the development and adoption of low-cost, fully differentiable diffraction surrogate models for XRD and ND.
The differentiable surrogates reproduce the key trends, peak shapes, and relative intensities of Debye-calculated spectra while reducing gradient evaluation times by more than an order of magnitude.
This improvement makes it feasible to evaluate diffraction-based guidance at every denoising step, enabling practical spectroscopic conditioning during score-based structure generation.

\subsection{Differentiable calculations of PDF, ADF, XRD, and ND}
\label{sec:diff-spectra}

To enable spectroscopic guidance during denoising, we implemented differentiable surrogate models for the radial distribution function (RDF; used as a PDF-like real-space descriptor), angular distribution function (ADF), X-ray diffraction (XRD), and neutron diffraction (ND) directly in \texttt{PyTorch}.
All observables were formulated as smooth functions of atomic coordinates, allowing gradients with respect to atomic positions to be obtained via automatic differentiation.

\paragraph{PDF (real space).}

Element-resolved partial pair distribution functions $g_{ij}(r)$ were computed using a fully differentiable formulation under periodic boundary conditions.
For a configuration of $N$ atoms with Cartesian positions $\{\mathbf{x}_n\}_{n=1}^{N} \subset \mathbb{R}^3$ and simulation cell matrix $\mathbf{H} \in \mathbb{R}^{3\times3}$ (volume $V = |\det \mathbf{H}|$), atomic species were provided as one-hot encodings and converted to discrete type labels $t_n$ via an $\arg\max$ operation.
For each unordered type pair $(i,j)$, directed partial correlation functions were first constructed.
Let $\mathcal{I}_i = \{n \mid t_n = i\}$ and $\mathcal{I}_j = \{m \mid t_m = j\}$ denote index sets of atoms of types $i$ and $j$, with cardinalities $N_i$ and $N_j$.
Pairwise displacement vectors were formed as $\mathbf{r}_{nm} = \mathbf{x}_n - \mathbf{x}_m$ for $n \in \mathcal{I}_i$, $m \in \mathcal{I}_j$.
Periodicity was enforced using the minimum-image convention in fractional coordinates: displacements were transformed via $\mathbf{s}_{nm} = \mathbf{r}_{nm}\mathbf{H}^{-T}$, wrapped as $\tilde{\mathbf{s}}_{nm} = \mathbf{s}_{nm} - \mathrm{round}(\mathbf{s}_{nm})$, and mapped back to Cartesian space $\tilde{\mathbf{r}}_{nm} = \tilde{\mathbf{s}}_{nm}\mathbf{H}^T$, yielding minimum-image distances $d_{nm} = \|\tilde{\mathbf{r}}_{nm}\|_2$.
Self-pairs were excluded using a small threshold $d_{nm} > 10^{-5}$~\AA, and distances were restricted to a fixed radial window $r \in [r_{\min}, r_{\max}]$ (default $0$--$8$~\AA).
A uniform radial grid $\{r_k\}_{k=1}^{K}$ with spacing $\Delta r$ was defined over this interval, and instead of hard histogram binning, a Gaussian kernel density estimator was employed.
The unnormalized smooth pair counts were evaluated as
\[
w_k^{(i\rightarrow j)} =
\sum_{n \in \mathcal{I}_i}
\sum_{m \in \mathcal{I}_j}
\exp\!\left(
-\frac{(d_{nm} - r_k)^2}{2\sigma^2}
\right)
\frac{\Delta r}{\sqrt{2\pi}\sigma},
\]
where $\sigma$ is the Gaussian bandwidth (default $0.15$~\AA).
The prefactor $\Delta r / (\sqrt{2\pi}\sigma)$ ensures correct normalization of the kernel contribution in the continuum limit.
The directed partial RDF was then obtained by normalizing against the ideal-gas expectation for a spherical shell,
\[
g_{i\rightarrow j}(r_k) =
\frac{
w_k^{(i\rightarrow j)}
}{
N_i \rho_j \, 4\pi r_k^2 \Delta r + \epsilon
},
\qquad
\rho_j = \frac{N_j}{V},
\]
where $\rho_j$ is the number density of species $j$ and $\epsilon$ is a small numerical constant added for stability near $r \to 0$.
For off-diagonal pairs ($i \neq j$), symmetry was enforced by averaging $g_{i\rightarrow j}(r_k)$ and $g_{j\rightarrow i}(r_k)$, ensuring consistency with the standard definition of partial pair correlation functions.
All operations were expressed using dense tensor algebra without neighbor-list approximations, leading to $\mathcal{O}(N_i N_j)$ scaling per type pair.
Because the full procedure consists of smooth tensor operations, gradients $\partial g_{ij}(r_k)/\partial \mathbf{x}_n$ were obtained directly via automatic differentiation, enabling stable integration of PDF-based loss functions within gradient-driven denoising dynamics.

\paragraph{ADF (real space).}

Element-resolved angular distribution functions (ADFs) were computed from three-body correlations under periodic boundary conditions.
Using the minimum-image displacement vectors $\tilde{\mathbf{r}}_{nm}$ defined as above, a neighbor mask was constructed by requiring interatomic distances $d_{nm} = \|\tilde{\mathbf{r}}_{nm}\|_2$ to satisfy $d_{nm} < r_{\mathrm{cut}}$ (default $5$~\AA), excluding self-pairs.
For each central atom $j$, all neighbors $i$ and $k$ within this cutoff were enumerated to form ordered triplets $(i,j,k)$ such that both $i$ and $k$ lie within the local coordination shell of $j$, without imposing any constraint on the $i$-$k$ separation.
The bond angle at atom $j$ was computed as
\[
\theta_{ijk}
=
\arccos\!\left(
\frac{
\tilde{\mathbf{r}}_{ji} \cdot \tilde{\mathbf{r}}_{jk}
}{
\|\tilde{\mathbf{r}}_{ji}\|_2 \, \|\tilde{\mathbf{r}}_{jk}\|_2 + \epsilon
}
\right),
\]
where $\epsilon$ is a small numerical constant to ensure stability, and the cosine argument was clipped to the interval $[-1,1]$ to prevent numerical overflow.
Atomic species were converted from one-hot encodings to discrete type labels, and triplets were grouped according to the canonical type ordering $(\min(t_i,t_k),\, t_j,\, \max(t_i,t_k))$, thereby enforcing symmetry with respect to exchange of the outer atoms $i$ and $k$.

Angles were accumulated onto a fixed grid $\{\theta_\ell\}$ spanning the interval $[0,\pi]$ with a specified number of bins.
To preserve differentiability and avoid discontinuities associated with hard angular binning, Gaussian kernel smoothing was applied.
For each triplet type and grid point $\theta_\ell$, the smooth histogram contribution was evaluated as
\[
h_\ell
=
\sum_{(i,j,k)}
\exp\!\left(
-\frac{(\theta_{ijk} - \theta_\ell)^2}{2\sigma^2}
\right),
\]
where $\sigma$ is the angular bandwidth (default $0.1$ rad).
Contributions were accumulated using tensor index addition to produce element-resolved ADF histograms.
Optionally, a global normalization was applied by dividing each histogram by its total sum to obtain a probability density over angles.
All operations were implemented using dense tensor algebra without neighbor-list approximations, ensuring that the mapping from atomic coordinates to ADF histograms is fully differentiable and compatible with automatic differentiation for gradient-based denoising dynamics.

\paragraph{XRD (reciprocal space).}

Differentiable X-ray diffraction (XRD) spectra were computed using a Debye-style pair summation expressed entirely in differentiable tensor operations.
For a configuration of $N$ atoms with Cartesian positions $\{\mathbf{x}_n\}_{n=1}^{N}$, the intensity was evaluated on a uniform reciprocal-space grid $\{q_m\}_{m=1}^{M}$ constructed via $q_m = q_{\min} + m\,\Delta q$.
Element-dependent atomic form factors were parameterized analytically as
\[
f_\alpha(q)
=
\sum_{k} a_{\alpha k} \exp(-b_{\alpha k} q) + c_\alpha ,
\]
where $\alpha$ denotes the chemical species and $\{a_{\alpha k}, b_{\alpha k}, c_\alpha\}$ are tabulated coefficients.
For multi-component systems, per-atom form factors were obtained via species-weighted mixing using one-hot encodings,
\[
f_n(q) = \sum_{\alpha} s_{n\alpha}\, f_\alpha(q),
\]
where $s_{n\alpha}$ is the one-hot species indicator.

Unique atomic pairs were constructed using upper-triangular indexing ($i<j$) to avoid double counting. Pairwise distances were computed as
\[
r_{ij} = \|\mathbf{x}_i - \mathbf{x}_j\|_2 ,
\]
with a small numerical floor $r_{ij} \ge \varepsilon$ applied to prevent singularities at $r \to 0$.
An alternative short-range threshold $r_{ij} \ge r_{\mathrm{thres}}$ could have also been imposed to exclude extremely small separations.
For each reciprocal-space point $q_m$, pair contributions were evaluated using the Debye kernel
\[
\mathrm{sinc}(q_m r_{ij})
=
\frac{\sin(q_m r_{ij})}{q_m r_{ij}},
\]
with a stabilized implementation such that $\mathrm{sinc}(x) \to 1$ as $x \to 0$ to avoid numerical instability.
The total pair contribution was therefore
\[
I_{\mathrm{pair}}(q_m)
=
\sum_{i<j}
2 f_i(q_m) f_j(q_m)
\frac{\sin(q_m r_{ij})}{q_m r_{ij}} .
\]

Self-scattering contributions were included explicitly as
\[
I_{\mathrm{self}}(q_m)
=
\frac{1}{2}
\sum_{i=1}^{N}
f_i(q_m)^2 ,
\]
yielding the undamped intensity
\[
I(q_m)
=
I_{\mathrm{pair}}(q_m)
+
I_{\mathrm{self}}(q_m).
\]

Thermal and static disorder were incorporated via an isotropic Debye-Waller damping factor
\[
D(q_m)
=
\exp\!\left(
-\frac{q_m^2 B_{\mathrm{iso}}}{8\pi^2}
\right),
\]
with default $B_{\mathrm{iso}} = 1.5~\text{\AA}^2$.
The final intensity was computed as
\[
I_{\mathrm{final}}(q_m)
=
D(q_m)\, I(q_m)
+
I_{\mathrm{self}}(q_m)\,[D(q_m) - 1],
\]
which preserves the correct treatment of self-terms under damping consistent with the implemented formulation.
All operations were performed using dense tensor contractions over pair indices, resulting in $\mathcal{O}(N^2)$ scaling.
Because the entire computation consists of smooth tensor operations (including stabilized $\mathrm{sinc}$ evaluation), gradients $\partial I(q_m)/\partial \mathbf{x}_n$ are obtained directly via automatic differentiation.

\paragraph{ND (reciprocal space).}

Neutron diffraction (ND) intensities were computed using the same Debye summation framework described above, replacing X-ray form factors with element-specific coherent neutron scattering lengths.
For each chemical species $\alpha$, a constant coherent scattering length $b_\alpha$ was used, and per-atom scattering amplitudes were obtained via species-weighted mixing from one-hot encodings,
\[
b_n = \sum_{\alpha} s_{n\alpha}\, b_\alpha .
\]
A site occupancy factor $o_n$ (defaulting to unity) multiplying $b_n$ was also included, allowing partial occupancy effects to be incorporated explicitly.

Using upper-triangular pair indexing ($i<j$), the pair contribution to the intensity at reciprocal-space point $q_m$ was evaluated as
\[
I_{\mathrm{pair}}(q_m)
=
\sum_{i<j}
2\, o_i o_j\, b_i b_j\,
\frac{\sin(q_m r_{ij})}{q_m r_{ij}},
\]
with the same stabilized $\mathrm{sinc}$ implementation and optional short-range distance threshold as described for XRD.
The self-scattering term was accumulated as
\[
I_{\mathrm{self}}
=
\frac{1}{2}
\sum_{i=1}^{N}
(o_i b_i)^2,
\]
and added uniformly across all $q_m$, yielding
\[
I(q_m)
=
I_{\mathrm{pair}}(q_m)
+
I_{\mathrm{self}}.
\]

When enabled, isotropic thermal or static disorder was incorporated via a Debye-Waller damping factor
\[
D(q_m)
=
\exp\!\left(
-\frac{q_m^2 B_{\mathrm{iso}}}{8\pi^2}
\right),
\]
applied to the total intensity,
\[
I_{\mathrm{final}}(q_m)
=
D(q_m)\, I(q_m).
\]

Because neutron scattering lengths are independent of $q$, the ND formulation differs from XRD only in the absence of $q$-dependent atomic form factors and in the explicit inclusion of occupancy weighting. As with the XRD implementation, all operations were expressed in differentiable tensor form, enabling direct computation of $\partial I(q_m)/\partial \mathbf{x}_n$ for use in spectroscopically guided denoising.

\subsection{Sensitivity of RMC to initialization}
\label{sec:rmc-init-sensitivity}

The comparatively poorer RMC performance observed for a-Si relative to a-C in the heat-map analysis arises primarily from the strong dependence of RMC refinement on the quality of the initial configuration.
Reverse Monte Carlo (RMC) refinement is a local, constraint-driven optimization procedure that explores configuration spaces on a step-wise, trial-and-error basis;
as a result, it converges reliably only when the starting structure already lies within the basin of attraction of the target amorphous state, or in the limit of long simulations.
When initialized too far from the target structure, RMC refinements can become trapped in metastable configurations that satisfy pair correlation constraints locally while failing to recover the correct short- and medium-range order.

\begin{figure}[htb!]
   \centering
    \includegraphics[width=0.8\textwidth]
    {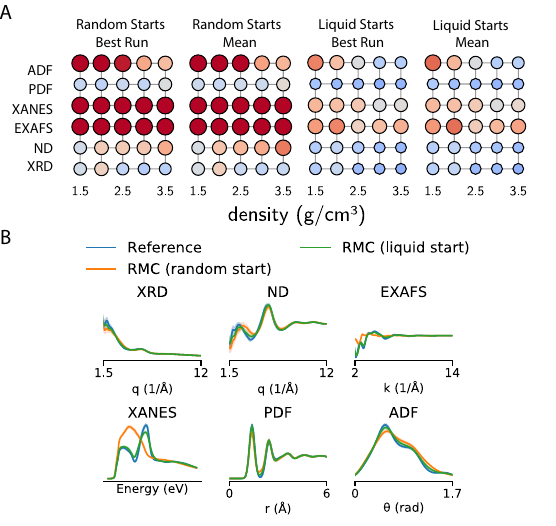}
    \caption{
    \textbf{Effect of initialization on Reverse Monte Carlo (RMC) refinement for amorphous carbon (a-C).}
    \textbf{A.} Heatmap summary of RMC performance across spectroscopic observables for different initialization strategies.
    For each condition, ten independent starting configurations were generated, and each was refined using ten independent RMC runs.
    Two evaluation modes are shown: \textit{best run}, which selects the run with the lowest PDF error from the best-fitting sample, and \textit{mean}, which averages over all runs and samples.
    Because the selection criterion is based on the PDF guidance signal, the best-run strategy does not necessarily yield lower errors across other spectroscopic observables, reflecting occasional overfitting to the PDF.
    The dominant effect instead arises from the initialization: starting from liquid configurations consistently improves agreement across spectra compared with random initializations. \textbf{B.} Representative spectra at $\rho = 2.5$ \gcm~ illustrating this behavior.
    Solid lines show the mean spectrum, and shaded regions indicate the standard deviation across ensembles (multiple RMC runs and samples for the refined structures, and multiple configurations for the reference). Liquid-initialized refinements produce spectra that more closely match the reference across the observables than those obtained from random starting structures.}
    \label{si:fig:rmc_C}
\end{figure}

\begin{figure}[htb!]
   \centering
    \includegraphics[width=0.8\textwidth]
    {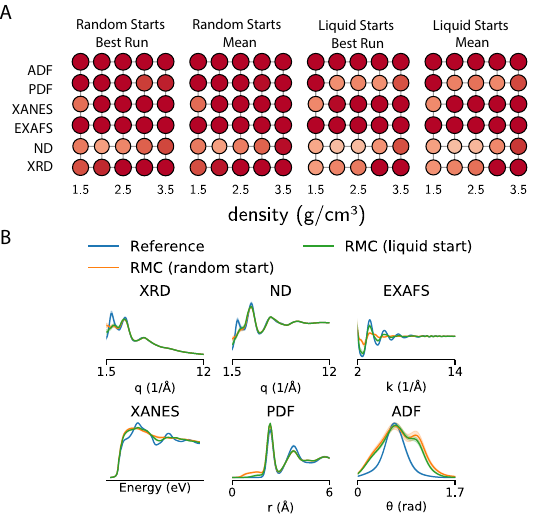}
    \caption{\textbf{Effect of initialization on Reverse Monte Carlo (RMC) refinement for amorphous silicon (a-Si).} \textbf{A.} Heatmap summary of RMC performance across spectroscopic observables for different initialization strategies. As in the a-C case, ten independent starting configurations were generated for each condition, and each was refined with ten independent RMC runs. Unlike a-C, neither random nor liquid initializations yield consistently good agreement with the reference structures across the observables. This behavior arises because the initial PDFs of both random and liquid configurations differ substantially from the target amorphous structure, making it difficult for RMC to reach the correct structural basin during refinement. \textbf{B.} Representative spectra illustrating this limitation. Both random and liquid starts remain noticeably offset from the reference spectra across multiple observables, highlighting the strong dependence of RMC on the quality of the initial configuration.
    }
    \label{si:fig:rmc_Si}
\end{figure}

\begin{figure}[htb!]
   \centering
    \includegraphics[width=0.5\textwidth]
    {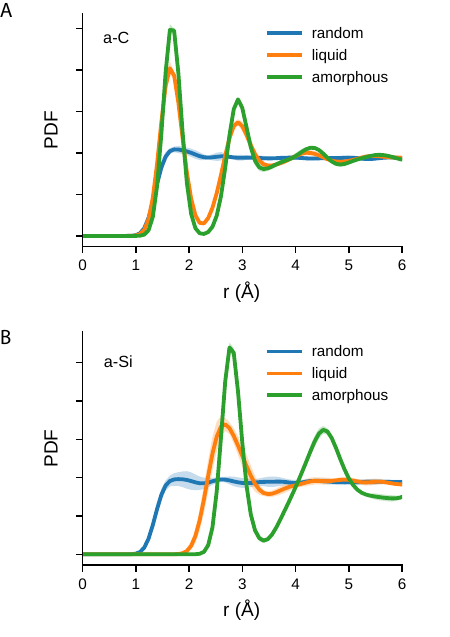}
    \caption{
    \textbf{Comparison of different structural states for \textbf{A.} C and \textbf{B.} Si at 1.5 \gcm.}
    Pair distribution functions (PDFs) are shown for three representative starting configurations: randomly placed atoms, liquid configurations, and amorphous structures. The liquid and amorphous configurations were obtained from LAMMPS molecular dynamics simulations. Solid lines denote the mean PDFs averaged over multiple configurations, while shaded regions indicate the corresponding standard deviations. For a-C, the liquid structure closely resembles the amorphous state, whereas random configurations differ substantially. This structural similarity explains why reverse Monte Carlo (RMC) refinements converge more reliably when initialized from liquid configurations. In contrast, for a-Si the liquid structure remains structurally distinct from the amorphous network, making liquid initialization less advantageous and highlighting the stronger dependence of the reconstruction on the choice of starting configuration.
    }
    \label{si:fig:pdf_compare_Si_C}
\end{figure}

This behavior is directly reflected in the RMC performance (Figs. \ref{si:fig:rmc_C}, \ref{si:fig:rmc_Si}, \ref{si:fig:pdf_compare_Si_C}), where successful matching of the PDF does not consistently translate into accurate agreement with other spectroscopic observables.
In other words, RMC can achieve a good fit to the imposed constraint while remaining structurally inconsistent with respect to unconstrained features.
This decoupling between PDF agreement and cross-spectral consistency is usually not observed in the denoiser-based approach, where spectroscopic guidance is applied continuously during structure generation, leading to coordinated improvements across multiple modalities rather than isolated optimization of a single observable.

The systematically poorer RMC performance for a-Si relative to a-C is primarily driven by initialization quality.
In our setup, the random initialization procedure for a-C uses a distance cutoff that happens to be close to the physically correct first-shell separation, so the starting configuration is already near the target short-range structure.
This is evident in the liquid vs.\ amorphous PDF comparison for a-C, where the first (and even second) coordination shells are already well aligned, indicating an initialization that is effectively near-converged before refinement.

In contrast, for a-Si the random initialization produces atoms that are systematically closer than physically expected (due to the chosen cutoff), yielding an incorrect first-shell peak position and local coordination environment.
Correcting this requires substantially larger collective rearrangements to reach the target PDF, which is difficult for RMC because it proceeds through local Monte Carlo moves and is highly sensitive to the starting basin.
As a result, even the PDF used as guidance can remain relatively more poorly matched for a-Si, and the mismatch propagates across other observables.
This explains the worse performance of RMC-generated a-Si across spectra modalities.
Liquid-derived starting structures improve convergence relative to purely random starts, but for a-Si the refinements can still remain far from the reference across spectra if the initial configuration is not already sufficiently close to the target amorphous state.

\subsection{Error and diversity metrics for spectral evaluation}
\label{sec:error-diversity-metrics}

To quantitatively assess denoising performance across different spectroscopic modalities, we defined and computed two complementary metrics: reference diversity and denoising error.
Reference diversity captures the intrinsic variability of spectra arising from structural fluctuations across an ensemble of independently generated amorphous replicas.
Denoising error, on the other hand, measures the deviation between denoised spectra and the corresponding reference spectra.

For each spectrum type (PDF, ADF, XRD, ND, EXAFS, and XANES), spectra were first grouped by feature type (e.g., pair types in PDF, triplet types in ADF, or element-resolved contributions in diffraction and absorption spectra).
A global normalization factor was computed per spectrum type as the maximum absolute intensity across all reference spectra, ensuring that error and diversity metrics are directly comparable across different spectral modalities and feature types.

Reference diversity was quantified as the mean standard deviation of spectra within the reference ensemble, averaged over the spectral axis and normalized by the global normalization factor.
This metric reflects the intrinsic spread of physically equivalent amorphous structures and provides a lower bound on the error that can be meaningfully resolved.
Denoising error was computed as the mean absolute deviation between the denoised spectrum and the ensemble-averaged reference spectrum, again normalized by the same global factor.
For denoised structures, spectra were first averaged over multiple independent denoising runs per sample to reduce stochastic noise before error evaluation.
Both error and diversity represent normalized fluctuations of intensities over the same spectral modality, and thus are dimensionless quantities on the same scale.
Even within the same spectral modality, normalization is performed separately for each combination of elemental pairs/triplets.

To enable direct comparison between denoising performance and intrinsic structural variability, we analyzed the difference $(\text{error} - \text{diversity})$ for each spectrum type and feature.
Negative or near-zero values indicate that the denoised spectra fall within the intrinsic variability of the reference ensemble, corresponding to ideal or near-ideal denoising behavior.
This framework allows meaningful cross-comparison across different spectroscopic probes and length scales, from short-range angular correlations (ADF) to long-range structural features (PDF, XRD, ND), and provides a consistent criterion for evaluating model generalization across synthesis conditions such as cooling rate and density.

\subsection{Physical constraints in the score model}
\label{sec:phys-constraints-score}

As one of the only constraints for the score model, namely preventing the formation of unphysical short-range contacts during stochastic exploration in denoising, a hard minimum-distance veto was applied.
Because GNNs are known to behave poorly in highly unrealistic conditions (e.g., overlapping atoms), we avoided this scenario without necessarily limiting the model's ability to explore the phase space.
After a displacement is proposed, the resulting trial configuration was examined for interatomic separations below a prescribed threshold.
Atoms participating in such pairs had their proposed displacements suppressed for that timestep, while all other atoms were allowed to evolve normally.
This procedure was optionally iterated within a single timestep to robustly eliminate overlaps.
The constraint operates purely at the level of the sampling dynamics and does not modify the learned score function or introduce explicit interatomic potentials.

Such physical constraints are particularly useful in multi-element systems where unconstrained stochastic updates can readily generate atomic overlaps, albeit for different physical reasons.
In dense metallic alloys such as Cu-Zr, high packing fractions make the system intrinsically sensitive to even moderate displacements, leading to frequent short-range overlaps during early denoising.
In molecular liquids such as water, overlaps occur every time that proposed structures along a denoising trajectories violate the expected chemical rules (e.g., short-range repulsion, hydrogen bond networks), which can steer the generation away from sensible, ``valid'' structures.
In both cases, the presence of unphysical short bonds can trap the denoising trajectory in configurations from which meaningful structural or spectroscopic guidance becomes ineffective.

An alternative mechanism for suppressing unphysical short-range behavior arises naturally when pair distribution function (PDF) guidance is employed.
Because the PDF directly encodes interatomic distance information, assigning sufficient weight at small $r$ strongly penalizes short bonds and eliminates overlaps even in the absence of other explicit constraints in the score model.
We observed that this implicit regularization, however, is specific to PDF-based guidance.
When other spectroscopic modalities (e.g., XRD, ND, EXAFS, or XANES) are used as guidance signals, no direct penalty exists for short-range interatomic distances, making the denoising process significantly more susceptible to atomic overlap and local collapse unless additional physical constraints are enforced.
This distinction explains why PDF guidance is consistently more effective at stabilizing denoising trajectories, especially in dense or multi-element systems, whereas other spectroscopic observables alone provide weaker constraints on short-range structure.

\subsection{Complementarity of spectral modalities}
\label{sec:spectral-comp}

The results in the main text showcase an asymmetry between spectroscopic modalities used to condition structure generation.
Rationalizing these differences requires describing how structural information is encoded in different observables.
PDFs directly constrain real-space interatomic distances across both short- and intermediate-length scales, imposing geometric constraints that propagate across modalities.
Reciprocal-space and absorption probes instead samples more limited or delocalized structural features.
Consistent with this interpretation, XANES exhibits the lowest reconstruction error regardless of the guidance signal, making it comparatively the ``easiest'' spectrum to reproduce (Fig. \ref{fig:benchmark}G of the main paper).
In contrast, ADF is consistently the most difficult modality to recover.
Defined with a short-range cutoff, ADF encodes only highly localized angular correlations and lacks sufficient sensitivity to intermediate-range order, making it insufficient as a standalone guidance signal.

\begin{figure}[htb!]
   \centering
   \includegraphics[width=\textwidth]{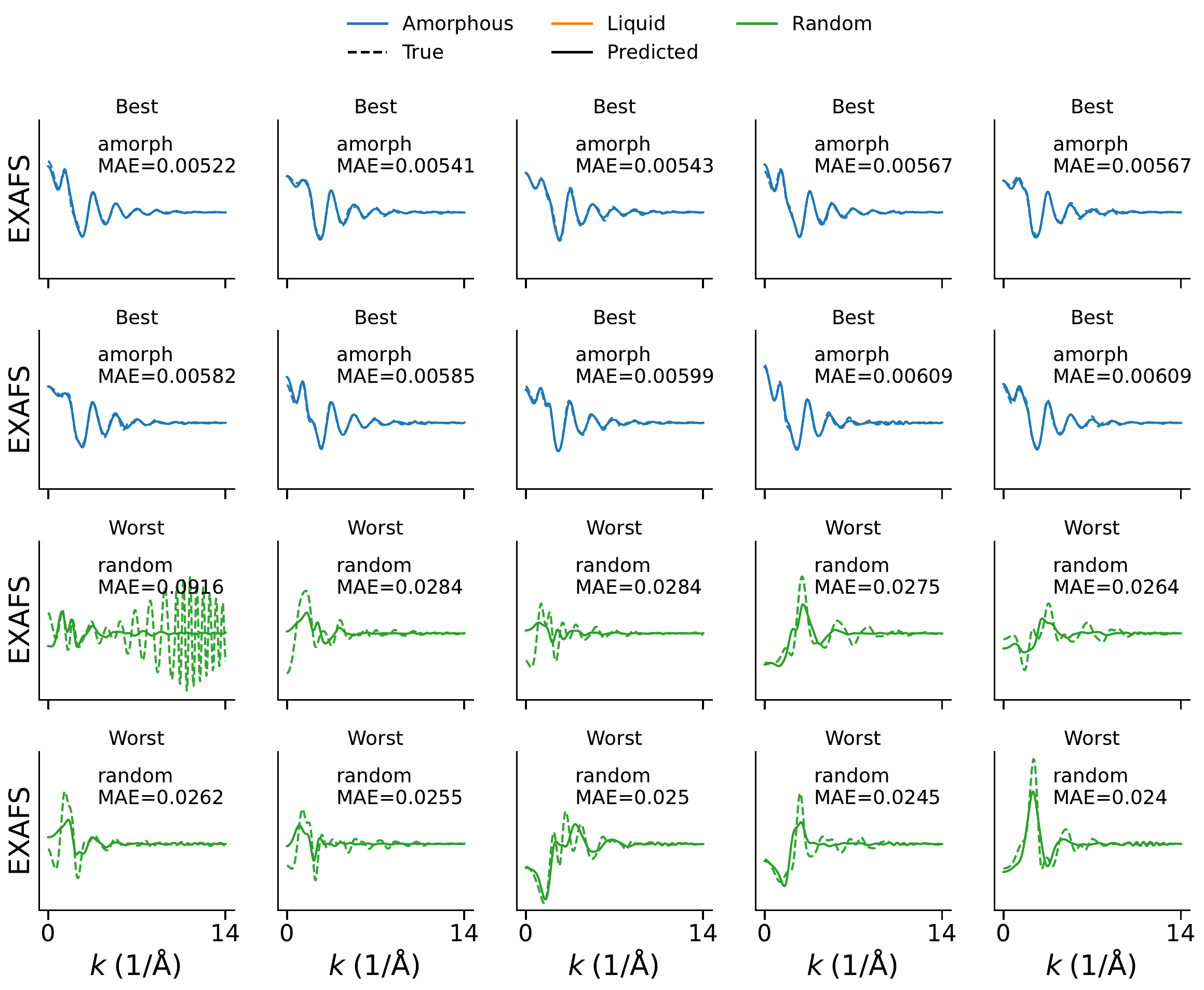}
   \caption{
    \textbf{Atom-wise EXAFS spectra for Si at $\rho = 2.5$ \gcm.}
    Atom-resolved EXAFS spectra predicted by the neural network model compared against reference spectra for validation atoms.
    Each panel shows the EXAFS signal $\chi(k)$ for an individual atom, with dashed lines indicating the reference spectrum and solid lines indicating the model prediction. Panels are ranked by per-atom spectral error, quantified as the mean absolute error over the full $k$-range (0--14 \AA$^{-1}$), computed across a validation set of 644 atom-wise spectra randomly sampled from amorphous and liquid configurations (obtained from LAMMPS molecular dynamics simulations), as well as randomly placed Si structures (which also serve as the initial configurations for denoising). The top two rows show the ten best-fitted atoms, while the bottom two rows show the ten worst-fitted atoms. Line color encodes the structural environment of the atom (blue: amorphous; orange: liquid; green: random). Notably, all best-predicted configurations originate from amorphous structures, reflecting their higher structural similarity to the training distribution, whereas all worst-predicted configurations correspond to randomly generated Si structures.} \label{si:fig:01_Si_exafs_atomwise_best10_worst10_2.5}
\end{figure}

One possible explanation for the worse performance of XANES and EXAFS is the fact that their surrogate approximations rely on GNNs rather than an analytical function such as PDF or XRD.
Some of our results showed that while model transferability was excellent (Fig. \ref{si:fig:01_Si_exafs_atomwise_best10_worst10_2.5}), it was not perfect for random structures.
To discard this explanation, we note that our reconstruction benchmarks were performed by first performing a pass of unconditional generation from a random initialization, then using conditional generation to refine the structure.
This initial step of unconditional generation drastically reduces the intrinsic randomness of the data, allowing the XANES and EXAFS models to provide reliable predictions of spectra.
While this still does not guarantee that the XAS predictions are going to be perfectly reconstructed across the conditional generation trajectory, it guarantees that the poor reconstruction quality is not necessarily due to the model, but also due to the spectroscopic modality.
Nevertheless, reliance on GNNs for reconstructing structures may not be practical as a scalable solution for the future, and are better employed for systems for which there is good control of the environments.

\subsection{Discussion on paracrystallinity in amorphous silicon}
\label{sec:paracrystallinity-modeling-a-si}

\textbf{Information content limits and medium-range underprediction.}
The systematic underprediction of crystallinity in the intermediate regime originates from intrinsic information-content limitations of the PDF.
Across the 1-–52\% crystallinity range, the variation in $g(r)$ is modest, particularly beyond the first coordination shell.
The short-range order remains largely invariant due to the robustness of tetrahedral Si bonding, and medium-range oscillations differ only subtly in amplitude and damping.
In contrast, real-space structural changes are substantial: crystalline domains grow, connect, and reorganize at mesoscopic length scales.
Thus, large differences in domain connectivity correspond to relatively small perturbations in the projected radial distribution.
This creates an ill-conditioned inverse problem in which significant structural variation must be inferred from weak spectral signals.

The intermediate crystallinity regime is additionally the most out-of-distribution region relative to the learned structural prior.
The score model was trained primarily on low-crystallinity configurations together with a crystalline endpoint.
While this enables interpolation between endpoints, the mid-range states are less densely represented in the training distribution.
When spectral gradients are weak and multiple configurations produce similar $g(r)$, the prior may be exerting a mild bias toward lower crystallinity.
Consequently, crystallinity is slightly underpredicted in the intermediate regime.
Importantly, however, the predicted trend remains monotonic and stable across independent runs, indicating that reconstruction fidelity is limited primarily by the information content of the observable rather than by model instability or insufficient capacity.

\noindent\textbf{Connectivity.}
A related concern of structure generation is with respect to the local bonding topology.
Although the framework reliably reconstructs medium-range ordering and domain morphology, the final structures retain a small fraction of three- and five-fold coordinated Si atoms.
In our study of 270 structures, the fraction of 4-coordinated Si atoms was $84 \pm 2$ \%, which is lower than simulated a-Si structures, but quite close to realistic models compared to some well-studied RMC-proposed structures.\citeSupp{cliffe2017structural-SI}
Generated structures also show a negligible or exactly zero number of atoms with coordination number higher than 5 or smaller than 3, as expected from typical distributions.
This behavior arises because the denoising process operates in continuous coordinate space without explicitly enforcing discrete bonding constraints.
While the prior statistically favors tetrahedral environments, strict four-fold coordination is not guaranteed at every refinement step.
Moreover, the PDF observable is primarily sensitive to radial correlations rather than bond topology; small fractions of coordination defects may remain spectroscopically compatible and therefore energetically neutral within the optimization landscape.
Addressing this limitation will likely require incorporating stronger physical constraints, such as connectivity-aware regularization, bond-order penalties, hierarchical priors, or topology checks during denoising.
Developing such mechanisms represents a longer-term methodological extension and lies beyond the scope of the present study.
Importantly, however, given that crystalline environments are observed in structures generated from the experimental PDF, the prediction of crystallinity in generated samples is a lower bound.
Thus, small deviations from 4-coordination environments does not change the conclusions of the crystallinity of samples.

\subsection{Training dataset construction for prior model and example with sulfur liquid-liquid transition}
\label{sec:training-dataset-sulfur}

The training configurations used for the sulfur score model were obtained from the dataset accompanying Ref.~\citeSupp{yang2024structure-SI}, available at \url{https://github.com/EnricoTrizio/sulfur_lambda_transition/tree/main/CV_training/EXAMPLE_training_data}, where they were originally provided as an example dataset for a collective-variable (CV) training workflow.
The dataset consists of configurations generated in 512-atom periodic cells at three temperatures (550 K, 650 K, and 750 K) for two limiting structural states relevant to the sulfur liquid--liquid transition: a pure S$_8$-ring liquid and a fully polymeric liquid network.
Each state contains 10 stored frames per temperature, yielding a total of 60 configurations (2 states $\times$ 3 temperatures $\times$ 10 frames).
Although the nominal dataset size is modest, the configurations densely sample the two dominant structural manifolds of liquid sulfur: ring-dominated and polymer-dominated networks.
Many frames correspond to small fluctuations within the same topological basin, resulting in substantial structural overlap across temperatures and trajectories.

Importantly, the training of the score-based model does not require knowledge of the physics or simulation details used to generate the training data, such as cooling rates, temperatures, and so on.
The score model does not learn time evolution but instead estimates gradients of the structural probability density across configurations.
As long as the dataset spans the relevant structural basins, the exact dynamical pathway connecting them does not materially affect the learned structural prior.
In retrospect, using all 60 configurations is not necessarily the most data-efficient choice, as much of the structural information is redundant, but it errs on the side of larger datasets and, thus, stronger priors.
However, this redundancy provides a useful robustness test: despite the limited diversity and shallow sampling depth, the trained score model is able to learn a meaningful structural prior sufficient for stable conditional denoising across densities and pressure conditions.

\begin{figure}[htb!]
   \centering
    \includegraphics[width=0.5\textwidth]
    {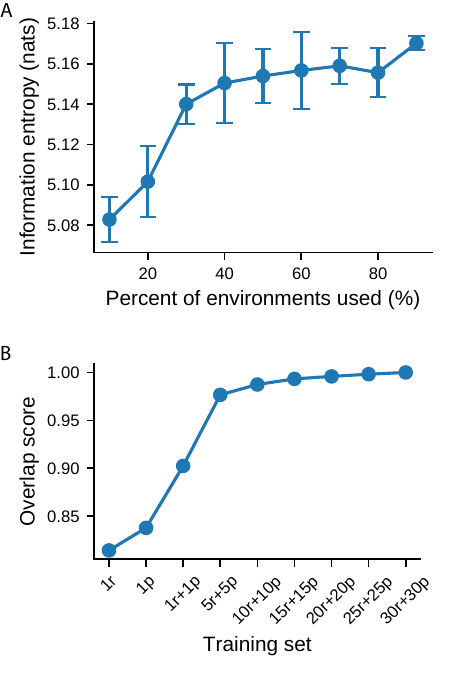}
    \caption{
    \textbf{Training-set sufficiency analysis for the sulfur score model.}
    \textbf{A.} Entropy-based learning curve obtained by randomly sampling fractions of the available configurations and computing the information entropy of the resulting environment distribution.
    Error bars denote the standard deviation across multiple sampling runs. The entropy rapidly saturates as additional environments are included, indicating that the structural diversity of the dataset is captured with relatively few configurations.
    \textbf{B.} Structural overlap score as a function of the number of configurations drawn from the two limiting structural manifolds of liquid sulfur: S$_8$ ring-dominated (r) and polymeric (p) networks. Training sets are constructed by progressively combining configurations from both manifolds (e.g., 1r, 1p, 1r+1p, 5r+5p, etc.). The overlap score quickly approaches unity once representative configurations from both structural states are included, demonstrating that only minimal manifold coverage is required to capture the relevant structural gradients governing the sulfur liquid–liquid transition.}
\label{si:fig:S_train_eval}
\end{figure}

To quantify data sufficiency, we performed systematic training-set scaling and entropy analyses (Fig. \ref{si:fig:S_train_eval}) based on our prior work.\citeSupp{schwalbekoda2025information-SI}
Both the structural overlap metric and the entropy-based learning curve show rapid saturation once representative configurations from both ring and polymer manifolds are included, confirming that only minimal manifold coverage is required to capture the structural gradients governing the LDL-HDL transition.
The full 60-configuration dataset therefore represents a conservative and redundant sampling rather than a necessary minimum.
Thus, this analysis demonstrates that the conditional denoising framework remains robust even when trained on structurally sparse data, highlighting the efficiency of score-based modeling in low-dimensional structural manifolds such as liquid sulfur.

\subsection{Limitations of multi-objective spectral optimization for amorphous structure generation}
\label{sec:spectral-opt-limitations}

The liquid sulfur results provide a concrete illustration of the broader interplay between the prior model, guidance signal, and initialization discussed above.
While guidance drives denoising toward configurations that reproduce the target experimental observables, the prior defines the region of configuration space considered physically plausible, and initialization determines whether the denoising trajectory begins within the overlapping knowledge space shared by both.
When this overlap is limited, matching the guidance signal alone is insufficient to ensure recovery of a meaningful amorphous structure.

\begin{figure}[htb!]
   \centering
    \includegraphics[width=\textwidth]
    {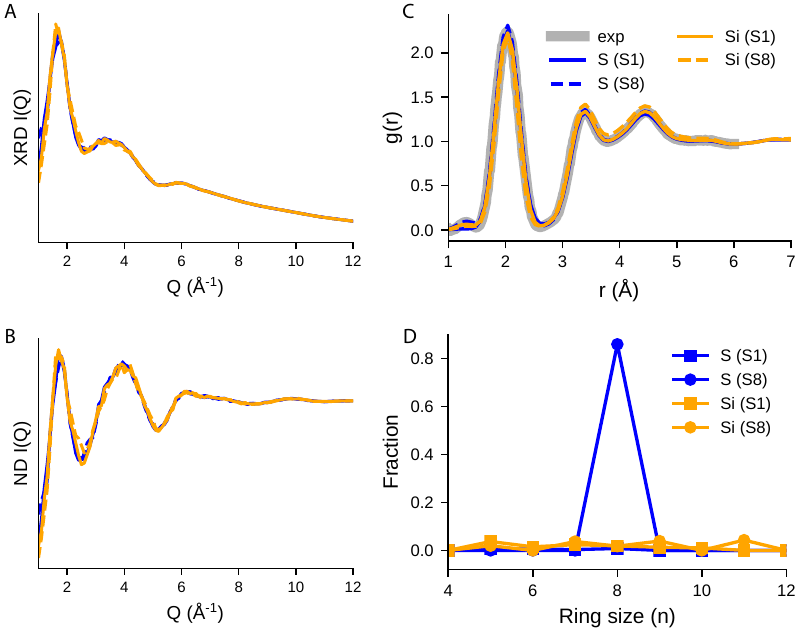}
    \caption{
    \textbf{Influence of prior model and initialization on sulfur structure generation}
    \textbf{A.} X-ray diffraction (XRD) intensity profiles for structures generated using different score-model priors and initial configurations.
    \textbf{B.} Neutron diffraction (ND) spectra for the same structures.
    \textbf{C.} Total pair distribution function $g(r)$ compared with the experimental reference at 0.11 GPa (gray band).
    \textbf{D.} Ring-size distribution showing the fraction of atoms participating in $n$-membered rings.
    Two different score-model priors are compared: an S-trained model (blue), trained on literature sulfur configurations containing both isolated S$_8$ rings and polymeric chains (Sec. \ref{sec:si:sulfur}), and a Si-trained model (orange), identical to the prior used in the paracrystalline silicon case.
    Two initialization strategies are used: \textbf{S1}, randomly distributed sulfur monomers (solid lines), and \textbf{S8}, randomly distributed S$_8$ rings (dashed lines).
    }
\label{si:fig:S_vary_model_init}
\end{figure}

In liquid sulfur, conditional denoising successfully reproduces the experimental total PDF across a wide range of models and starting configurations, demonstrating the strong constraining power of PDF guidance.
However, this apparent success masks structural ambiguity.
We compared denoising outcomes using two distinct priors:
a Si-trained model, which has no explicit knowledge of sulfur ring formation or polymeric connectivity, but only 4-connected networks;
and an S-trained model trained on literature sulfur configurations including isolated S$_8$ rings and polymeric chains across multiple temperatures.
Despite their vastly different prior knowledge spaces, both models converge to nearly identical $g(r)$ curves under spectral guidance (Fig. \ref{si:fig:S_vary_model_init}A).
Even higher-order observables such as XRD and ND become nearly indistinguishable once the PDF is matched, underscoring that guidance alone does not uniquely constrain the underlying structure (Figs. \ref{si:fig:S_vary_model_init}B,C).

Structural analysis in Fig. \ref{si:fig:S_vary_model_init}D reveals that the differences instead emerge from the interaction between the prior and the initialization.
Ring-tracking results show that the Si-trained model rapidly collapses into unphysical configurations lacking stable S$_8$ motifs, despite excellent agreement with experimental spectra.
The S-trained model, although more physically informed, still struggles to form stable crown-shaped S$_8$ rings when initialized from random sulfur monomers, reflecting the difficulty of transitioning from a high-entropy solution to a low-entropy one even with strong priors.
In contrast, initializing from randomly distributed S$_8$ rings places the system closer to the prior's learned knowledge manifold, preserving realistic connectivity and yielding significantly more stable and physically meaningful structures.

These results demonstrate that understanding the interplay between multi-objective spectral optimization is required for amorphous structure generation.
Across different priors and initializations, all cases can reproduce the experimental PDF and even higher-order spectra while converging to fundamentally different atomic networks (Fig. \ref{si:fig:S_vary_model_init}A--C).
This behavior reflects the trade-offs discussed earlier: guidance can pull structures toward the target observation, but without a sufficiently transferable prior and physically meaningful initialization, amorphous structure generation can converge to spectrally correct yet structurally degenerate solutions, even in multi-modal cases.
Our GLASS approach is able to efficiently navigate these solutions, from which structural analysis can be performed to further interpret the results.

\subsection{Selection of initial densities via grid search and shell-weighted PDF loss}
\label{sec:grid-search-shell-weighted-pdf}

For the sulfur systems studied here, the experimental density was not explicitly reported, introducing an additional degree of freedom in the denoising setup.
Since the score model was trained at a fixed density ($1.78$ g/cm$^3$), and density directly affects the medium-range features of the pair distribution function, we explicitly treated density as a hyperparameter to be benchmarked prior to analyzing liquid--liquid transition trends from the denoised results.

For each pressure-temperature that the sample is subject to, we performed a density grid search over the range $1.80$--$1.92$ \gcm (1.80, 1.82, 1.84, 1.86, 1.88, 1.90, 1.92 \gcm).
At each density point, we generated 10 random S$_8$-ring initial configurations and performed 5 independent denoising runs per initialization, yielding 50 candidate configurations per density.

Density selection was based on minimizing a shell-weighted $g(r)$ loss evaluated over the range $3.8$--$5.0$~\AA, corresponding to the third and fourth coordination shells.
This region captures the medium-range structural rearrangements associated with the LDL--HDL transition in liquid sulfur and is more sensitive to density variations than the first coordination shell, which is strongly constrained by the S$_8$-ring initialization and local bonding.

Across all pressure conditions and repeated trials, the same optimal density was consistently identified.
For all systems considered, the best-matching density was found to be $1.82$ g/cm$^3$, independent of pressure.
This consistency indicates that density selection is not dominated by stochastic effects in the denoising process, but instead reflects a genuine structural preference imposed by the experimental PDF in the medium-range regime.

Importantly, this density optimization step highlights a key aspect of the denoising framework: even when the score model and guidance are fixed, global structural parameters such as density must be chosen to place the initialization within the overlapping information space of the prior and the guidance.
Performing a targeted density search prior to final denoising significantly improves convergence stability and ensures that subsequent structural interpretations are not confounded by an implicit density mismatch.
On the other hand, while pressure is well-defined in simulations that compute potential energy surfaces, it is not defined in the context of structural fitting.
Therefore, a treatment of structure generation where density is the variable parameter is more appropriate as it allows multiple solutions across variable densities as opposed to potential collapse of cell parameters in solutions that are not proven to be unique.

\subsection{Accounting for scattering weights and partial correlations in multi-component spectroscopic guidance}
\label{sec:scattering-weights-correlations}

For multi-component systems such as amorphous ice, the effectiveness of spectroscopic guidance is governed by how the experimental observable weights different atomic correlations.
Using medium-density amorphous ice (MDA) as a representative example, we find that number-weighted total PDFs provide an incomplete guidance signal, as they are dominated by strong intramolecular first-shell O-H and H-H correlations.
As a result, excellent agreement in the total, number-weighted $g(r)$ does not imply correct recovery of the underlying partial correlations, particularly the medium-range O-O network that distinguishes amorphous ice polyamorphism.

Incorporating element-dependent scattering weights into the guidance fundamentally changes this behavior.
X-ray-weighted PDFs enhance sensitivity to O-involving correlations (O-O and O-H), improving reconstruction of the oxygen network but still underrepresenting H-H correlations due to weak sensitivity to light elements.
Neutron-weighted PDFs, by contrast, introduce strong contrast between H and O, most notably through the opposite signs of the neutron scattering lengths, thereby providing balanced sensitivity across all partials.
This neutron-weighted guidance pushes H-H correlations substantially closer to the target and yields a more faithful reconstruction of the underlying hydrogen-bond network.

Direct guidance using partial PDFs provides the strongest and most unambiguous constraint, yielding the best overall reconstruction of both total and partial correlations.
However, such information is rarely available experimentally.
In practice, neutron-weighted guidance represents a crucial intermediate solution, supplying richer and more balanced structural constraints than number- or X-ray-weighted totals alone.

These results show that for medium-range-dominated, multi-element systems like MDA ice, denoising success can be limited not only by the expressiveness of the prior, but also by the information content and self-consistency of the guidance signal.
Even with a transferable prior and reasonable initialization, inadequate guidance can trap the optimization in structurally incorrect yet spectroscopically compatible configurations, reinforcing that good guidance is as essential as a good prior within the denoising framework.


\section{Supplementary Figures}

\begin{figure}[htb!]
   \centering
    \includegraphics[width=0.8\textwidth]{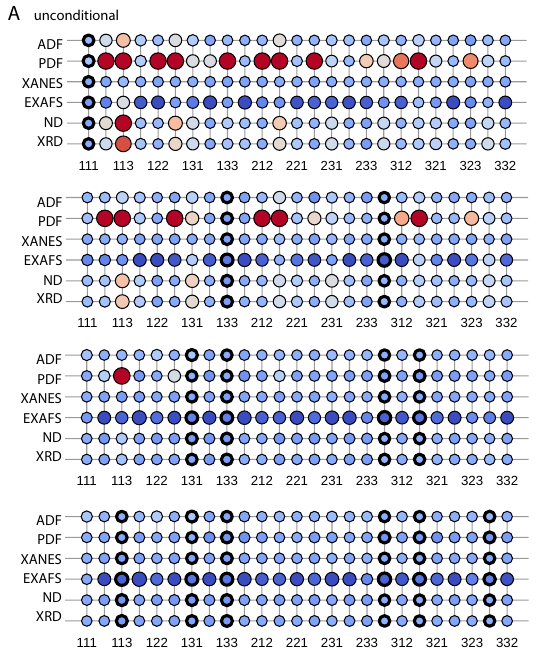}
    \label{si_figs/ZrNiAl_heatmaps_A.pdf}
\end{figure}

\clearpage
\begin{thebibliography}{10}
\expandafter\ifx\csname url\endcsname\relax
  \def\url#1{\texttt{#1}}\fi
\expandafter\ifx\csname urlprefix\endcsname\relax\def\urlprefix{URL }\fi
\providecommand{\bibinfo}[2]{#2}
\providecommand{\eprint}[2][]{\url{#2}}

\bibitem{Elliott1990Amorphous}
\bibinfo{author}{Elliott, S.~R.}
\newblock \emph{\bibinfo{title}{{Physics of Amorphous Materials}}} (\bibinfo{publisher}{Longman}, \bibinfo{year}{1990}).

\bibitem{Billinge2007Science}
\bibinfo{author}{Billinge, S. J.~L.} \& \bibinfo{author}{Levin, I.}
\newblock \bibinfo{title}{{The Problem with Determining Atomic Structure at the Nanoscale}}.
\newblock \emph{\bibinfo{journal}{{Science}}} \textbf{\bibinfo{volume}{316}}, \bibinfo{pages}{561--565} (\bibinfo{year}{2007}).

\bibitem{kaduk2021Powder}
\bibinfo{author}{Kaduk, J.~A.} \emph{et~al.}
\newblock \bibinfo{title}{{Powder Diffraction}}.
\newblock \emph{\bibinfo{journal}{{Nature Reviews Methods Primers}}} \textbf{\bibinfo{volume}{1}}, \bibinfo{pages}{77} (\bibinfo{year}{2021}).

\bibitem{Treacy1996ActaCryst}
\bibinfo{author}{Treacy, M. M.~J.} \& \bibinfo{author}{Gibson, J.~M.}
\newblock \bibinfo{title}{{Fluctuation Electron Microscopy: A Probe of Medium-Range Order in Disordered Materials}}.
\newblock \emph{\bibinfo{journal}{{Acta Crystallographica Section A}}} \textbf{\bibinfo{volume}{52}}, \bibinfo{pages}{212--220} (\bibinfo{year}{1996}).

\bibitem{Voyles2001Nature}
\bibinfo{author}{Voyles, P.~M.}, \bibinfo{author}{Muller, D.~A.}, \bibinfo{author}{Grazul, J.~L.}, \bibinfo{author}{Citrin, P.~H.} \& \bibinfo{author}{Gossmann, H.-J.~L.}
\newblock \bibinfo{title}{{Medium-Range Order in Amorphous Silicon Observed by Fluctuation Electron Microscopy}}.
\newblock \emph{\bibinfo{journal}{{Nature}}} \textbf{\bibinfo{volume}{416}}, \bibinfo{pages}{826--829} (\bibinfo{year}{2001}).

\bibitem{Sheng2006Nature}
\bibinfo{author}{Sheng, H.~W.}, \bibinfo{author}{Luo, W.~K.}, \bibinfo{author}{Alamgir, F.~M.}, \bibinfo{author}{Bai, J.~M.} \& \bibinfo{author}{Ma, E.}
\newblock \bibinfo{title}{{Atomic Packing and Short-to-Medium-Range Order in Metallic Glasses}}.
\newblock \emph{\bibinfo{journal}{{Nature}}} \textbf{\bibinfo{volume}{439}}, \bibinfo{pages}{419--425} (\bibinfo{year}{2006}).

\bibitem{deringer2021origins}
\bibinfo{author}{Deringer, V.~L.} \emph{et~al.}
\newblock \bibinfo{title}{{Origins of structural and electronic transitions in disordered silicon}}.
\newblock \emph{\bibinfo{journal}{{Nature}}} \textbf{\bibinfo{volume}{589}}, \bibinfo{pages}{59--64} (\bibinfo{year}{2021}).

\bibitem{McGreevy1988RMC}
\bibinfo{author}{McGreevy, R.~L.} \& \bibinfo{author}{Pusztai, L.}
\newblock \bibinfo{title}{{Reverse Monte Carlo Simulation: A New Technique for the Determination of Disordered Structures}}.
\newblock \emph{\bibinfo{journal}{{Molecular Simulation}}} \textbf{\bibinfo{volume}{1}}, \bibinfo{pages}{359--367} (\bibinfo{year}{1988}).

\bibitem{McGreevy2001RMCReview}
\bibinfo{author}{McGreevy, R.~L.}
\newblock \bibinfo{title}{{Reverse Monte Carlo Modelling}}.
\newblock \emph{\bibinfo{journal}{{Journal of Physics: Condensed Matter}}} \textbf{\bibinfo{volume}{13}}, \bibinfo{pages}{R877--R913} (\bibinfo{year}{2001}).

\bibitem{Cliffe2010RMC}
\bibinfo{author}{Cliffe, M.~J.} \& \bibinfo{author}{Goodwin, A.~L.}
\newblock \bibinfo{title}{{Reverse Monte Carlo Refinement of Amorphous Materials}}.
\newblock \emph{\bibinfo{journal}{{Journal of Physics: Condensed Matter}}} \textbf{\bibinfo{volume}{22}}, \bibinfo{pages}{404210} (\bibinfo{year}{2010}).

\bibitem{Soper2001EPSR}
\bibinfo{author}{Soper, A.~K.}
\newblock \bibinfo{title}{{Empirical Potential Monte Carlo Simulation of Fluid Structure}}.
\newblock \emph{\bibinfo{journal}{{Chemical Physics}}} \textbf{\bibinfo{volume}{258}}, \bibinfo{pages}{121--137} (\bibinfo{year}{2001}).

\bibitem{opletal2008hrmc}
\bibinfo{author}{Opletal, G.} \emph{et~al.}
\newblock \bibinfo{title}{{HRMC: Hybrid Reverse Monte Carlo method with silicon and carbon potentials}}.
\newblock \emph{\bibinfo{journal}{{Computer Physics Communications}}} \textbf{\bibinfo{volume}{178}}, \bibinfo{pages}{777--787} (\bibinfo{year}{2008}).

\bibitem{Kwon2024SpectroscopyGuided}
\bibinfo{author}{Kwon, H.} \emph{et~al.}
\newblock \bibinfo{title}{{Spectroscopy-Guided Discovery of Three-Dimensional Structures of Disordered Materials with Diffusion Models}}.
\newblock \emph{\bibinfo{journal}{{Machine Learning: Science and Technology}}} \textbf{\bibinfo{volume}{5}}, \bibinfo{pages}{045037} (\bibinfo{year}{2024}).

\bibitem{guo2025ab}
\bibinfo{author}{Guo, G.} \emph{et~al.}
\newblock \bibinfo{title}{{Ab initio structure solutions from nanocrystalline powder diffraction data via diffusion models}}.
\newblock \emph{\bibinfo{journal}{{Nature Materials}}} \textbf{\bibinfo{volume}{24}}, \bibinfo{pages}{1726--1734} (\bibinfo{year}{2025}).

\bibitem{anker2025autonomous}
\bibinfo{author}{Anker, A.~S.}, \bibinfo{author}{Gardner, J.~L.}, \bibinfo{author}{Rosset, L.~A.}, \bibinfo{author}{Goodwin, A.~L.} \& \bibinfo{author}{Deringer, V.~L.}
\newblock \bibinfo{title}{{Autonomous interpretation of atomistic scattering data}}.
\newblock \emph{\bibinfo{journal}{{arXiv:2510.05938}}}  (\bibinfo{year}{2025}).

\bibitem{yang2025generative}
\bibinfo{author}{Yang, K.} \& \bibinfo{author}{Schwalbe-Koda, D.}
\newblock \bibinfo{title}{{A generative diffusion model for amorphous materials}}.
\newblock \emph{\bibinfo{journal}{{npj Computational Materials}}} \textbf{\bibinfo{volume}{12}}, \bibinfo{pages}{29} (\bibinfo{year}{2026}).

\bibitem{song2021scorebased}
\bibinfo{author}{Song, Y.} \emph{et~al.}
\newblock \bibinfo{title}{{Score-Based Generative Modeling through Stochastic Differential Equations}}.
\newblock In \emph{\bibinfo{booktitle}{International Conference on Learning Representations}} (\bibinfo{year}{2021}).
\newblock \urlprefix\url{https://openreview.net/forum?id=PxTIG12RRHS}.

\bibitem{Elliott1991Amorphous}
\bibinfo{author}{Elliott, S.~R.}
\newblock \emph{\bibinfo{title}{{Physics of Amorphous Materials}}} (\bibinfo{publisher}{Longman Scientific}, \bibinfo{year}{1991}).

\bibitem{Zallen1983AmorphousStructure}
\bibinfo{author}{Zallen, R.}
\newblock \emph{\bibinfo{title}{{The Physics of Amorphous Solids}}} (\bibinfo{publisher}{Wiley}, \bibinfo{year}{1983}).

\bibitem{biswas2004reverse}
\bibinfo{author}{Biswas, P.}, \bibinfo{author}{Atta-Fynn, R.} \& \bibinfo{author}{Drabold, D.}
\newblock \bibinfo{title}{{Reverse Monte Carlo modeling of amorphous silicon}}.
\newblock \emph{\bibinfo{journal}{{Physical Review B}}} \textbf{\bibinfo{volume}{69}}, \bibinfo{pages}{195207} (\bibinfo{year}{2004}).

\bibitem{cliffe2017structural}
\bibinfo{author}{Cliffe, M.~J.} \emph{et~al.}
\newblock \bibinfo{title}{Structural simplicity as a restraint on the structure of amorphous silicon}.
\newblock \emph{\bibinfo{journal}{Physical Review B}} \textbf{\bibinfo{volume}{95}}, \bibinfo{pages}{224108} (\bibinfo{year}{2017}).

\bibitem{liu2023reliability}
\bibinfo{author}{Liu, C.}, \bibinfo{author}{Zhang, Z.}, \bibinfo{author}{Ding, J.} \& \bibinfo{author}{Ma, E.}
\newblock \bibinfo{title}{{On the reliability of using reverse Monte Carlo simulations to construct the atomic structure model of metallic glasses}}.
\newblock \emph{\bibinfo{journal}{{Scripta Materialia}}} \textbf{\bibinfo{volume}{225}}, \bibinfo{pages}{115159} (\bibinfo{year}{2023}).

\bibitem{carbone2020machine}
\bibinfo{author}{Carbone, M.~R.}, \bibinfo{author}{Topsakal, M.}, \bibinfo{author}{Lu, D.} \& \bibinfo{author}{Yoo, S.}
\newblock \bibinfo{title}{{Machine-learning X-ray absorption spectra to quantitative accuracy}}.
\newblock \emph{\bibinfo{journal}{{Physical Review Letters}}} \textbf{\bibinfo{volume}{124}}, \bibinfo{pages}{156401} (\bibinfo{year}{2020}).

\bibitem{treacy1998paracrystallites}
\bibinfo{author}{Treacy, M.}, \bibinfo{author}{Gibson, J.} \& \bibinfo{author}{Keblinski, P.}
\newblock \bibinfo{title}{{Paracrystallites found in evaporated amorphous tetrahedral semiconductors}}.
\newblock \emph{\bibinfo{journal}{{Journal of Non-Crystalline Solids}}} \textbf{\bibinfo{volume}{231}}, \bibinfo{pages}{99--110} (\bibinfo{year}{1998}).

\bibitem{treacy2012local}
\bibinfo{author}{Treacy, M.} \& \bibinfo{author}{Borisenko, K.}
\newblock \bibinfo{title}{{The local structure of amorphous silicon}}.
\newblock \emph{\bibinfo{journal}{{Science}}} \textbf{\bibinfo{volume}{335}}, \bibinfo{pages}{950--953} (\bibinfo{year}{2012}).

\bibitem{roorda2012comment}
\bibinfo{author}{Roorda, S.} \& \bibinfo{author}{Lewis, L.~J.}
\newblock \bibinfo{title}{{Comment on ``the Local Structure of Amorphous Silicon''}}.
\newblock \emph{\bibinfo{journal}{{Science}}} \textbf{\bibinfo{volume}{338}}, \bibinfo{pages}{1539--1539} (\bibinfo{year}{2012}).

\bibitem{lewis2022fifty}
\bibinfo{author}{Lewis, L.~J.}
\newblock \bibinfo{title}{{Fifty years of amorphous silicon models: the end of the story?}}
\newblock \emph{\bibinfo{journal}{{Journal of Non-Crystalline Solids}}} \textbf{\bibinfo{volume}{580}}, \bibinfo{pages}{121383} (\bibinfo{year}{2022}).

\bibitem{rosset2025signatures}
\bibinfo{author}{Rosset, L.~A.}, \bibinfo{author}{Drabold, D.~A.} \& \bibinfo{author}{Deringer, V.~L.}
\newblock \bibinfo{title}{{Signatures of paracrystallinity in amorphous silicon from machine-learning-driven molecular dynamics}}.
\newblock \emph{\bibinfo{journal}{{Nature Communications}}} \textbf{\bibinfo{volume}{16}}, \bibinfo{pages}{2360} (\bibinfo{year}{2025}).

\bibitem{laaziri1999high}
\bibinfo{author}{Laaziri, K.} \emph{et~al.}
\newblock \bibinfo{title}{{High resolution radial distribution function of pure amorphous silicon}}.
\newblock \emph{\bibinfo{journal}{{Physical Review Letters}}} \textbf{\bibinfo{volume}{82}}, \bibinfo{pages}{3460} (\bibinfo{year}{1999}).

\bibitem{borisenko2012medium}
\bibinfo{author}{Borisenko, K.~B.} \emph{et~al.}
\newblock \bibinfo{title}{{Medium-range order in amorphous silicon investigated by constrained structural relaxation of two-body and four-body electron diffraction data}}.
\newblock \emph{\bibinfo{journal}{{Acta Materialia}}} \textbf{\bibinfo{volume}{60}}, \bibinfo{pages}{359--375} (\bibinfo{year}{2012}).

\bibitem{henry2020Liquid}
\bibinfo{author}{Henry, L.} \emph{et~al.}
\newblock \bibinfo{title}{{Liquid--Liquid Transition and Critical Point in Sulfur}}.
\newblock \emph{\bibinfo{journal}{{Nature}}} \textbf{\bibinfo{volume}{584}}, \bibinfo{pages}{382--386} (\bibinfo{year}{2020}).

\bibitem{tanaka2020Liquid}
\bibinfo{author}{Tanaka, H.}
\newblock \bibinfo{title}{{Liquid--Liquid Transition and Polyamorphism}}.
\newblock \emph{\bibinfo{journal}{The Journal of Chemical Physics}} \textbf{\bibinfo{volume}{153}} (\bibinfo{year}{2020}).

\bibitem{schwalbekoda2025information}
\bibinfo{author}{Schwalbe-Koda, D.}, \bibinfo{author}{Hamel, S.}, \bibinfo{author}{Sadigh, B.}, \bibinfo{author}{Zhou, F.} \& \bibinfo{author}{Lordi, V.}
\newblock \bibinfo{title}{{Model-free estimation of completeness, uncertainties, and outliers in atomistic machine learning using information theory}}.
\newblock \emph{\bibinfo{journal}{{Nature Communications}}} \textbf{\bibinfo{volume}{16}}, \bibinfo{pages}{4014} (\bibinfo{year}{2025}).

\bibitem{rosu2023medium}
\bibinfo{author}{Rosu-Finsen, A.} \emph{et~al.}
\newblock \bibinfo{title}{{Medium-density amorphous ice}}.
\newblock \emph{\bibinfo{journal}{{Science}}} \textbf{\bibinfo{volume}{379}}, \bibinfo{pages}{474--478} (\bibinfo{year}{2023}).

\bibitem{almeidaribeiro2024medium}
\bibinfo{author}{de~Almeida~Ribeiro, I.} \emph{et~al.}
\newblock \bibinfo{title}{{Medium-density amorphous ice unveils shear rate as a new dimension in water’s phase diagram}}.
\newblock \emph{\bibinfo{journal}{{Proceedings of the National Academy of Sciences}}} \textbf{\bibinfo{volume}{121}}, \bibinfo{pages}{e2414444121} (\bibinfo{year}{2024}).

\bibitem{pfaff2020learning}
\bibinfo{author}{Pfaff, T.}, \bibinfo{author}{Fortunato, M.}, \bibinfo{author}{Sanchez-Gonzalez, A.} \& \bibinfo{author}{Battaglia, P.}
\newblock \bibinfo{title}{Learning mesh-based simulation with graph networks}.
\newblock In \emph{\bibinfo{booktitle}{International Conference on Learning Representations}} (\bibinfo{year}{2020}).

\bibitem{Tancik2020FourierFeatures}
\bibinfo{author}{Tancik, M.} \emph{et~al.}
\newblock \bibinfo{title}{{Fourier Features Let Networks Learn High Frequency Functions in Low Dimensional Domains}}.
\newblock In \emph{\bibinfo{booktitle}{NeurIPS}} (\bibinfo{year}{2020}).

\bibitem{Kingma2015Adam}
\bibinfo{author}{Kingma, D.~P.} \& \bibinfo{author}{Ba, J.}
\newblock \bibinfo{title}{{Adam: A Method for Stochastic Optimization}}.
\newblock \emph{\bibinfo{journal}{{International Conference on Learning Representations (ICLR)}}}  (\bibinfo{year}{2015}).

\bibitem{Waasmaier1995}
\bibinfo{author}{Waasmaier, D.} \& \bibinfo{author}{Kirfel, A.}
\newblock \bibinfo{title}{New analytical scattering-factor functions for free atoms and ions}.
\newblock \emph{\bibinfo{journal}{Acta Crystallographica Section A}} \textbf{\bibinfo{volume}{51}}, \bibinfo{pages}{416--431} (\bibinfo{year}{1995}).

\bibitem{Egami2003Underneath}
\bibinfo{author}{Egami, T.} \& \bibinfo{author}{Billinge, S. J.~L.}
\newblock \emph{\bibinfo{title}{{Underneath the Bragg Peaks: Structural Analysis of Complex Materials}}} (\bibinfo{publisher}{Pergamon}, \bibinfo{year}{2003}).

\bibitem{Sears1992}
\bibinfo{author}{Sears, V.~F.}
\newblock \bibinfo{title}{Neutron scattering lengths and cross sections}.
\newblock \emph{\bibinfo{journal}{Neutron News}} \textbf{\bibinfo{volume}{3}}, \bibinfo{pages}{26--37} (\bibinfo{year}{1992}).

\bibitem{Rehr2010FEFF9}
\bibinfo{author}{Rehr, J.~J.}, \bibinfo{author}{Kas, J.~J.}, \bibinfo{author}{Vila, F.~D.}, \bibinfo{author}{Prange, M.~P.} \& \bibinfo{author}{Jorissen, K.}
\newblock \bibinfo{title}{{Parameter-free calculations of X-ray spectra with FEFF9}}.
\newblock \emph{\bibinfo{journal}{{Physical Chemistry Chemical Physics}}} \textbf{\bibinfo{volume}{12}}, \bibinfo{pages}{5503--5513} (\bibinfo{year}{2010}).

\bibitem{hsu2024score}
\bibinfo{author}{Hsu, T.} \emph{et~al.}
\newblock \bibinfo{title}{Score-based denoising for atomic structure identification}.
\newblock \emph{\bibinfo{journal}{npj Computational Materials}} \textbf{\bibinfo{volume}{10}}, \bibinfo{pages}{155} (\bibinfo{year}{2024}).

\bibitem{thompson2022lammps}
\bibinfo{author}{Thompson, A.~P.} \emph{et~al.}
\newblock \bibinfo{title}{{LAMMPS} - a flexible simulation tool for particle-based materials modeling at the atomic, meso, and continuum scales}.
\newblock \emph{\bibinfo{journal}{Computer Physics Communications}} \textbf{\bibinfo{volume}{271}}, \bibinfo{pages}{108171} (\bibinfo{year}{2022}).

\bibitem{Tersoff1988_1}
\bibinfo{author}{Tersoff, J.}
\newblock \bibinfo{title}{{Empirical Interatomic Potential for Silicon with Improved Elastic Properties}}.
\newblock \emph{\bibinfo{journal}{{Physical Review B}}} \textbf{\bibinfo{volume}{38}}, \bibinfo{pages}{9902--9905} (\bibinfo{year}{1988}).

\bibitem{Tersoff1988_2}
\bibinfo{author}{Tersoff, J.}
\newblock \bibinfo{title}{{New Empirical Approach for the Structure and Energy of Covalent Systems}}.
\newblock \emph{\bibinfo{journal}{{Physical Review B}}} \textbf{\bibinfo{volume}{37}}, \bibinfo{pages}{6991--7000} (\bibinfo{year}{1988}).

\bibitem{Munetoh2007SiO}
\bibinfo{author}{Munetoh, S.}, \bibinfo{author}{Motooka, T.}, \bibinfo{author}{Moriguchi, K.} \& \bibinfo{author}{Shintani, A.}
\newblock \bibinfo{title}{{Interatomic potential for Si--O systems using Tersoff parameterization}}.
\newblock \emph{\bibinfo{journal}{{Computational Materials Science}}} \textbf{\bibinfo{volume}{39}}, \bibinfo{pages}{334--339} (\bibinfo{year}{2007}).

\bibitem{Mendelev2009CuZr}
\bibinfo{author}{Mendelev, M.~I.} \emph{et~al.}
\newblock \bibinfo{title}{{Development of suitable interatomic potentials for simulation of liquid and amorphous Cu--Zr alloys}}.
\newblock \emph{\bibinfo{journal}{{Philosophical Magazine}}} \textbf{\bibinfo{volume}{89}}, \bibinfo{pages}{967--987} (\bibinfo{year}{2009}).

\bibitem{Cheng2009BMG}
\bibinfo{author}{Cheng, Y.~Q.}, \bibinfo{author}{Ma, E.} \& \bibinfo{author}{Sheng, H.~W.}
\newblock \bibinfo{title}{{Atomic-Level Structure in Multicomponent Bulk Metallic Glass}}.
\newblock \emph{\bibinfo{journal}{{Physical Review Letters}}} \textbf{\bibinfo{volume}{102}}, \bibinfo{pages}{245501} (\bibinfo{year}{2009}).

\bibitem{Nose1984}
\bibinfo{author}{Nos{\'e}, S.}
\newblock \bibinfo{title}{{A Unified Formulation of the Constant Temperature Molecular Dynamics Methods}}.
\newblock \emph{\bibinfo{journal}{{The Journal of Chemical Physics}}} \textbf{\bibinfo{volume}{81}}, \bibinfo{pages}{511--519} (\bibinfo{year}{1984}).

\bibitem{Hoover1985}
\bibinfo{author}{Hoover, W.~G.}
\newblock \bibinfo{title}{{Canonical Dynamics: Equilibrium Phase-Space Distributions}}.
\newblock \emph{\bibinfo{journal}{{Physical Review A}}} \textbf{\bibinfo{volume}{31}}, \bibinfo{pages}{1695--1697} (\bibinfo{year}{1985}).

\bibitem{aoun2016fullrmc}
\bibinfo{author}{Aoun, B.}
\newblock \bibinfo{title}{{FullRMC, a rigid body reverse Monte Carlo modeling package enabled with machine learning and artificial intelligence}}.
\newblock \emph{\bibinfo{journal}{{Journal of Computational Chemistry}}} \textbf{\bibinfo{volume}{37}}, \bibinfo{pages}{1102--1111} (\bibinfo{year}{2016}).

\end{thebibliography}

\clearpage
\begin{thebibliography}{10}
\expandafter\ifx\csname url\endcsname\relax
  \def\url#1{\texttt{#1}}\fi
\expandafter\ifx\csname urlprefix\endcsname\relax\def\urlprefix{URL }\fi
\providecommand{\bibinfo}[2]{#2}
\providecommand{\eprint}[2][]{\url{#2}}

\bibitem{Rehr2010FEFF9-SI}
\bibinfo{author}{Rehr, J.~J.}, \bibinfo{author}{Kas, J.~J.}, \bibinfo{author}{Vila, F.~D.}, \bibinfo{author}{Prange, M.~P.} \& \bibinfo{author}{Jorissen, K.}
\newblock \bibinfo{title}{{Parameter-free calculations of X-ray spectra with FEFF9}}.
\newblock \emph{\bibinfo{journal}{{Physical Chemistry Chemical Physics}}} \textbf{\bibinfo{volume}{12}}, \bibinfo{pages}{5503--5513} (\bibinfo{year}{2010}).

\bibitem{schwalbekoda2025information-SI}
\bibinfo{author}{Schwalbe-Koda, D.}, \bibinfo{author}{Hamel, S.}, \bibinfo{author}{Sadigh, B.}, \bibinfo{author}{Zhou, F.} \& \bibinfo{author}{Lordi, V.}
\newblock \bibinfo{title}{{Model-free estimation of completeness, uncertainties, and outliers in atomistic machine learning using information theory}}.
\newblock \emph{\bibinfo{journal}{{Nature Communications}}} \textbf{\bibinfo{volume}{16}}, \bibinfo{pages}{4014} (\bibinfo{year}{2025}).

\bibitem{yang2025generative-SI}
\bibinfo{author}{Yang, K.} \& \bibinfo{author}{Schwalbe-Koda, D.}
\newblock \bibinfo{title}{{A generative diffusion model for amorphous materials}}.
\newblock \emph{\bibinfo{journal}{{npj Computational Materials}}} \textbf{\bibinfo{volume}{12}}, \bibinfo{pages}{29} (\bibinfo{year}{2026}).

\bibitem{song2021scorebased-SI}
\bibinfo{author}{Song, Y.} \emph{et~al.}
\newblock \bibinfo{title}{{Score-Based Generative Modeling through Stochastic Differential Equations}}.
\newblock In \emph{\bibinfo{booktitle}{International Conference on Learning Representations}} (\bibinfo{year}{2021}).
\newblock \urlprefix\url{https://openreview.net/forum?id=PxTIG12RRHS}.

\bibitem{Kwon2024SpectroscopyGuided-SI}
\bibinfo{author}{Kwon, H.} \emph{et~al.}
\newblock \bibinfo{title}{{Spectroscopy-Guided Discovery of Three-Dimensional Structures of Disordered Materials with Diffusion Models}}.
\newblock \emph{\bibinfo{journal}{{Machine Learning: Science and Technology}}} \textbf{\bibinfo{volume}{5}}, \bibinfo{pages}{045037} (\bibinfo{year}{2024}).

\bibitem{graphite_llnl-SI}
\bibinfo{author}{Hsu, T.}
\newblock \bibinfo{title}{{Graphite: Graph Neural Network Models for Atomic Structures}}.
\newblock \bibinfo{howpublished}{\url{https://github.com/LLNL/graphite}} (\bibinfo{year}{2022}).
\newblock \bibinfo{note}{Accessed: 2026-03-17}.

\bibitem{pfaff2020learning-SI}
\bibinfo{author}{Pfaff, T.}, \bibinfo{author}{Fortunato, M.}, \bibinfo{author}{Sanchez-Gonzalez, A.} \& \bibinfo{author}{Battaglia, P.}
\newblock \bibinfo{title}{Learning mesh-based simulation with graph networks}.
\newblock In \emph{\bibinfo{booktitle}{International Conference on Learning Representations}} (\bibinfo{year}{2020}).

\bibitem{Kingma2015Adam-SI}
\bibinfo{author}{Kingma, D.~P.} \& \bibinfo{author}{Ba, J.}
\newblock \bibinfo{title}{{Adam: A Method for Stochastic Optimization}}.
\newblock \emph{\bibinfo{journal}{{International Conference on Learning Representations (ICLR)}}}  (\bibinfo{year}{2015}).

\bibitem{Plimpton1995LAMMPS-SI}
\bibinfo{author}{Plimpton, S.}
\newblock \bibinfo{title}{{Fast Parallel Algorithms for Short-Range Molecular Dynamics}}.
\newblock \emph{\bibinfo{journal}{{Journal of Computational Physics}}} \textbf{\bibinfo{volume}{117}}, \bibinfo{pages}{1--19} (\bibinfo{year}{1995}).

\bibitem{thompson2022lammps-SI}
\bibinfo{author}{Thompson, A.~P.} \emph{et~al.}
\newblock \bibinfo{title}{Lammps - a flexible simulation tool for particle-based materials modeling at the atomic, meso, and continuum scales}.
\newblock \emph{\bibinfo{journal}{Computer Physics Communications}} \textbf{\bibinfo{volume}{271}}, \bibinfo{pages}{108171} (\bibinfo{year}{2022}).

\bibitem{Tersoff1988_1-SI}
\bibinfo{author}{Tersoff, J.}
\newblock \bibinfo{title}{{Empirical Interatomic Potential for Silicon with Improved Elastic Properties}}.
\newblock \emph{\bibinfo{journal}{{Physical Review B}}} \textbf{\bibinfo{volume}{38}}, \bibinfo{pages}{9902--9905} (\bibinfo{year}{1988}).

\bibitem{Tersoff1988_2-SI}
\bibinfo{author}{Tersoff, J.}
\newblock \bibinfo{title}{{New Empirical Approach for the Structure and Energy of Covalent Systems}}.
\newblock \emph{\bibinfo{journal}{{Physical Review B}}} \textbf{\bibinfo{volume}{37}}, \bibinfo{pages}{6991--7000} (\bibinfo{year}{1988}).

\bibitem{Munetoh2007SiO-SI}
\bibinfo{author}{Munetoh, S.}, \bibinfo{author}{Motooka, T.}, \bibinfo{author}{Moriguchi, K.} \& \bibinfo{author}{Shintani, A.}
\newblock \bibinfo{title}{{Interatomic potential for Si--O systems using Tersoff parameterization}}.
\newblock \emph{\bibinfo{journal}{{Computational Materials Science}}} \textbf{\bibinfo{volume}{39}}, \bibinfo{pages}{334--339} (\bibinfo{year}{2007}).

\bibitem{Mendelev2009CuZr-SI}
\bibinfo{author}{Mendelev, M.~I.} \emph{et~al.}
\newblock \bibinfo{title}{{Development of suitable interatomic potentials for simulation of liquid and amorphous Cu--Zr alloys}}.
\newblock \emph{\bibinfo{journal}{{Philosophical Magazine}}} \textbf{\bibinfo{volume}{89}}, \bibinfo{pages}{967--987} (\bibinfo{year}{2009}).

\bibitem{Cheng2009BMG-SI}
\bibinfo{author}{Cheng, Y.~Q.}, \bibinfo{author}{Ma, E.} \& \bibinfo{author}{Sheng, H.~W.}
\newblock \bibinfo{title}{{Atomic-Level Structure in Multicomponent Bulk Metallic Glass}}.
\newblock \emph{\bibinfo{journal}{{Physical Review Letters}}} \textbf{\bibinfo{volume}{102}}, \bibinfo{pages}{245501} (\bibinfo{year}{2009}).

\bibitem{Nose1984-SI}
\bibinfo{author}{Nos{\'e}, S.}
\newblock \bibinfo{title}{{A Unified Formulation of the Constant Temperature Molecular Dynamics Methods}}.
\newblock \emph{\bibinfo{journal}{{The Journal of Chemical Physics}}} \textbf{\bibinfo{volume}{81}}, \bibinfo{pages}{511--519} (\bibinfo{year}{1984}).

\bibitem{Hoover1985-SI}
\bibinfo{author}{Hoover, W.~G.}
\newblock \bibinfo{title}{{Canonical Dynamics: Equilibrium Phase-Space Distributions}}.
\newblock \emph{\bibinfo{journal}{{Physical Review A}}} \textbf{\bibinfo{volume}{31}}, \bibinfo{pages}{1695--1697} (\bibinfo{year}{1985}).

\bibitem{Trott2022Kokkos}
\bibinfo{author}{Trott, C.~R.} \emph{et~al.}
\newblock \bibinfo{title}{Kokkos 3: Programming model extensions for the exascale era}.
\newblock \emph{\bibinfo{journal}{IEEE Transactions on Parallel and Distributed Systems}} \textbf{\bibinfo{volume}{33}}, \bibinfo{pages}{805--817} (\bibinfo{year}{2022}).

\bibitem{Tancik2020FourierFeatures-SI}
\bibinfo{author}{Tancik, M.} \emph{et~al.}
\newblock \bibinfo{title}{{Fourier Features Let Networks Learn High Frequency Functions in Low Dimensional Domains}}.
\newblock In \emph{\bibinfo{booktitle}{NeurIPS}} (\bibinfo{year}{2020}).

\bibitem{Trizio2023DebyeCalculator}
\bibinfo{author}{Trizio, E.} \emph{et~al.}
\newblock \bibinfo{title}{{DebyeCalculator: A fast and efficient implementation of the Debye scattering equation for total scattering analysis}}.
\newblock \emph{\bibinfo{journal}{Journal of Applied Crystallography}} \textbf{\bibinfo{volume}{56}}, \bibinfo{pages}{1430--1438} (\bibinfo{year}{2023}).

\bibitem{aoun2016fullrmc-SI}
\bibinfo{author}{Aoun, B.}
\newblock \bibinfo{title}{{FullRMC, a rigid body reverse Monte Carlo modeling package enabled with machine learning and artificial intelligence}}.
\newblock \emph{\bibinfo{journal}{{Journal of Computational Chemistry}}} \textbf{\bibinfo{volume}{37}}, \bibinfo{pages}{1102--1111} (\bibinfo{year}{2016}).

\bibitem{rosset2025signatures-SI}
\bibinfo{author}{Rosset, L.~A.}, \bibinfo{author}{Drabold, D.~A.} \& \bibinfo{author}{Deringer, V.~L.}
\newblock \bibinfo{title}{{Signatures of paracrystallinity in amorphous silicon from machine-learning-driven molecular dynamics}}.
\newblock \emph{\bibinfo{journal}{{Nature Communications}}} \textbf{\bibinfo{volume}{16}}, \bibinfo{pages}{2360} (\bibinfo{year}{2025}).

\bibitem{Larsen2016PTM}
\bibinfo{author}{Larsen, P.~M.}, \bibinfo{author}{Schmidt, S.} \& \bibinfo{author}{Schi{\o}tz, J.}
\newblock \bibinfo{title}{Robust structural identification via polyhedral template matching}.
\newblock \emph{\bibinfo{journal}{Modelling and Simulation in Materials Science and Engineering}} \textbf{\bibinfo{volume}{24}}, \bibinfo{pages}{055007} (\bibinfo{year}{2016}).

\bibitem{Stukowski2010OVITO}
\bibinfo{author}{Stukowski, A.}
\newblock \bibinfo{title}{Visualization and analysis of atomistic simulation data with ovito—the open visualization tool}.
\newblock \emph{\bibinfo{journal}{Modelling and Simulation in Materials Science and Engineering}} \textbf{\bibinfo{volume}{18}}, \bibinfo{pages}{015012} (\bibinfo{year}{2010}).

\bibitem{SI-henry2020Liquid}
\bibinfo{author}{Henry, L.} \emph{et~al.}
\newblock \bibinfo{title}{{Liquid--Liquid Transition and Critical Point in Sulfur}}.
\newblock \emph{\bibinfo{journal}{{Nature}}} \textbf{\bibinfo{volume}{584}}, \bibinfo{pages}{382--386} (\bibinfo{year}{2020}).

\bibitem{rosu2023medium-SI}
\bibinfo{author}{Rosu-Finsen, A.} \emph{et~al.}
\newblock \bibinfo{title}{{Medium-density amorphous ice}}.
\newblock \emph{\bibinfo{journal}{{Science}}} \textbf{\bibinfo{volume}{379}}, \bibinfo{pages}{474--478} (\bibinfo{year}{2023}).

\bibitem{Shanks2024Bayesian}
\bibinfo{author}{Shanks, B.~L.}, \bibinfo{author}{Sullivan, H.~W.} \& \bibinfo{author}{Hoepfner, M.~P.}
\newblock \bibinfo{title}{Bayesian analysis reveals the key to extracting pair potentials from neutron scattering data}.
\newblock \emph{\bibinfo{journal}{The Journal of Physical Chemistry Letters}} \textbf{\bibinfo{volume}{15}}, \bibinfo{pages}{12608--12618} (\bibinfo{year}{2024}).

\bibitem{kim_sevennet_mf_2024}
\bibinfo{author}{Kim, J.} \emph{et~al.}
\newblock \bibinfo{title}{Data-efficient multifidelity training for high-fidelity machine learning interatomic potentials}.
\newblock \emph{\bibinfo{journal}{Journal of the American Chemical Society}} \textbf{\bibinfo{volume}{147}}, \bibinfo{pages}{1042--1054} (\bibinfo{year}{2024}).

\bibitem{Egami2003Underneath-SI}
\bibinfo{author}{Egami, T.} \& \bibinfo{author}{Billinge, S. J.~L.}
\newblock \emph{\bibinfo{title}{{Underneath the Bragg Peaks: Structural Analysis of Complex Materials}}} (\bibinfo{publisher}{Pergamon}, \bibinfo{year}{2003}).

\bibitem{Rehr2000RMP}
\bibinfo{author}{Rehr, J.~J.} \& \bibinfo{author}{Albers, R.~C.}
\newblock \bibinfo{title}{{Theoretical Approaches to X-ray Absorption Fine Structure}}.
\newblock \emph{\bibinfo{journal}{{Reviews of Modern Physics}}} \textbf{\bibinfo{volume}{72}}, \bibinfo{pages}{621--654} (\bibinfo{year}{2000}).

\bibitem{SI_Billinge2007Science}
\bibinfo{author}{Billinge, S. J.~L.} \& \bibinfo{author}{Levin, I.}
\newblock \bibinfo{title}{{The Problem with Determining Atomic Structure at the Nanoscale}}.
\newblock \emph{\bibinfo{journal}{{Science}}} \textbf{\bibinfo{volume}{316}}, \bibinfo{pages}{561--565} (\bibinfo{year}{2007}).

\bibitem{Terban2021ChemRev}
\bibinfo{author}{Terban, M.~W.} \& \bibinfo{author}{Billinge, S. J.~L.}
\newblock \bibinfo{title}{{Structural Analysis of Molecular Materials Using the Pair Distribution Function}}.
\newblock \emph{\bibinfo{journal}{{Chemical Reviews}}} \textbf{\bibinfo{volume}{121}}, \bibinfo{pages}{1208--1272} (\bibinfo{year}{2021}).

\bibitem{SI_Koningsberger1988XAFS}
\bibinfo{author}{Koningsberger, D.~C.} \& \bibinfo{author}{Prins, R.}
\newblock \emph{\bibinfo{title}{{X-ray Absorption: Principles, Applications, Techniques of EXAFS, SEXAFS and XANES}}} (\bibinfo{publisher}{Wiley}, \bibinfo{year}{1988}).

\bibitem{SI_McGreevy2001RMCReview}
\bibinfo{author}{McGreevy, R.~L.}
\newblock \bibinfo{title}{{Reverse Monte Carlo Modelling}}.
\newblock \emph{\bibinfo{journal}{{Journal of Physics: Condensed Matter}}} \textbf{\bibinfo{volume}{13}}, \bibinfo{pages}{R877--R913} (\bibinfo{year}{2001}).

\bibitem{SI_Soper2001EPSR}
\bibinfo{author}{Soper, A.~K.}
\newblock \bibinfo{title}{{Empirical Potential Monte Carlo Simulation of Fluid Structure}}.
\newblock \emph{\bibinfo{journal}{{Chemical Physics}}} \textbf{\bibinfo{volume}{258}}, \bibinfo{pages}{121--137} (\bibinfo{year}{2001}).

\bibitem{cliffe2017structural-SI}
\bibinfo{author}{Cliffe, M.~J.} \emph{et~al.}
\newblock \bibinfo{title}{Structural simplicity as a restraint on the structure of amorphous silicon}.
\newblock \emph{\bibinfo{journal}{Physical Review B}} \textbf{\bibinfo{volume}{95}}, \bibinfo{pages}{224108} (\bibinfo{year}{2017}).

\bibitem{yang2024structure-SI}
\bibinfo{author}{Yang, M.}, \bibinfo{author}{Trizio, E.} \& \bibinfo{author}{Parrinello, M.}
\newblock \bibinfo{title}{{Structure and polymerization of liquid sulfur across the $\lambda$-transition}}.
\newblock \emph{\bibinfo{journal}{{Chemical Science}}} \textbf{\bibinfo{volume}{15}}, \bibinfo{pages}{3382--3392} (\bibinfo{year}{2024}).

\end{thebibliography}
\end{document}